\documentclass[letterpaper,twocolumn,10pt]{article}
\pdfoutput=1

\usepackage{usenix-2020-09}

\usepackage[english]{babel}
\usepackage{blindtext}

\usepackage{tikz}
\usepackage{amsfonts}
\usepackage{amsthm}
\usepackage{amsmath}
\usepackage{subfigure}
\usepackage{hyperref}
\usepackage{xspace}
\usepackage{url}
\usepackage{algorithm}
\usepackage{algpseudocode}
\usepackage{setspace}
\microtypecontext{spacing=nonfrench}

\newtheorem{proposition}{Proposition}
\newtheorem{theorem}{Theorem}
\newtheorem{assumption}{Assumption}
\newtheorem{definition}{Definition}
\newtheorem{lemma}{Lemma}
\newcommand{\cut}[1]{}

\usepackage[normalem]{ulem}
\usepackage{authblk}

\newcommand{\ignore}[1]{}

\newcommand{\red}[1]{{\color{red}{#1}}}

\newcommand{\sysname}{\emph{DOTE\xspace}}
\newcommand{\rs}{$\mathcal{R}$\xspace}
\newcommand{\rsm}{\mathcal{R}}

\newcommand{\rst}[1]{$\mathcal{R}^{(#1)}$}

\newcommand{\xref}[1]{\S\ref{#1}}

\newcommand{\obj}{\mathcal{L}}
\newcommand{\hor}{H}

\definecolor{cadmiumred}{rgb}{0.89, 0.0, 0.13}

\algdef{SE}[VARIABLES]{Variables}{EndVariables}
   {\algorithmicvariables}
   {\algorithmicend\ \algorithmicvariables}
\algnewcommand{\algorithmicvariables}{\textbf{global variables}}

\newcommand{\project}{Proj}
\DeclareMathOperator*{\argmin}{arg\,min}

\newcommand{\pwan}{{$\mbox{\sf PWAN}$}}
\newcommand{\pwandc}{${\mbox{\sf PWAN}}_{\mbox{\sf DC}}$}

\newenvironment{Itemize}%
{\begin{itemize}%
\setlength{\leftmargin}{1em}%
\setlength{\itemsep}{0in}%
\setlength{\topsep}{-.1in}%
\setlength{\partopsep}{-.1in}%
\setlength{\parsep}{-.1in}%
\setlength{\parskip}{0in}}%
{\end{itemize}}

\usepackage{multirow}

\begin{document}

\title{A Deep Learning Perspective on Network Routing}

\makeatletter
\renewcommand\AB@affilsepx{, \protect\Affilfont}
\makeatother
\author[1]{Yarin Perry}
\author[2]{Felipe Vieira Frujeri}
\author[1]{Chaim Hoch}
\author[2]{Srikanth Kandula}
\author[2]{Ishai Menache}
\author[1]{Michael Schapira}
\author[3]{Aviv Tamar}
\affil[1]{Hebrew University of Jerusalem}
\affil[2]{Microsoft Research}
\affil[3]{Technion}

\date{To appear at NSDI 2023}

\maketitle

\noindent{\bf Abstract--} Routing is, arguably, the most fundamental task in computer networking, and the most extensively studied one. A key challenge for routing in real-world environments is the need to contend with uncertainty about future traffic demands. We present a new approach to routing under demand uncertainty: tackling this challenge as stochastic optimization, and employing deep learning to learn complex patterns in traffic demands. We show that our method provably converges to the global optimum in well-studied theoretical models of multicommodity flow. We exemplify the practical usefulness of our approach by zooming in on the real-world challenge of traffic engineering (TE) on wide-area networks (WANs). Our extensive empirical evaluation on real-world traffic and network topologies establishes that our approach's TE quality almost matches that of an (infeasible) omniscient oracle, outperforming previously proposed approaches, and also substantially lowers runtimes.

\section{Introduction}

To meet the constant rise in traffic, service providers invest huge effort into traffic engineering (TE)---optimizing traffic flow across their backbone WANs~\cite{SWAN, B4, calendaring, zhang2017guaranteeing, FFC, SMORE, TeaVaR}, which interconnect their datacenters with each other and with external networks. The production state-of-the-art involves periodically solving a (logically centralized) optimization problem to determine how to best split traffic across network paths. Changes to TE configurations are realized using software-defined control of network hardware~\cite{SWAN, B4, FFC, BWE, EBB, TeaVaR}.

A key challenge for WAN TE is uncertainty regarding future traffic demands. The standard approach for contending with this is twofold. For time-sensitive traffic, providers measure application-specific usage data from switches~(e.g., using sampled netflow or ipfix counters) and attempt to \textit{predict} future usage. For bandwidth-hungry, scavenger-class traffic~\cite{SWAN}, providers deploy so called agents/shims in the OS of hosts from which traffic originates. These agents explicitly signal applications' traffic demands to ``brokers'' that, in turn, aggregate demands, relay them to the centralized optimizer, and enforce the resulting rate allocations~\cite{SWAN,B4}. 

Both of the above approaches for handling traffic uncertainty have drawbacks. Demand predictions can naturally be erroneous and, more importantly, there is an objective mismatch between the loss functions to predict future traffic demands (e.g., mean-squared-error, L1 norm error) and the \textit{end-to-end objective} of producing high-performance TE configurations. For example, mean-squared-error would weight error in {\em any demand} equally, yet errors on demands that are more problematic to carry on a given topology will exert a disproportionately large effect on TE quality. The other approach -- brokering and explicitly specifying demands -- entails nontrivial operational overheads, including changes to end-hosts and applications. This can increase the lag experienced by application requests (which is why this approach is used in practice only for bandwidth-hungry, scavenger-class traffic~\cite{SWAN}). 

The demand uncertainty challenge is further amplified for \textit{customer-facing traffic} (web, images, e-mails, videos, etc.), which constitutes a large and growing share of the total traffic traversing some providers' backbones. For such traffic, which originates in unmodified apps or clients, brokering in the host OS is not applicable. Moreover (see~\S\ref{subsec:traffic}), such traffic exhibits high variability and is difficult to predict accurately.

We explore a new design point for WAN TE: training a TE decision model on \textit{historical} data about traffic demands to \textit{directly} output high-quality TE configurations. We present the \textit{DOTE} (Direct Optimization for Traffic Engineering) TE framework. \textit{DOTE} applies \textit{stochastic optimization} to \textit{learn} how to map recently observed traffic demands (e.g., empirically-derived traffic demands from the last hour) to the next choice of TE configuration. Using \textit{DOTE}, providers need only \textit{passively} monitor traffic to/from datacenters and do not have to onboard applications onto brokers. Directly predicting TE outcomes that optimize TE performance also resolves the objective mismatch between demand prediction and TE performance, yielding TE outcomes that are more robust to traffic unpredictability. We show how \emph{DOTE} can scale to handle large WANs and real-world traffic by harnessing the expressiveness of deep learning.
 
We evaluate \emph{DOTE} both analytically and empirically. Our theoretical results establish that if the TE optimization objective satisfies desirable convexity/concavity properties, \emph{DOTE} \textit{provably} converges to the optimum. We prove that this is indeed the case for standard TE optimization objectives such as minimizing the maximum-link-utilization (MLU)~\cite{TExCP,mate,Cohen2003}, maximizing network throughput~\cite{NC,SWAN,B4,SMORE}, and maximizing concurrent-flow~\cite{karakostas,TeaVaR}.

Our empirical evaluation compares \textit{DOTE}, in terms of both quality and runtimes, to TE with explicit demand estimates from end-hosts, demand-prediction-based TE, demand-oblivious TE, deep-reinforcement-learning-based TE, and more. Evaluating data-driven TE schemes like \emph{DOTE} requires substantial empirical data regarding traffic conditions for both training and performance analysis. We conduct a large-scale empirical study using both publicly available datasets and historical data from Microsoft's private WAN. These datasets span months of traffic demands at few-minutes granularity, amounting to tens of thousands of demand snapshots. Our evaluation covers small ($10$s of nodes) and large ($100$s of nodes) WANs, different types of traffic (including inter-datacenter and customer-facing), and different TE optimization objectives. To facilitate reproducibility, our code is available at~\cite{repo_code}. 

Our evaluation results show that:
\begin{Itemize} 
\item {\bf \emph{DOTE} achieves TE quality almost matching that of an \textit{infeasible} oracle with \textit{perfect knowledge} of future demands}. Across all evaluated network topologies, traffic traces, and considered TE objectives, \emph{DOTE} compares favorably to all other evaluated TE schemes. We also demonstrate \emph{DOTE}'s robustness to changes in traffic conditions and to network failures. 

\vspace{0.05in}

\item {\bf By invoking a DNN for the online computation of TE configurations, \emph{DOTE} achieves runtimes 1-2 orders of magnitude faster} than solving a linear program (LP), even for large WANs, matching the gains from recent proposals for fast (approximate) LP optimizations~\cite{NC,POP}. Our approach thus also holds promise for expediting decision making for TE.
\end{Itemize}

We view our investigation of direct optimization for WAN TE as a first step and discuss current limitations of our approach that we hope future research can address.

\noindent{\bf This work does not raise any ethical concerns.\footnote{In particular, the measured traffic demands, used in our evaluation, are aggregate counters between pairs of datacenters at the granularity of minutes (or coarser).  They do not contain user IP addresses or packet contents.}}

\section{Motivation and Key Insights}\label{s:prelim}

\subsection{Inter-DC \emph{vs.} Customer-Facing Traffic}\label{subsec:traffic}
Enterprise WANs carry traffic between the provider's own datacenters (e.g., geo-replication of datasets, newly computed search indices) as well as traffic traversing the backbone towards/from customers (e.g., web traffic, videos).

\begin{figure}[t!]
\centering
\subfigure[Inter-data-center traffic]{ 
\label{fig:std:bb1}\includegraphics[width=0.48\linewidth]{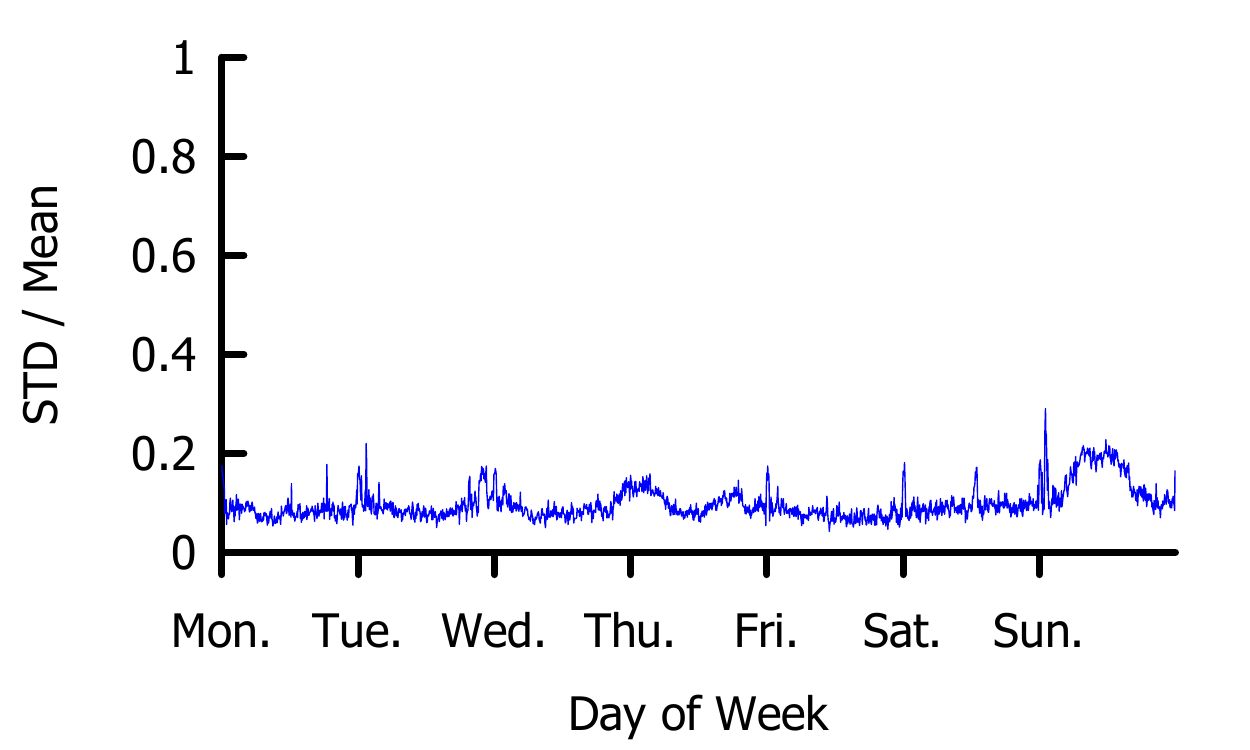}}
\hfill
\subfigure[Customer-facing traffic]{ 
\label{fig:std:bb2}\includegraphics[width=0.48\linewidth]{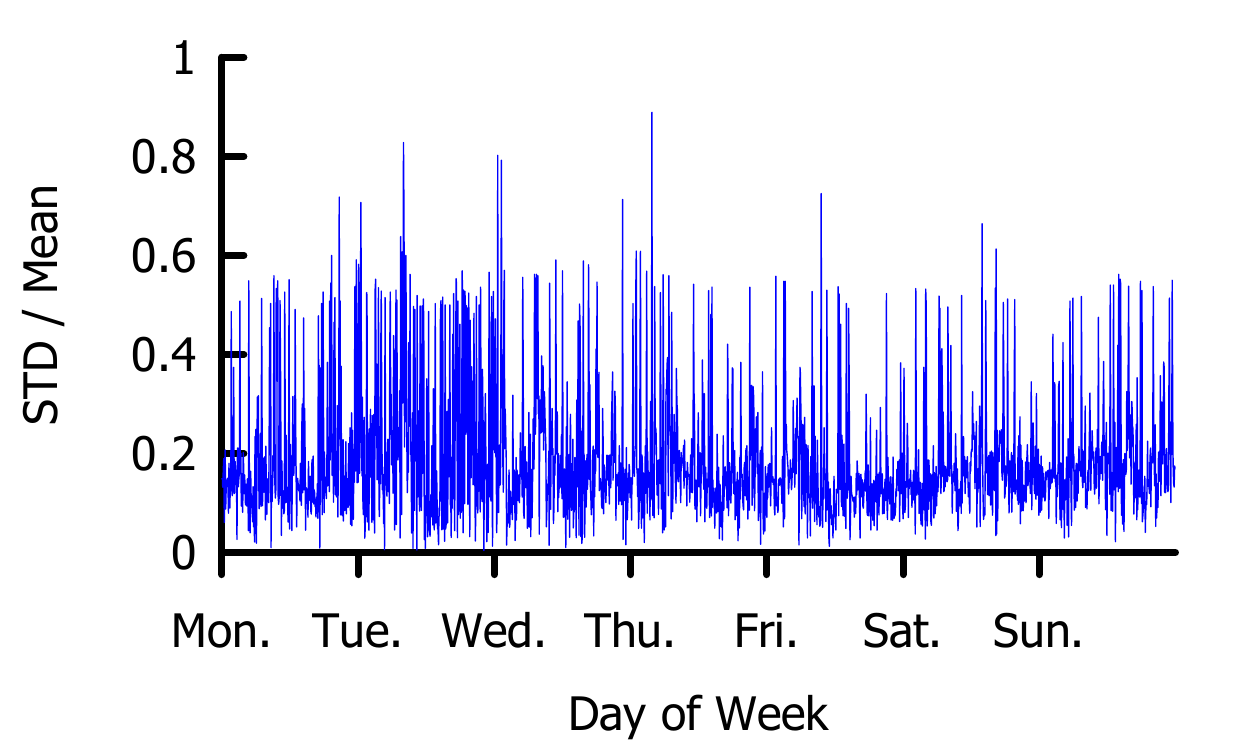}}
\vspace{-0.1in}
\caption{Variability in traffic demands for inter-datacenter traffic and customer-facing traffic across different weeks.}
\label{fig:std_demand2}
\end{figure}

To motivate our direct optimization approach, we present analyses of traffic on Microsoft's production WAN. \autoref{fig:std:bb1} plots the standard deviation in \textit{inter-datacenter traffic} demands, normalized by the mean, across $11$ consecutive weeks, for the pair of datacenters with the highest average demand. Demands are collected at $5$-minute granularity. Similarly, \autoref{fig:std:bb2} plots the normalized standard deviation in \textit{custo\-mer-facing traffic} demands over $4$ consecutive weeks for the pair of nodes with the highest average demand. Observe the substantial difference; in the inter-datacenter traffic trace, demands are \textit{significantly} less variable. 

High variability in customer-facing traffic demands can accrue from different sources, e.g., (1) flash-crowds that may cause a surge in search requests, e-mail volume, etc., (2) congestion on the WAN's peering links with ISP networks, and (3) route changes and outages that cause traffic to ingress or egress the WAN at different sites. We have observed that customer-facing demands can exceed $100\times$ the average value for extended stretches of time. Thus, customer-facing traffic is harder to accurately predict than inter-datacenter traffic. See \autoref{fig:bb1_predict}--\autoref{fig:bb2_predict} in the appendix for differences in demand-prediction accuracy between the above discussed two traffic traces. 

To summarize: for customer-facing traffic, which is a large and growing share of overall WAN traffic, not only is direct inference of traffic demands by the host OS infeasible, but accurate demand prediction also appears elusive. We seek a method that can achieve nearly optimal TE outcomes even for the unpredictable traffic demands.

\subsection{Demand Prediction \emph{vs.} Direct Optimization}

\begin{figure*}[htbp]
\subfigure[Network topology]{
  \includegraphics[width=.22\linewidth]{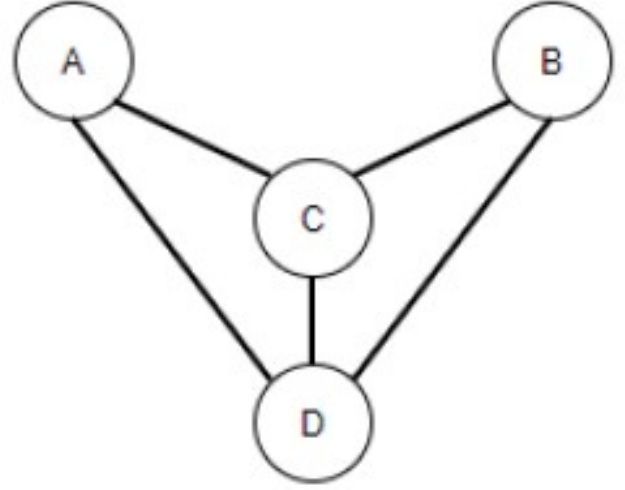}
  \label{fig:small_example_topology}
  }
\subfigure[Induced splitting ratios for a demand-predictor]{
  \includegraphics[width=.22\linewidth]{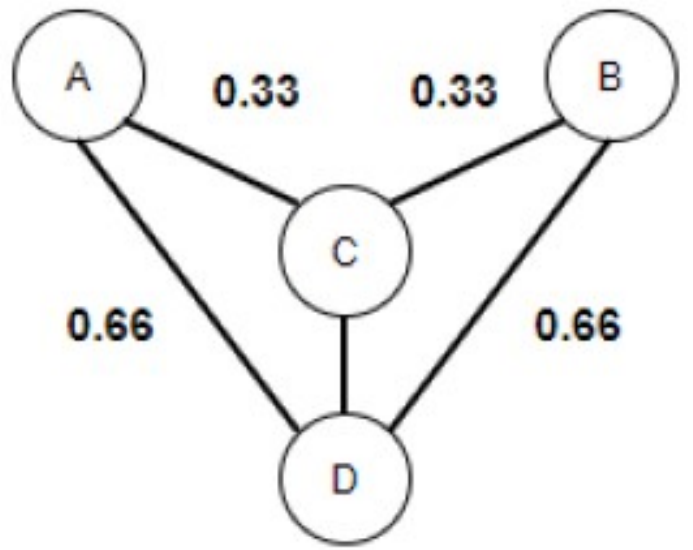}
  \label{fig:small_example_opt_pred}
  }
\subfigure[Optimal splitting ratios]{
  \includegraphics[width=.22\linewidth]{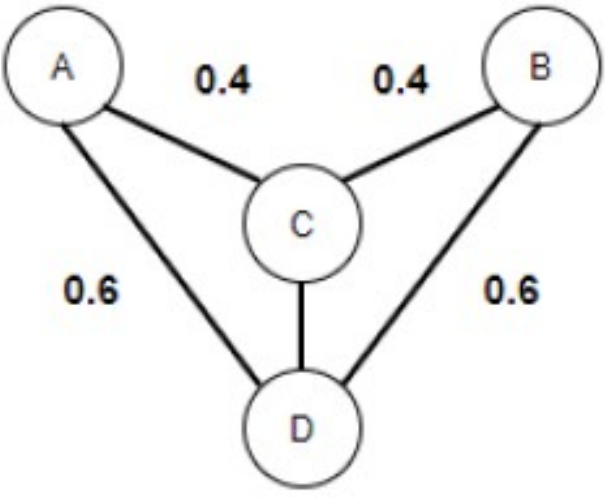}
  \label{fig:small_example_opt_dist}
  }
\subfigure[Expected MLU as a function of the splitting ratios]{
  \includegraphics[width=.31\linewidth]{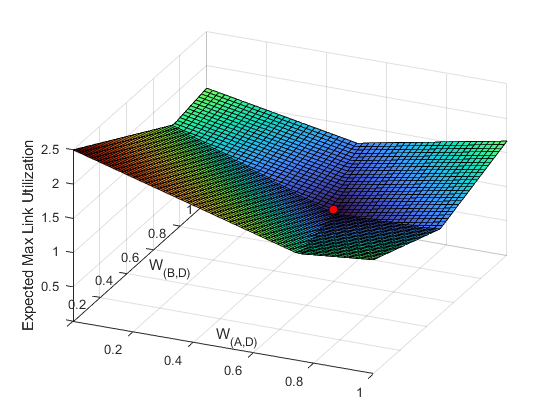}
  \label{fig:expected_MLU}
  }
\caption{Simple WAN TE example}
\label{fig:small_example}
\end{figure*}

We illustrate key insights underlying \emph{DOTE} using the example in \autoref{fig:small_example_topology}. Each of nodes $A$ and $B$ wishes to send traffic to node $D$, and can do so either via its direct link to $D$ or its $2$-hop path to $D$ through node $C$. All link capacities are $1$. Every fixed time interval (say, $5$ minutes), the TE system must determine, for each of the two source nodes, $A$ and $B$, traffic splitting ratios specifying which fraction of its demand is forwarded along each of its assigned two paths to $D$. $A$ and $B$'s traffic demands for each time interval are drawn (i.i.d) at the beginning of each time interval from a \textit{fixed} probability distribution: with probability $\frac{1}{2}$ node $A$'s demand is $\frac{5}{3}$ and node $B$'s demand is $\frac{5}{6}$ and with probability $\frac{1}{2}$ node $B$'s demand is $\frac{5}{3}$ and node $A$'s demand is $\frac{5}{6}$. The TE system has no a priori knowledge of the realization of the traffic demands; splitting ratios must be determined before actual traffic demands are revealed.

\vspace{0.05in}\noindent{\bf Demand-prediction-based TE and its shortcomings.} A natural solution is training a predictor on empirical data containing past demands for $A$ and $B$ to predict the combination of demands closest (in expectation) to the realized combination of demands (\textit{e.g.}, in terms of mean-squared-error), and then performing global optimization with respect to the predicted demands. In our simple example, this leads to the predicted demand-combination being $(\frac{5}{4},\frac{5}{4})$ and the induced splitting ratios presented in \autoref{fig:small_example_opt_pred}. Under these splitting ratios, \emph{regardless} of the realization of the demands, either link $(A,D)$ or link $(B,D)$ will carry more traffic than its capacity can accommodate. In the optimal solution shown in \autoref{fig:small_example_opt_dist}, however, \textit{regardless of the realized demands, no link carries more traffic than its capacity can support}.

Of course, instead of predicting a single demand-combination, one could have predicted a \emph{probability distribution} over the traffic demands and optimized splitting ratios with respect to that. This entails two nontrivial challenges, which are significantly amplified for large WANs and real-world traffic: (1) We must impose a specific structure on the probability distribution to be predicted (e.g., Gaussian, bimodal), which might not be a good fit for actual WAN traffic. This is particularly true when there hidden correlations between demands (as in our example); (2) Optimizing an LP with respect to a distribution over \textit{multiple} demand-combinations can be prohibitively time consuming for large WANs.

\vspace{0.05in}\noindent{\bf On \textit{direct} optimization of traffic splitting ratios and why it might do better.} An alternative approach, which avoids presuppositions regarding the traffic, and also LP optimizations, is training a decision model on past realizations of $A$ and $B$'s traffic demands to \textit{directly} output traffic splitting ratios that are close to the global optimum. This approach can outperform the demand-prediction-based approach in scenarios where traffic is volatile and hard to predict but a certain configuration of splitting ratios performs well on most traffic realizations. Directly inferring the splitting ratios also obviates the need for solving an LP to optimize splitting ratios with respect to predicted traffic. As our evaluation results in \S\ref{sec:eval} show, this significantly accelerates TE runtimes for large WANs. In our example, after sufficient training, the model is expected to learn the splitting ratios in \autoref{fig:small_example_opt_dist} (the unique global optimum). Indeed, \emph{DOTE}, which is a manifestation of this approach, quickly converges to this global outcome. 

\vspace{0.05in}\noindent{\bf Exploiting convexity/concavity for direct optimization of splitting ratios via gradient descent.} A key insight is that for classical TE optimization objectives, the function mapping splitting ratios to \textit{expected} performance scores satisfies desirable properties, namely, \textit{convexity/concavity}. This facilitates utilizing elegant direct optimization methods, like (stochastic) gradient descent, circumventing explicit demand prediction. 

To illustrate this, we consider the classical TE objective of minimizing maximum-link-utilization (MLU). We visualize in \autoref{fig:expected_MLU} the impact of different choices of splitting ratios on MLU, i.e., the maximum ratio, across all network links, between the traffic traversing a link and the link capacity. x-axis values specify the fraction of $A$'s traffic sent on the direct path $(A,D)$. Since $A$ only has two available paths, this value also uniquely determines the fraction of $A$'s traffic sent on the indirect path $(A,C,D)$. Similarly, y-axis values specify the fraction of $B$'s traffic sent on $(B,D)$ and so also on $(B,C,D)$. $z$-axis values represent the \textit{expected} MLU for different choices of splitting ratios for $A$ (x-axis) and $B$ (y-axis) for the underlying demand distribution described above. For instance, the scenario where $A$ and $B$ send all of their traffic on $(A,D)$ and $(B,D)$, respectively, is captured by $w_{(A,D)}=1$ (x-axis) and $w_{(B,D)}=1$ (y-axis), and the derived expected MLU is $\frac{5}{3}$ (z-axis). Indeed, in this scenario, regardless of which of the two demand combinations is realized, the traffic injected into either link $(A,D)$ or link $(B,D)$ will be $\frac{5}{3}$x its capacity. The \emph{unique} global minimum for MLU, in which no link capacity is exceeded, is achieved for $w_{(A,D)}=0.6$ and $w_{(B,D)}=0.6$ (the red dot in \autoref{fig:expected_MLU}, which corresponds to the splitting ratios in \autoref{fig:small_example_opt_dist}).

As seen in \autoref{fig:expected_MLU}, the expected MLU exhibits a desirable structure---\textit{convexity} in the traffic splitting ratios. This suggests the following procedure for converging to the optimum: start with \emph{arbitrary} splitting ratios, and adapt the splitting ratios in the direction of the steepest slope of the (expected) MLU (i.e., the opposite direction of the \emph{gradient} with respect to the splitting ratios) until converging to the global minimum. We show (in \S\ref{subsec:DOTE-theory}) that the convexity of the expected MLU extends to \emph{any} network topology, \emph{any} choice of network paths (tunnels), and \emph{any} underlying demand distribution, and so, this elegant optimization procedure is guaranteed to converge to the global optimum in general.

\subsection{TE as Stochastic Optimization}

\noindent{\bf How to estimate the gradient of the \textit{expected} MLU?} Executing gradient descent on the \textit{expected} MLU requires repeatedly evaluating the gradient for different traffic splitting configurations. However, \textit{exact knowledge of the gradient is impossible without exact knowledge of the underlying demand distribution}. Once again, the specific structure of the TE setting gives rise to opportunities for effective optimization. We show how the gradient can be closely approximated from data samples of past realizations of the demands. Our approach builds on the following two observations that, while illustrated using our toy example, generalize to arbitrary network topologies, tunneling schemes, and distributions over traffic demands (see \S\ref{sec:DOTE}).

\begin{itemize}
    \item  {\bf For any realized demand-combination, the MLU gradient \textit{with respect to these specific demands} can be expressed \textit{in closed form}.} Suppose that the realized demands in our simple example are $\frac{5}{3}$ for $A$ and $\frac{5}{6}$ for $B$. The MLU as a function of $A$'s splitting ratios, $w_{(A,D)}$ and $(1-w_{(A,D)})$, and $B$'s splitting ratios, $w_{(B,D)}$ and $(1-w_{(B,D)})$, can be expressed as: \[\max\{\frac{5}{3}w_{(A,D)},\frac{5}{3}(1-w_{(A,D)})+\frac{5}{6}(1-w_{(B,D)}),\frac{5}{6}w_{(B,D)}\}\] (i.e., the maximum load across the links $(A,D)$, $(C,D)$, and $(B,D)$, respectively\footnote{Observe that the load on $(A,C)$ and $(B,C)$ is always dominated by the load on $(C,D)$, and so we disregard these links.}). This representation of the MLU for the realized demands as a convex function of the splitting ratios enables deriving a closed form expression of the (sub)\textit{gradient} of the MLU\footnote{Note that even though this function is not differentiable for all inputs due to the maximum operator, the subgradient always exists and can be explicitly derived.}, as shall be discussed in \S\ref{sec:DOTE}.

    \item {\bf Averaging over the MLU gradients \textit{for past realized demands} closely approximates the gradient of the \textit{expected} MLU.} Exact knowledge of the underlying probability distribution over demands is elusive in most real-world scenarios. Hence, the gradient of the \textit{expected} MLU for a given configuration of splitting ratios cannot be precisely derived. However, this gradient can be well-approximated by averaging over the gradients \emph{for realized demands} at those splitting ratios. In our example, deriving the expected MLU gradient for specific traffic splitting ratios for $A$ and $B$ can be achieved by sampling sufficiently many past realizations of $A$ and $B$'s demand-combinations, deriving the MLU gradient with respect to each such realized demand combination (at these splitting ratios), and averaging over these.

\end{itemize}

\noindent{\bf Why is reinforcement learning (RL) \emph{not} a good fit?} (Deep) RL methods have been applied to many networking domains, including routing~\cite{l2r}. Similarly to \emph{DOTE}, RL approaches to TE also replace explicit demand prediction with end-to-end optimization, mapping recent traffic demands to TE configurations~\cite{l2r}. However, while RL can be applied to essentially any sequential decision making context, RL suffers from higher data-sample complexity, notorious sensitivity to noisy training, and a brittle optimization process that necessitates painstakingly sweeping hyperparameters~\cite{henderson2018deep}. A key observation underlying \emph{DOTE} is that WAN TE exhibits a desirable structure that gives rise to opportunities for much simpler and more robust optimization, rendering RL an ``overkill''.

\subsection{Harnessing Deep Learning}

In our simple example, traffic demands were repeatedly drawn from the \emph{same} probability distribution. Real-world traffic exhibits intricate temporal (hourly, diurnal, weekly), and other, patterns. To pick up on such regularities, the TE system could take into account the recent history of observed traffic demands (e.g., traffic demands from the last hour). However, there are infinitely many possible recent histories of traffic demands the TE system might observe. To address this, \emph{DOTE} trains a deep neural network (DNN) to \textit{approximate the optimal mapping from traffic histories to TE configurations}, exploiting the capability of DNNs to automatically identify complex patterns in large, high-dimensional data (\S\ref{sec:DOTE_DNN}). \emph{DOTE} builds on recent developments in large-scale optimization, namely, the ADAM stochastic gradient descent optimizer~\cite{adam}, to accommodate efficient training on extensive empirical data ($10$s of thousands of traffic demand snapshots in our experiments).

\section{Direct Optimization for TE (\emph{DOTE})}\label{sec:DOTE}

Below, we present our model for WAN TE with uncertain traffic demands, which extends the classical WAN TE model. We then delve into the the \emph{DOTE} stochastic optimization framework, provide theoretical guarantees, and discuss how \emph{DOTE} can be implemented in practice.

\subsection{Modeling WAN TE}\label{subsec:model}

\noindent{\bf Network.} The network is modeled as a capacitated graph $G=(V, E, c)$. $V$ and $E$ are the vertex and edge (link) sets, respectively, and $c:E \to \mathbb{R^+}$ assigns a capacity to each edge. %

\vspace{0.05in}\noindent{\bf Tunnels.} Each source vertex $s$ communicates with each destination vertex $t$ via a set of network paths, or ``\textit{tunnels}'', $P_{st}$. 

\vspace{0.05in}\noindent{\bf Traffic demands.} A \emph{demand matrix} (DM) $D$ is an $|V|\times |V|ß$ matrix whose $(i,j)$'th entry $D_{i,j}$ specifies the traffic demand between source $i$ and destination $j$.

\vspace{0.05in}\noindent{\bf Optimization objective.} To simplify exposition, we first describe \emph{DOTE} for the case of one classical TE objective: minimizing maximum-link utilization (MLU)~\cite{Fortz,coyote,Azar2003obliv}. We discuss other optimization objectives (maximum network throughput and maximum-concurrent-flow) in \S\ref{subsec:other_objectives}.

\vspace{0.05in}\noindent{\bf TE configurations.} We focus on how traffic should be split across a \textit{given} set of tunnels so as to achieve the optimization objective. \emph{DOTE} is compatible with any tunnel-selection method. We discuss an extension that incorporates data-driven tunnel selection in \S\ref{discussion}. 

Given a network graph and demand matrix, a \emph{TE configuration} \rs specifies for each source vertex $s$ and destination vertex $t$ how the $D_{s,t}$ traffic from $s$ to $t$ is split across the tunnels in $P_{st}$. Thus, a TE configuration specifies for each tunnel $p\in P_{st}$ a value $x_p$, where $x_p$ is the fraction of the traffic demand from $s$ to $t$ forwarded along tunnel $p$ (and so $\sum_{p\in P_{st}} x_p=1$).

Given a demand matrix $D$ and TE configuration \rs, the total amount of flow traversing edge $e$ is $f_e = \sum_{s,t\in V,p\in P_{st},e\ni p} D_{s,t}\times x_p$. The objective is minimizing the maximum link utilization induced by \rs and $D$, $\max_{e\in E}\frac{f_e}{c(e)}$, which we will refer to as MLU and represent as $\obj(\rsm,D)$. WAN operators seek to reduce the MLU to keep more headroom open for unplanned failures and traffic spikes. Typically, operators spend to increase link capacities when MLU exceeds a threshold value, and so reducing MLU can reduce CAPEX~\cite{TExCP,mate}. 

In this work, we aim to select TE configurations without a priori knowledge of the traffic demands. To do so, we augment the above model as follows:

\vspace{0.05in}\noindent{\bf WAN TE under traffic uncertainty.} Time is divided into consecutive intervals, called ``epochs'', of length $\delta_t$. $\delta_t$ is determined by the network operator (e.g., at some large service providers~\cite{SWAN,B4}, $\delta_t$ is a few minutes). At the beginning of each epoch $t$, the TE configuration \rst{t} for that epoch is decided based only on the history of \emph{past} demand matrices and TE configurations. We also assume that the demand matrix is fixed within an epoch and can be approximately estimated after the fact.\footnote{For e.g., by sampling ipfix (or equivalent) data at each node in the WAN, as is done in production in SWAN~\cite{SWAN} and B4~\cite{B4}. This data contains source and destination nodes and volume of bytes exchanged. Alternatively, traditional ISP backbones use network tomography on measured link usage data (see, e.g.,~\cite{Roughan-TE,estimating-DMs}).} Such periodic changes to TE configuration reflect the current practice in private WANs~\cite{SWAN,B4,b4_and_after}.

After selecting the TE configuration \rst{t} for epoch $t$, the demand matrix $D_t$ is revealed. To minimize MLU, the goal for direct optimization is to devise a \textit{TE function} $\pi(D_{t-1},\ldots D_1)$ that, for every $t>0$, maps the history of DMs from the previous $t-1$ time epochs to a TE configuration \rst{t} for the upcoming time epoch $t$ so as to minimize $\frac{1}{T}\Sigma_{x=1}^{t} \obj(\rsm^{(x)},D_x)$, where $T$ represents the length of time in which TE configurations are computed according to $\pi$.

To reason about WAN TE in the presence of traffic uncertainty, we assume that the demand matrix $D_t$ at each epoch $t$ is generated from some probability distribution. We also make the following two assumptions, which are fundamental to any data-driven approach to WAN TE. First, we assume that there is some sufficiently large $\hor>0$ such that the finite window of $\hor$ recent historical observations of DMs is sufficient for informing the decision of the next TE configuration. (Our empirical results in \S\ref{sec:eval} suggest that $\hor=12$ suffices for attaining high performance on our datasets.) Formally, we model the demand matrix $D_t$ as generated according to an unknown $\hor$-Markov process with transition probabilities such that $P(D_t|D_{t-1},\dots,D_{t-\hor}) = P(D_t|D_{t-1},\dots,D_{1})$. Second, we assume that the probability of observing a particular sequence of $\hor$ DMs in the training data and during real-time system execution is the same. This formally translates to the Markov process being in a steady state. Let $P(D_{t-1},\dots,D_{t-\hor})$ denote the Markov process' stationary distribution, which determines the probability for any specific $\hor$-long recent history of DMs. The \textit{expected} MLU for a TE configuration $\rsm$ at epoch $t$ is therefore $\mathbb{E}_{D_t} \left[ \obj(\rsm,D_t)\right]$, 
where the expectation is with respect to the (unknown) probability distributions $P(D_{t-1},\dots,D_{t-\hor})$ and $P(D_t|D_{t-1},\dots,D_{t-\hor})$ defined above.

\subsection{The \emph{DOTE} TE Framework}\label{sec:DOTE_overview}

\textit{DOTE} leverages stochastic optimization to compute a TE function $\pi_{\theta}(D_{t-1},\dots,D_{t-\hor})$, parametrized by $\theta$, which maps the $\hor$-long recent history of DMs to the TE configuration for the next time epoch, \rst{t}. If the TE function is sufficiently expressive, there should exist parameters that closely approximate the optimal TE function. As we shall discuss in \S\ref{sec:DOTE_DNN}, in \emph{DOTE}, $\pi_{\theta}$ is realized by a deep neural network (DNN), and the parameters $\theta$ correspond to the DNN's link weights. We thus consider the optimization problem of seeking parameters $\theta$ for which the following expression is minimized: $\mathbb{E}\left[ \obj(\pi_{\theta}(D_{t-1},\dots,D_{t-\hor}),D_t)\right]$, where the expectation is with respect to choosing $t$ uniformly at random from $\{1,\ldots,T\}$, and the probability distributions $P(D_{t-1},\dots,D_{t-\hor})$ and $P(D_t|D_{t-1},\dots,D_{t-\hor})$ defined above. Observe that by the linearity of expectation and the above equation,
$\mathbb{E}\left[ \obj(\pi_{\theta}(D_{t-1},\dots,D_{t-\hor}),D_t)\right] 
     =\frac{1}{T}\Sigma_{t=1}^{T} \mathbb{E}_{D_t}\left[ \obj(\rsm^{(t)},D_t)\right],$ which is precisely our optimization objective in \emph{DOTE}.

The training data for \emph{DOTE} is a trace of historical DMs, consisting of $N$ sequences of DMs of the form $\left\{D_t^i, D_{t-1}^i,\dots,D_{t-\hor}^i\right\}$, where each sequence consists of $\hor+1$ DMs and captures a specific realization of a $\hor$-long history of DMs and the subsequent realized DM. We assume that these $N$ observations of DM sequences are sampled i.i.d.~from $t\in [1,\dots,T]$, $P(D_{t-1},\dots,D_{t-\hor})$ and $P(D_t|D_{t-1},\dots,D_{t-\hor})$.\footnote{When the data is a long trace of historical DMs, the samples are not necessarily independent. However, we assume that the mixing time of the Markov process is fast enough such that correlations between the data samples are negligible. This is a common assumption in time series prediction.} 

\emph{DOTE} executes \textit{stochastic gradient descent} (SGD)~\cite{shapiro2021lectures} to optimize the parameters $\theta$ by sequentially sampling $m$-sized mini-batches of data, where each data point in the mini-batch is drawn from the data uniformly at random. For each mini-batch of sampled data points, the parameters $\theta$ are updated as follows:

\begin{equation*}
 \theta := \theta - \alpha \frac{1}{m}\sum_{i \textrm{ in batch} }\nabla_\theta \obj(D_t^i,\pi_{\theta}(D_{t-1}^i,\dots,D_{t-\hor}^i)),
\end{equation*}
where $\alpha$ is a step size parameter and $\nabla_\theta \obj(D_t^i,\pi_{\theta}(D_{t-1}^i,\dots,D_{t-\hor}^i))$ is the gradient of the loss function with respect to $\theta$. Our realization of stochastic optimization in \emph{DOTE} follows the ADAM~\cite{adam} method, which incorporates momentum and an adaptive step size.

\vspace{0.05in}\noindent{\bf A closer look at \emph{DOTE}'s parameter update step.} Recall that our objective is to reach a performant TE configuration with respect to the \emph{expected} loss (MLU). The success of \emph{DOTE}'s SGD is thus crucially dependent on \emph{DOTE}'s ability to well-approximate the gradient with respect to the expected loss. Unfortunately, in most real-world TE environments, exact knowledge of the underlying distribution over traffic demands is unattainable. To address this, \emph{DOTE}'s parameter update step (see above) incorporates the expression
$\frac{1}{m}\sum_{i \textrm{ in batch} }\nabla_\theta \obj(D_t^i,\pi_{\theta}(D_{t-1}^i,\dots,D_{t-\hor}^i))$. As discussed above, each sequence of $\hor+1$ demand matrices $\left\{D_t^i, D_{t-1}^i,\dots,D_{t-\hor}^i\right\}$ in the batch is assumed to be independently drawn from the underlying stationary distribution of the Markov process. Hence, $\frac{1}{m}\sum_{i \textrm{ in batch} }\nabla_\theta \obj(D_t^i,\pi_{\theta}(D_{t-1}^i,\dots,D_{t-\hor}^i))$ is an \textit{unbiased} estimate of the gradient of the expected loss, and closely approximates the gradient of the expected loss for a large enough $m$. Approximating the gradient of the expected loss in this manner is termed Sample Average Approximation (SAA) in stochastic optimization literature~\cite{shapiro2021lectures}. Relying on unbiased stochastic gradients for SGD guarantees convergence to a \emph{global} optimum with respect to the \textit{expected} loss~\cite{shalev2014understanding} when the loss function is concave (as in our context, see \S\ref{subsec:DOTE-theory}).

We are left with the challenge of deriving $\frac{1}{m}\sum_{i \textrm{ in batch} }\nabla_\theta \obj(D_t^i,\pi_{\theta}(D_{t-1}^i,\dots,D_{t-\hor}^i))$. An important technical observation is that each data point $i$ in the batch, $\obj(D_t^i,\pi_{\theta}(D_{t-1}^i,\dots,D_{t-\hor}^i))$ is a composition of \textit{differentiable} computations. \emph{DOTE} capitalizes on this for calculating 
the gradient $\nabla_\theta \obj(D_t^i,\pi_{\theta}(D_{t-1}^i,\dots,D_{t-\hor}^i))$ in closed form via backpropagation. We revisit this point in \S\ref{sec:DOTE_DNN}.

\newcommand{\histset}{\mathbf{H}}
\newcommand{\policy}{\pi}
\newcommand{\teset}{\mathbf{R}}
\newcommand{\pmax}{p_{max}}
\newcommand{\dmax}{D_{max}}
\newcommand{\cmin}{c_{min}}
\newcommand{\pibound}{B}
\newcommand{\grad}{v}

\subsection{Analytical Optimality Results}\label{subsec:DOTE-theory}

We prove that, for a \textit{perfectly expressive} TE function, i.e., when the TE function can be \textit{any} mapping from demand histories to TE configurations, and in the limit of infinite empirical data sampled from the underlying Markov process' stationary distribution, \emph{DOTE} attains optimal performance. In practice, we relax both assumptions: in \emph{DOTE}, we sample from a large, but \textit{finite}, dataset of historical demands, and use a \textit{parametric} model (specifically, a neural network) to map from the set of possible histories to valid TE configurations. Our theoretical result below, however, establishes that our approach is \textit{fundamentally sound}, and so high performance in practice can be achieved by acquiring sufficient empirical data and employing a sufficiently expressive decision model (e.g., a deep enough neural network). Our empirical results in \S\ref{sec:eval} corroborate this.

For the sake of analysis, we assume that the set of possible history realizations, which we denote by $\histset$, is finite. 
Let $\policy:\histset \to \teset$ denote a mapping from history to TE configuration\footnote{Note that we dropped the subscript $\theta$ in $\pi$, as in our analysis we consider the space of all possible TE configurations, and not a specific parametrization.}. We consider an idealized stochastic gradient descent (SGD) algorithm that, at each iteration $k$ samples \textit{a single} data point $D_t, D_{t-1},\dots,D_{t-\hor}$ from the probability distributions $P(D_{t-1},\dots,D_{t-\hor})$ and $P(D_t|D_{t-1},\dots,D_{t-\hor})$, 
and updates $\pi_{k+1} = \project \left\{ \pi_{k} - \eta \grad_k \right\}$, 
where $\grad_k \in \partial \obj(\pi_{k}(D_{t-1},\dots,D_{t-\hor}),D_t)$ denotes the subgradient of the objective function, and $\project$ denotes a projection onto the simplex for each $(s,d)$ pair. The final output after $K$ iterations is $\bar{\pi} = \frac{1}{K}\sum_{k=1}^K \pi_k$. Let $\bar{\obj}(\pi) =  \mathbb{E}\left[{\obj(\pi(D_{t-1},\dots,D_{t-\hor}),D_t)}\right]$ denote the expected MLU of a TE function, and let
$\pi^* \in \argmin_{\pi} \bar{\obj}(\pi)$ denote the optimal TE function. We prove the following theorem:

\begin{theorem}\label{thm:sgd}
For any $\epsilon > 0$, there exists $\eta > 0$ and finite $K$ such that $\left|\mathbb{E}\left[\bar{\obj}(\bar{\pi})\right] - \bar{\obj}(\pi^*)\right| \leq \epsilon$, where the expectation is w.r.t. the sampling by the algorithm.
\end{theorem}

The proof of Thoerem~\ref{thm:sgd}, which crucially relies on the convexity of the MLU objective, appears in Appendix~\ref{apx:DOTE-theory}.

\subsection{Scalability and Real-World Traffic}\label{sec:DOTE_DNN}

Direct TE optimization aims at computing a mapping from the history of recent traffic demands to a TE configuration that optimizes expected performance for the next demands. A key insight is that with real-world traffic, one may expect certain \textit{patterns} in this mapping; for example, if two histories of traffic conditions are very similar, their corresponding optimal TE configurations should also be similar. However, measuring and explicitly quantifying such similarities is nontrivial. Our approach is to exploit deep neural networks, which have demonstrated remarkable success in identifying complex patterns in high dimensional data, for this task.

\emph{DOTE} employs a DNN to realize the TE function $\pi_{\theta}(D_{t-1},\dots,D_{t-\hor})$. Specifically, \emph{DOTE}'s DNN maps an input of $\hor$ ($12$ in our experiments) most recent DMs into an output vector specifying the splitting ratios across tunnels for all source-destination pairs. In our implementation of \emph{DOTE}, we use the popular Fully Connected DNN architecture. See Appendix~\ref{sec:description_loss} for a formal exposition of how the DNN's output and the realized DM are fed into the loss function to derive the induced MLU. Importantly, the sequence of steps for mapping the DNN output to the MLU value $\obj(D_t^i,\pi_{\theta}(D_{t-1}^i,\dots,D_{t-\hor}^i))$ involves only \textit{differentiable} computations; the loss as a function of the TE configuration is a composition of a $\max$ and a linear function, and the neural network is differentiable by design. Hence, the gradient $\nabla_\theta \obj(D_t^i,\pi_{\theta}(D_{t-1}^i,\dots,D_{t-\hor}^i))$ can be calculated in closed form via backpropagation. In our implementation, the Pytorch~\cite{paszke2019pytorch} auto-differentiation package is used to calculate the gradients.

\subsection{On Maximum and Concurrent Flow} \label{subsec:other_objectives}

We next explain how \emph{DOTE} extends to two other central TE objectives: maximizing network throughput~\cite{garg2007faster,multicommodity,B4,SWAN} (maximum multicommodity flow) and maximum concurrent-flow~\cite{Shahrokhi1990TheMC,karakostas,TeaVaR}.

\vspace{0.05in}\noindent{\bf TE configurations for flow maximization.} TE objectives that capture different notions of flow maximization require that the outputs of the TE mechanism satisfy strict capacity constraints. To address this, we revise our definition of TE configuration $\rsm$ from \S\ref{subsec:model}: for each source-destination pair $s,t\in V$, $\rsm$ now specifies (1) traffic splitting ratios $x_p$ over the paths (tunnels) $p\in P_{st}$ (as in \S\ref{subsec:model}), (2) for each path (tunnel) $p\in P_{st}$, a ``\textit{cap}'' $\omega_p\geq 0$. $\omega_p$ represents the maximum permissible flow between $s$ to $t$ along the path $p$ (enforced via rate limiting). $\rsm$ must satisfy that no link capacity is exceeded (\textit{regardless of the realized demands}), i.e., that for each link $e\in E$, $\Sigma_{s,t\in V, p\in P_{st}, e\ni p}\omega_p\leq c(e)$. 
A TE configuration $\rsm$ and demand matrix $D$ induce, for each tunnel $p$ a flow $f_p(\rsm,D)=\min\{x_p\times D_{s,t},\omega_p\}$. The total flow between $s$ and $t$ is thus $f_{st}(\rsm,D)=\Sigma_{p\in P_{st}}f_p(\rsm,D)$.

\vspace{0.05in}\noindent{\bf The maximum-multicommodity-flow and maximum-concurrent-flow objectives.} In maximum-multicommodity-flow~\cite{garg2007faster,multicommodity,B4,SWAN}, the performance objective $\obj(\rsm,D)$ is to compute, for a given demand matrix $D$, a TE configuration $\rsm$ that maximizes the expression $\obj(\rsm,D)=\Sigma_{s,t\in V}f_{st}(\rsm,D)$ (the total network throughput). For a TE configuration $\rsm$ and demand matrix $D$, let $\alpha(\rsm,D)$ denote the maximum value $\alpha\in [0,1]$ for which at least an $\alpha$-fraction of each $D_{s,t}$ is routed \textit{concurrently}, i.e., such that for all $s,t\in V$, $f_{st}(\rsm,D)\geq \alpha D_{s,t}$. The goal in maximimum-concurrent-flow is to compute, for an input DM $D$, the TE configuration $\rsm$ for which $\obj(\rsm,D)=\alpha(\rsm,D)$ is maximized. Relative to maximum-multicommodity-flow above, the maximum-concurrent-flow objective enhances fairness. Practical TE systems~\cite{SWAN,B4} use a sequence of optimizations wherein they employ different objectives for different priority classes. For example, they may use maximum-multicommodity-flow or minimizing MLU for high priority traffic and maximum-concurrent-flow for scavenger-class traffic.

\vspace{0.05in}\noindent{\bf \emph{DOTE} for maximum-multicommodity-flow and maximum-concurrent-flow.} Adapting \emph{DOTE} to the above two flow-maximization objectives is accomplished along the lines described in \S\ref{sec:DOTE_DNN}. In particular, a DNN is again utilized to map the recent observations of DMs to the next TE configuration. Recall from the above discussion that the (revised) TE configuration consists of both traffic splitting ratios across tunnels and a ``flow cap'' for each tunnel. In our design, the DNN outputs $w_p\geq 0$ for each tunnel $p$. The $w_p$'s are used to derive traffic splitting ratios and flow caps as follows. We set $\omega_p=\frac{w_p}{\gamma}$, where $\gamma=\max\left(\max_{e\in E}\frac{\Sigma_{p:e\in p}w_p}{c(e)},1\right)$. Observe that this guarantees that no link capacity can be exceeded even if each tunnel $p$ carries its maximum permissible flow $\omega_p$ (i.e., that $\Sigma_{s,t\in V, p\in P_{st}^e}\omega_p\leq c(e)$). We then set the traffic split share on tunnel $p$ to simply be its proportional weight: $x_p=\frac{\omega_p}{\Sigma_{q\in P_{st}}\omega_q}$. Since the objective is now \textit{maximizing} a performance metric, \emph{DOTE} now involves stochastic gradient \emph{ascent}. 

\vspace{0.05in}\noindent{\bf Optimality via stochastic quasi-concave optimization.} In Appendix~\ref{apx:DOTE-theory}, we prove the analogues of Theorem~\ref{thm:sgd} for maximum-multicommodity-flow and for maximum-concurrent-flow, establishing \emph{DOTE}'s optimality for these two objectives. Similarly to Theorem~\ref{thm:sgd} (for MLU), this implies that with sufficient training data and a sufficiently expressive decision model, \emph{DOTE} attains near-optimal TE configurations. Our evaluation results for maximum-multicommodity-flow and for maximum-concurrent-flow exemplify this (\S\ref{ss:e_generalization}).

Our proofs for maximum-multicommodity-flow and for maximum-concurrent-flow are considerably more subtle than that of Theorem~\ref{thm:sgd}, as both objectives are not concave (the analogue of convexity for maximization problems). Instead, we show that the average maximum-multicommodity-flow / maximum-concurrent-flow score of a TE configuration over any set of DMs is \textit{quasi}-concave. This result, which may be of independent interest, allows us to leverage the analytical arguments in~\cite{hazan2015beyond} to show convergence of a suitable stochastic gradient ascent algorithm to the global optimum, and bound the number of required iterations.

\cut{

\noindent{\bf Additional performance metrics.} We also consider the $\alpha$-fairness optimization objective for resource allocation~\ref{x}. Now, the goal is to compute, for a given DM $D$, the TE configuration $\rsm$ for which $\obj(\rsm,D)=\Sigma_{s,t\in V}\frac{f_{st}^{1-\alpha}}{1-\alpha}$ is maximized. $\alpha$-fairness spans a broad spectrum of \emph {concave} optimization objectives~\cite{x}. In particular, as $\alpha$ tends to infinity, the $\alpha$-fairness objective tends to max-min-fairness~\cite{x,y,z}, whereas as $\alpha$ tends to $0$, $\alpha$-fairness tends to maximum-multicommodity-flow. For $\alpha=1$ (for which the above formula is not defined), $\alpha$-fairness is set to be proportional fairness~\cite{x}, i.e., maximizing $\obj(\rsm,D)=\Sigma_{s,t\in V} \log(f_{st})$. We establish that \textit{DOTE} provably attains the optimum for \emph{any} $\alpha$-fairness objective by proving the analogue of Theorem~\ref{thm:sgd} for $\alpha$-fairness in Appendix~\ref{apx:DOTE-theory}.}

\begin{figure*}
\centering
\subfigure[Architecture of prior SD-WAN TE schemes~\cite{SWAN,B4}.]
{\label{f:prior_arch} \includegraphics*[height=1.6in]
{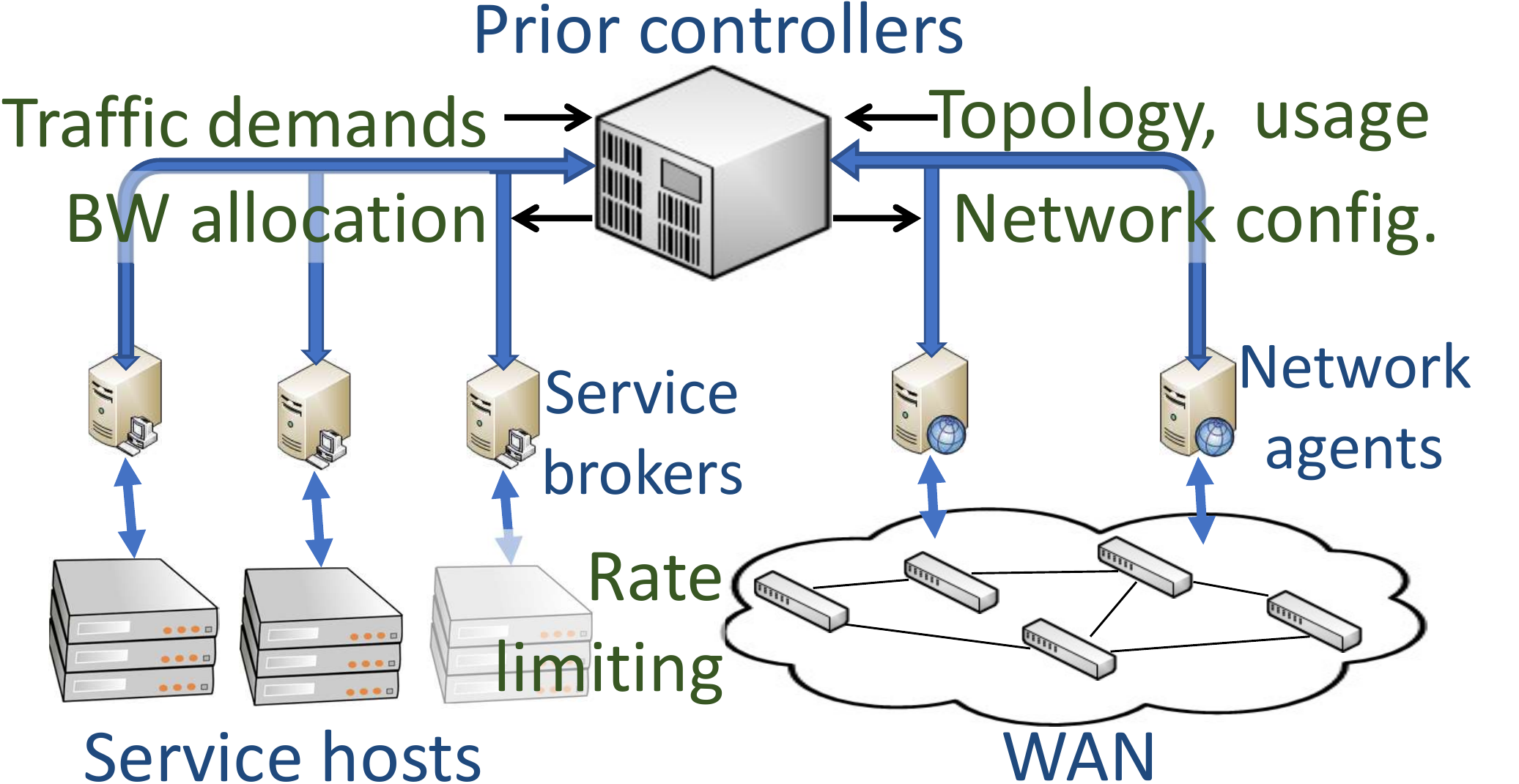}
}
\hfill
\subfigure[{\sysname} with differences shown in \textcolor{cadmiumred}{red}.]
{\label{f:stote_arch}\includegraphics*[height=1.6in]
{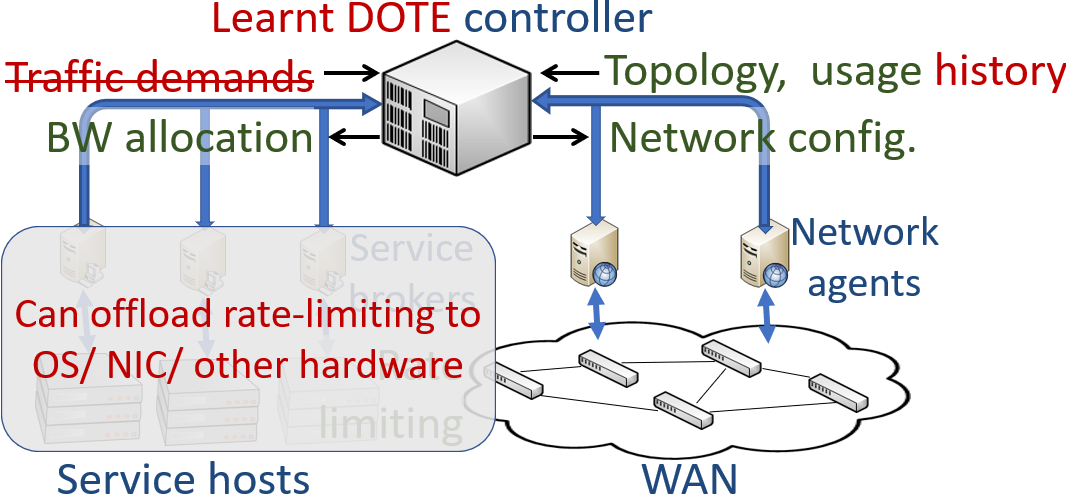}
}
\vspace{-0.15in}
\caption{Illustration of the key differences from previous SD-WAN TE schemes.\label{f:arch}}
\end{figure*}
\subsection{Realizing \emph{DOTE}}
\label{ss:arch}

\autoref{f:arch} illustrates key differences between \emph{DOTE} and prior software-defined WAN TE schemes. One key difference is the use of historical traffic demands and a learnt controller instead of running an optimization solver, leading to substantial decrease in deployment overheads and runtimes. In particular, bandwidth brokers are no longer needed to estimate application demands. Furthermore, rate allocations can, if necessary, be enforced by piggy-backing on novel traffic shaping techniques that are deployed in modern cloud servers at the OS-level as well as in NIC/FPGA offloads~\cite{carousel,gc_ratelimits,catapult1,senic}.

\vspace{0.05in}\noindent{\bf Training \emph{DOTE}.} Since \emph{DOTE}'s decision model is trained \textit{offline} on historical data, its operational model can be periodically replaced by a model trained in the background on more recent and up to date data, to gracefully adapt to \textit{planned} changes in WAN topology (adding capacity, planned addition/removal of nodes or links) and to temporal drifts in traffic demand distributions. Our evaluation results (\S\ref{sec:eval}) indicate that \emph{DOTE} produces high performance TE configurations even weeks after being deployed, and even if the network topology changes during this time (e.g., due to failures). This provides ample time for training substitute TE functions (a process that requires less than a day on large networks for our empirical datasets without code and hardware optimizations).

\vspace{0.05in}\noindent{\bf Handling network failures.} Tunnelling protocols (e.g., MPLS) identify tunnels with failed nodes/links. A traditional approach in TE to rerouting traffic around failed tunnels is to let traffic sources redistribute traffic proportionally among their remaining tunnels~\cite{SWAN,B4,FFC,Joint-failure-TE}.\footnote{Traffic split $(0.6,0.3,0.1)$ becomes $(0, 0.75, 0.25)$ if the first path goes down.} We incorporate this simple approach into \emph{DOTE} and evaluate its effectiveness in \S\ref{sec:eval}), showing that it achieves high resiliency to failures. We discuss other possible approaches in \S\ref{discussion}.

\section{Evaluation} \label{sec:eval}

Using actual traffic demands from three different production WANs~(Abilene, GEANT, and Microsoft's WAN), we ask the following questions: (1) How does {\sysname} compare against an omniscient oracle with perfect knowledge of future demands? (2) How does {\sysname} compare with state-of-the-art prediction-based TE~\cite{SWAN,B4,yates,NC,POP}, demand-oblivious TE~\cite{Kodialam2011,Cohen2003}, and RL-based TE~\cite{l2r}? (3) Can {\sysname} support different TE objectives (e.g., MLU~\cite{Cohen2003}, maximum-multicommodity-flow~\cite{SWAN,B4,NC})? (4) How long does {\sysname} take to train and to apply online at each solver activation? (5) How does {\sysname} perform under network faults and drift in traffic patterns?

\subsection{Methodology}\label{subsec:eval-framework}
\label{ss:e_method}

\begin{table}[t!]
    \centering
{\scriptsize
    \begin{tabular}{c||c|c|c|c}
          & {\bf \#Nodes} & {\bf \#Edges} & {\bf Length} & {\bf Granularity}\\
    \hline\hline
         {\bf Abilene} & $11$ & $14$ & $4.5$months & $5$ min. \\
    \hline
         {\bf GEANT} & $23$ & $37$ & $4$ months & $15$ min.\\
    \hline
         {\bf \emph{PWAN}} & O($100$) & O($100$) & O($1$) months & minutes\\
    \hline
         \emph{PWAN$_{DC}$} & O($10$)  & O($10$)  & O($1$) months & minutes\\
    \hline
         GtsCe & $149$ & $193$ & \multicolumn{2}{c}{\multirow{3}{*}{Synthetic}}\\\cline{1-3}
         Cogentco & $197$ & $243$ & \multicolumn{2}{c}{} \\ \cline{1-3}
         KDL & $754$ & $895$ & \multicolumn{2}{c}{}\\
    \end{tabular}
}
    \caption{Datasets used to evaluate {\sysname}}
    \label{tab:network_sizes}
\end{table}

\noindent{\bf Datasets:} Data-driven TE is best evaluated on \textit{real-world} datasets; we use the production topology and the traffic demands from GEANT~\cite{geant}, Abilene~\cite{abilene}, and {\pwan}, a private WAN at Microsoft. Traffic traces were collected at a few-minute granularity over several months. We also use three topologies~(GtsCe, Cogentco and KDL) from Topology Zoo~\cite{topozoo} with synthetic traffic (generated using the gravity model~\cite{Cohen2003,gravity}). \autoref{tab:network_sizes} lists the topology sizes and traffic demands. Nodes in these WAN topologies are datacenters, edge sites, or peering points. Traffic on {\pwan} includes both traffic between datacenters and traffic to/from end users. To better understand how {\sysname} performs for each traffic class, we consider a subset--{\pwandc}--which only contains large datacenters and the traffic between them. For each WAN, we use the earlier $75$\% of demand matrices~(DMs) for training and the later $25$\% DMs as the test set. 

\vspace{0.05in}\noindent{\bf Tunnel choices} are $k$-shortest-paths, edge-disjoint paths, and SMORE trees. More specifically, we use (1) Yen's algorithm for $k$-shortest-paths, with $k=8$ per commodity~(pair of nodes), (2) edge-disjoint shortest-paths where, for each commodity, we iteratively compute a shortest-path in the network and remove all links on that path from consideration until no more paths exist for that commodity, and (3) tunnels from SMORE~\cite{SMORE} generated using Yates~\cite{yates}. 

\vspace{0.05in}\noindent{\bf Comparables} to {\sysname} include: (1) {\bf Omniscient oracle}, which is an optimization with perfect knowledge of future demands and bounds the quality of \emph{any} WAN TE scheme. (2) {\bf Demand-prediction-based TE} methods~\cite{SWAN,B4,yates,NC,POP}, which are in production today~\cite{SWAN,B4}. We consider a rich collection of possible predictors of future demands: linear regression, ridge regression, random forest, DNN models, and autoregressive models (\xref{sec:demand-prediction}). (3) {\bf RL-based} WAN TE~\cite{l2r}, which leverages a neural network of the same size as {\sysname}'s (see below). (4) {\bf Demand-oblivious TE}~\cite{Cohen2003}, which optimizes the \textit{worst-case performance} over \emph{all} traffic demands. (5) {\bf SMORE}~\cite{SMORE}, which picks source-rooted trees for worst-case demands but adapts splitting ratios over the chosen trees based on \textit{predicted} future demands. (6) {\bf COPE}~\cite{cope}, which enhances demand-oblivious schemes by also optimizing over a set of predicted traffic demands.

\vspace{0.05in}\noindent{\bf Metrics:} Our TE quality metric is the ratio between the value obtained by the evaluated TE scheme and the performance obtained by the omniscient oracle, which has perfect information about future traffic demands. We consider three TE objectives: minimize maximum link utilization~(MLU)~\cite{TExCP,mate,Cohen2003}, maximize multicommodity flow~\cite{NC,SWAN,B4,SMORE} and maximize concurrent-flow~\cite{karakostas,TeaVaR}. Note that this ratio is $\geq 1$ for MLU~(because lower max-link utilization is better) and $\leq 1$ for the other metrics~(because carrying more flow is better). We refer to the relative gap from $1$ as the \textit{optimality gap}. We also measure the runtimes~(latency) of the evaluated TE schemes on the same physical machine.\footnote{VM with $8$ vCPUs and $256$GB RAM.}

\vspace{0.05in}\noindent{\bf DNN architecture:} Unless otherwise specified, results for {\sysname} use five fully connected NN layers with 128 neurons each and $ReLU(x)$ activation except for the output layer which uses $Sigmoid(x)$. For different TE objectives, {\sysname} uses a similar architecture with small changes. We chose this architecture because it empirically outperformed other investigated architectures.

\vspace{0.05in}\noindent{\bf Infrastructure and code:} We ran our experiments in cloud VMs and made use of cloud ML training systems. To enable further research, we have released our code at~\cite{repo_code}.

\vspace{0.05in}\noindent{\bf Fault model:} To examine TE behaviour under network faults, we randomly bring down a certain number of links~(e.g., $1$ to $20$ while ensuring network is not partitioned), and compare the performance of  \emph{DOTE} (see \emph{DOTE}'s failure-recovery scheme at the end of \xref{ss:arch}) and alternatives with an omniscient oracle with perfect knowledge of both future failures and future traffic demands.

\begin{figure*}[t!]
\subfigure[$PWAN_{DC}$ \& PWAN]{
  \includegraphics[width=.49\linewidth]{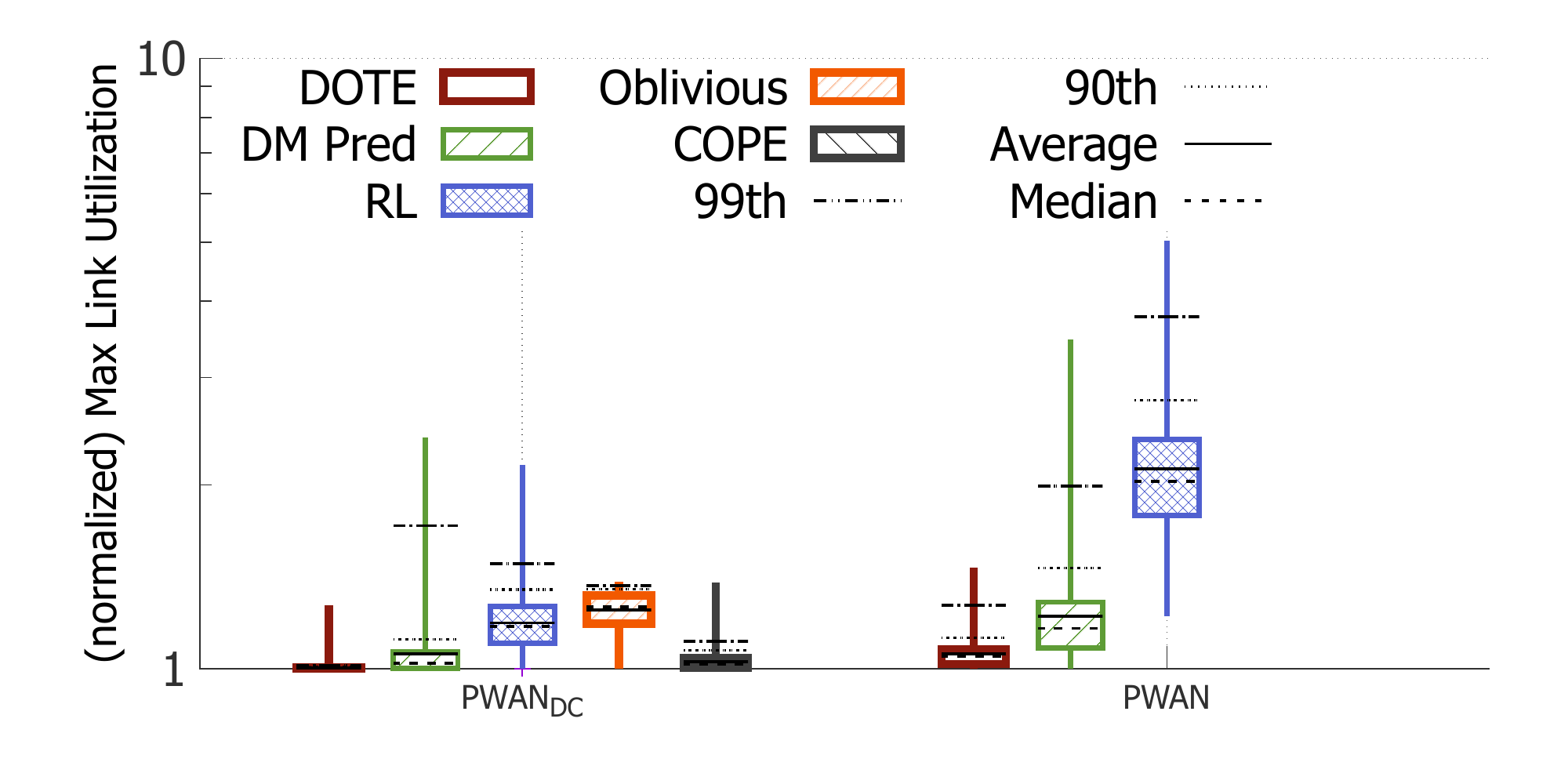}
 \label{f:mlu_pwans_8sp}
\vspace{-0.1in}
  }
\subfigure[Abilene \& GEANT]{
  \includegraphics[width=.49\linewidth]{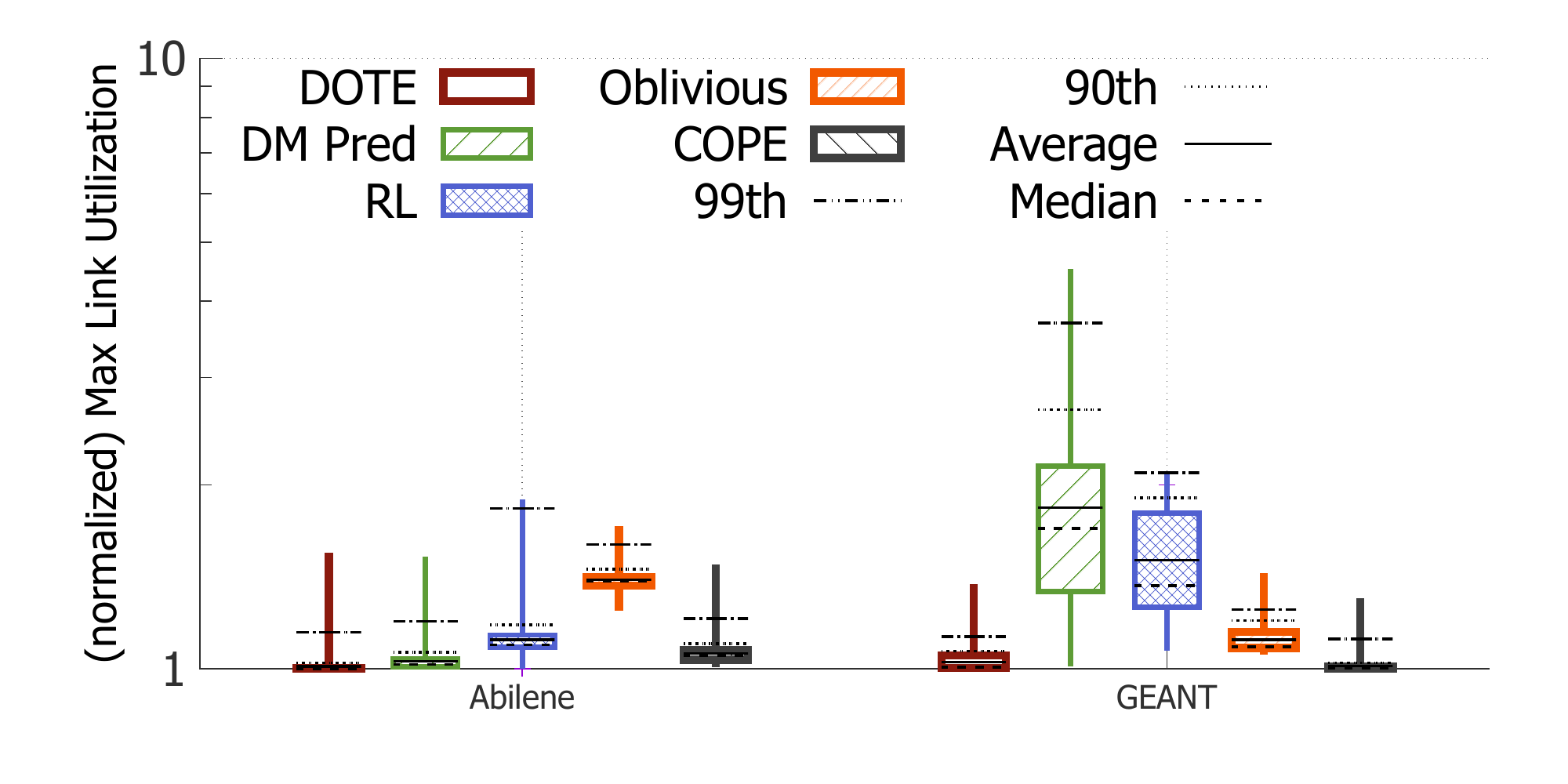}
\vspace{-0.1in} \label{f:mlu_abilene_geant_8sp}
  }
\vspace{-0.15in}
\caption{TE quality when aiming to minimize the maximum link utilization with $8$ shortest paths per demand. Candlesticks depict results across hundreds of demands; the boxes are from the $25$th to the $75$th percentile, the whiskers range from min to max value, dashed lines capture other percentiles of interest. {\sysname} achieves much lower MLU compared to the alternatives.\label{f:money}}
\end{figure*}
\begin{table}[t!]
\centering
{\scriptsize
\begin{tabular}{c|cc|ccc}
\multirow{2}{*}{\bf WAN} & \multicolumn{2}{c|}{Online Lat. (s)} & \multicolumn{3}{c}{Precomp. Lat. (s)}\\
& {\sysname} & {LP} & {\sysname} & COPE & Oblivious \\\hline
Abilene & $0.0005$ & $0.02$ & $1800$ & $180$ & $1$\\
{\pwandc} & $0.003$ & $0.05$ & $1200$ & $7200$ & $15$ \\
Geant & $0.002$ & $0.04$ & $2400$ & $10800$ & $180$\\
{\pwan} & $0.2$ & $1$ & $36000$ & \red{{\bf $>345600$}} & \red{$\sim86400$} \\
\hline
\end{tabular}
}
\caption{\label{t:runtimes} Comparing the online latency (to compute a TE configuration for a demand matrix) as well as the precomputation latency (to train models, to compute demand-oblivious configurations, etc.) for various TE schemes. $8$ shortest paths are used per demand across all WANs and TE schemes.}
\end{table}

\subsection{Comparing {\sysname} with Other TE Schemes}
\label{ss:e_mlu_one_tunnel_choice}

\noindent{\bf TE quality.} \autoref{f:money} compares {\sysname} with the other TE schemes described in~\xref{ss:e_method}, with the exception of SMORE (to be discussed in \xref{ss:e_generalization}). The values plotted here are the maximum link utilization~(MLU) normalized by that of the ominiscient oracle with perfect knowledge of future demands. The figure shows results on four different topologies. Each candlestick shows the distribution of MLUs achieved on the various demand matrices with the boxes ranging from 25th to 75th percentile and the whiskers going from minimum to maximum value. The figure also plots values achieved at various other percentile values. We note a few findings. 

\begin{Itemize}

\item
First, optimizing for \textit{predicted} demands can lead to poor TE quality (see results for GEANT and {\pwan}). Note that the y axis is in log scale. A value of $y=2$ indicates that the link most utilized by the TE scheme is twice as utilized as the most utilized link in the optimal solution (produced by the oracle). Optimizing with respect to predicted demands performs well only on Abilene and {\pwandc}, where the traffic demands are predictable. These results are for a linear-regression-based predictor that outperforms all other considered predictors on our real-world traffic datasets (see Appendix~\ref{sec:demand-prediction}).

\item
Next, we observe that the RL-based TE scheme~\cite{l2r} has extremely poor TE quality even on Abilene. This could be due to the infamous training complexity of RL.
\item 
Third, demand-oblivious TE~\cite{Cohen2003} results in somewhat decent TE quality on GEANT but not on any of the other WANs. This could be because optimizing \textit{worst-case} performance across \emph{all} possible demands fails to take advantage of the specific characteristics of real-world traffic demands.
\item 
Fourth, COPE~\cite{cope}, which explicitly accounts for historically observed demands, significantly outperforms demand-oblivious TE. The key issue with COPE is its extremely high runtimes. Our analysis (see~\autoref{t:runtimes} and discussion below) suggests that COPE's applicability does not extend beyond topologies with tens of nodes.
\item 
Finally, note that {\sysname} achieves TE quality that is almost always significantly better than the alternatives' and nearly as good as the omniscient oracle's.  The difference in TE quality is especially stark at the higher percentiles. Relative to the compared TE schemes, {\sysname} offers MLU up to $25$\% better at the median and $170$\% better at the 99th percentile.
\end{Itemize} 

\begin{figure*}[t!]
\subfigure[{\pwandc} \& {\pwan}]{
  \includegraphics[width=.48\linewidth]{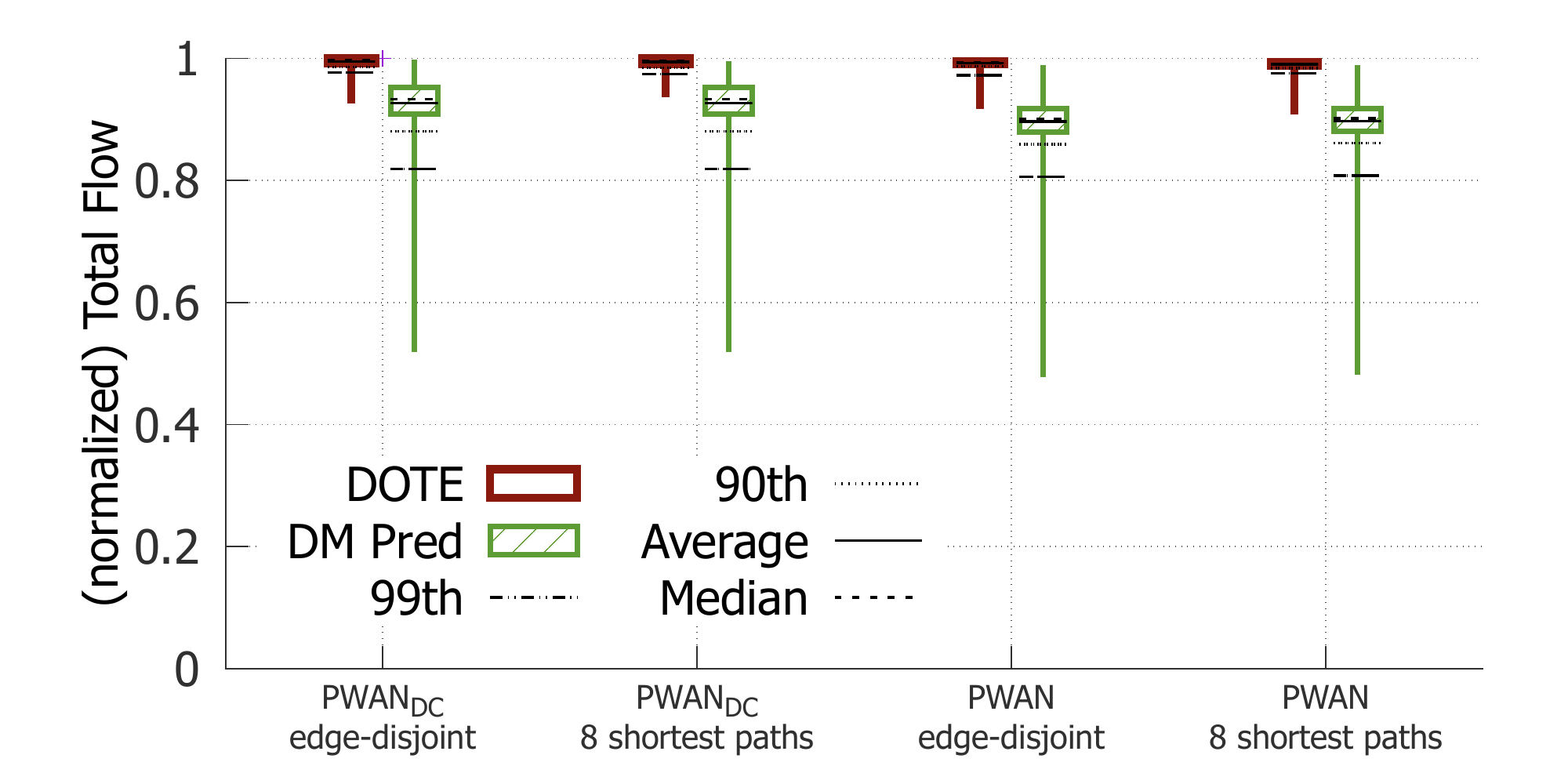}
\label{f:mcf_pwans}
  }
\subfigure[Abilene \& GEANT]{
  \includegraphics[width=.48\linewidth]{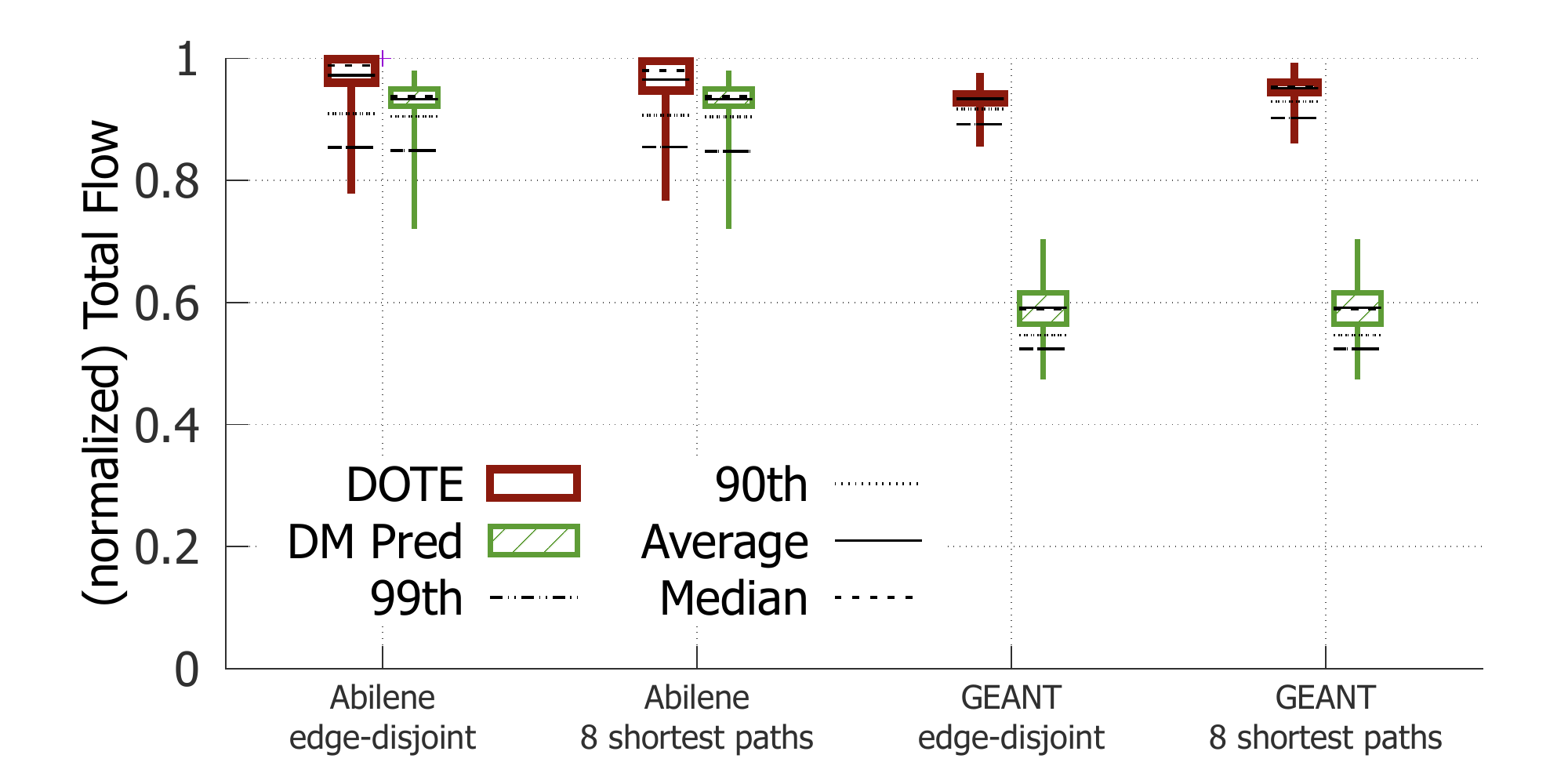}
 \label{f:mcf_abilene_geant}
  }
\vspace{-0.15in}
\caption{TE quality when aiming to maximize total flow with two different tunnel choices.\label{f:maxflow}}
\end{figure*}

\begin{figure*}[htb]
\subfigure[{\pwandc} \& {\pwan}]{
  \includegraphics[width=.48\linewidth]{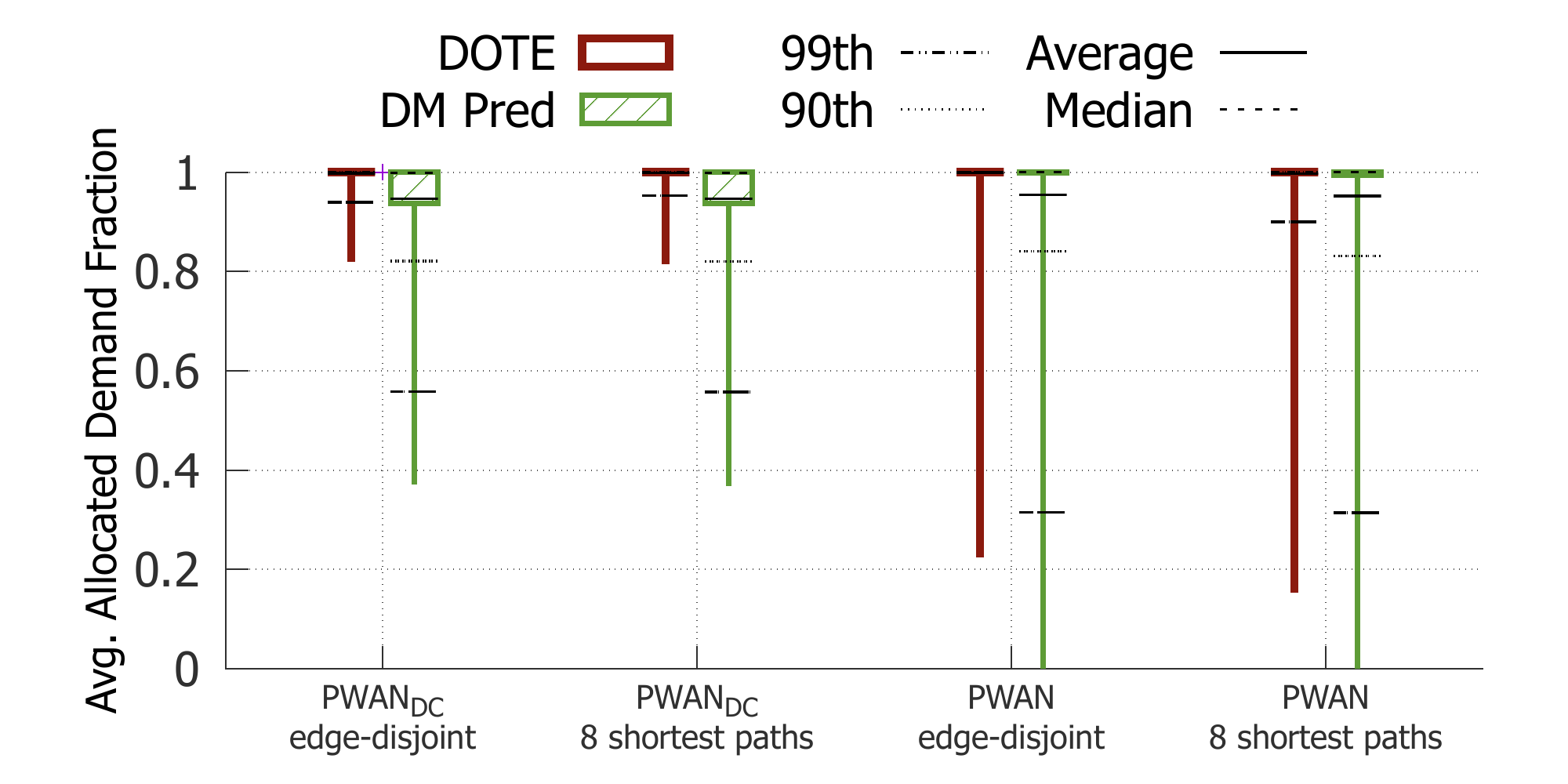}
  }
\subfigure[Abilene \& GEANT]{
  \includegraphics[width=.48\linewidth]{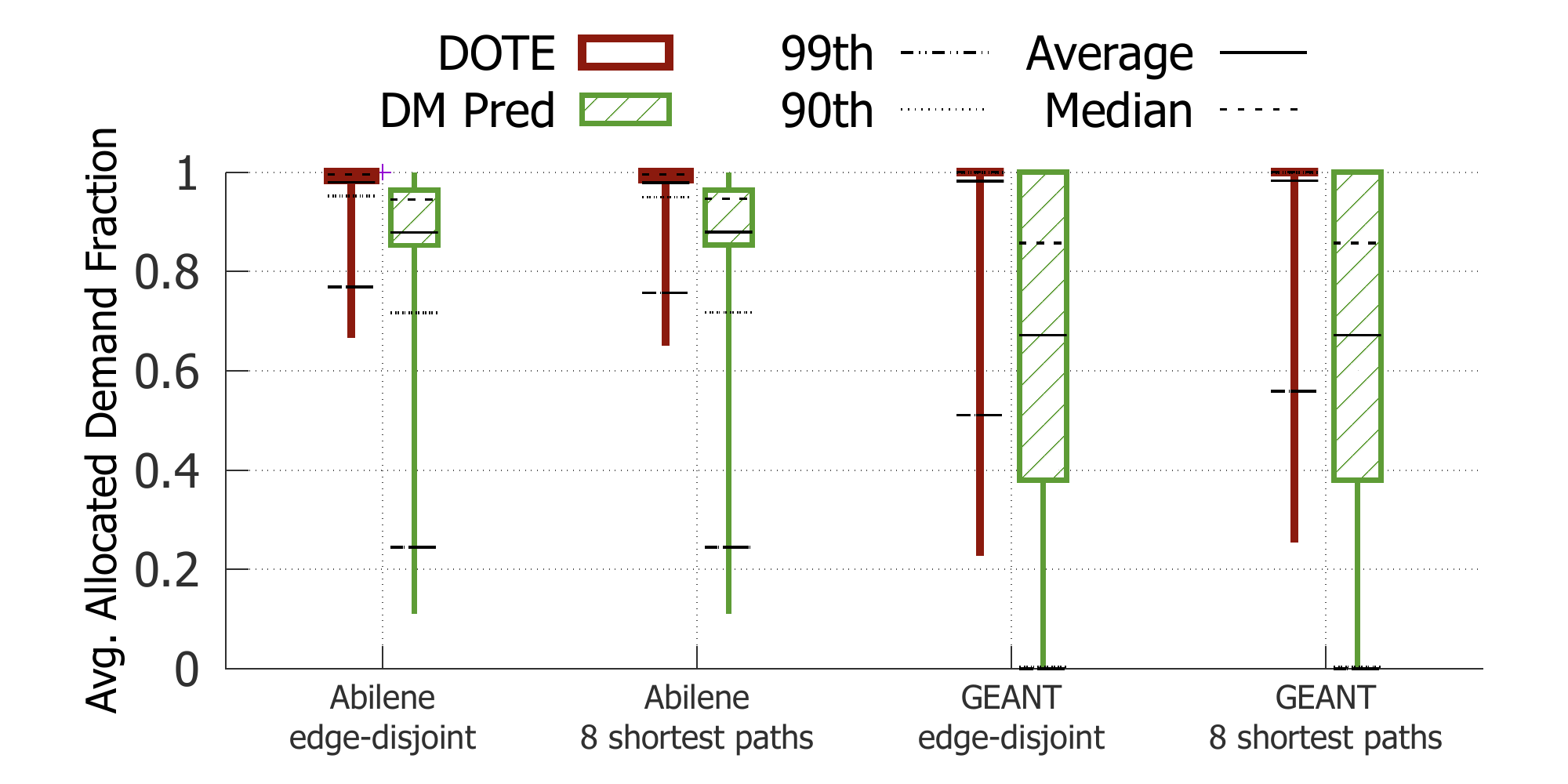}
  }
\vspace{-0.15in}
\caption{TE quality when aiming to maximize the concurrent flow for two different tunnel choices. For each demand matrix, we compute the fraction of demand satisfied for each source and destination, and sort these values into a vector. Across many hundreds of demand matrices, the candlesticks plot the average over all such allocation vectors. Note: allocating more flow is better. The box in each candlestick is the 25th and 75th percentile (fractional allocation) and the whiskers go from min to max value.\label{f:concurrentflow}}
\end{figure*}
\begin{figure*}[t!]
\subfigure[{\pwandc} \& {\pwan}]{
  \includegraphics[width=.48\linewidth]{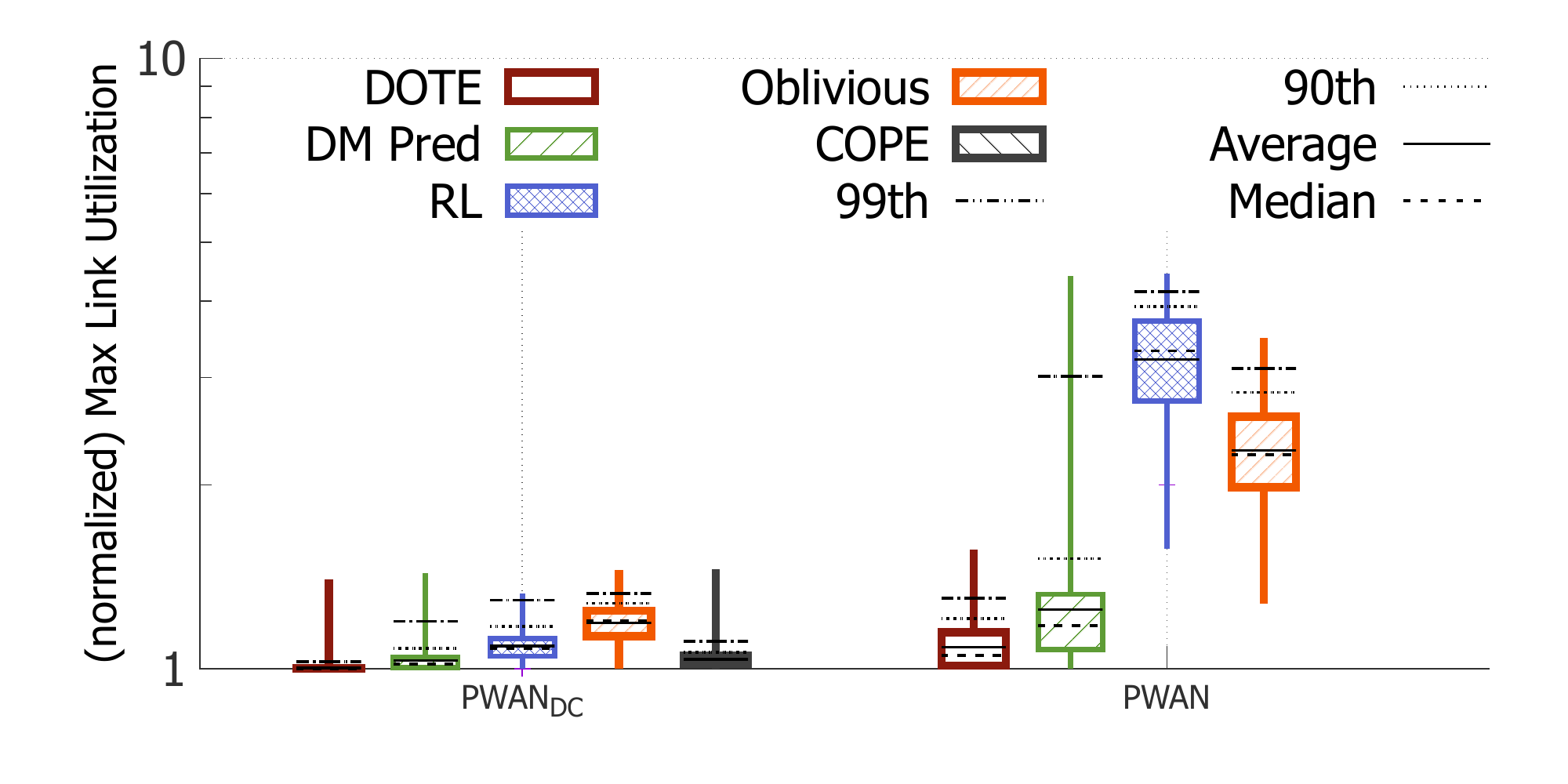}
\label{f:mlu_pwans_disjoint}
  }
\subfigure[Abilene \& GEANT]{
  \includegraphics[width=.48\linewidth]{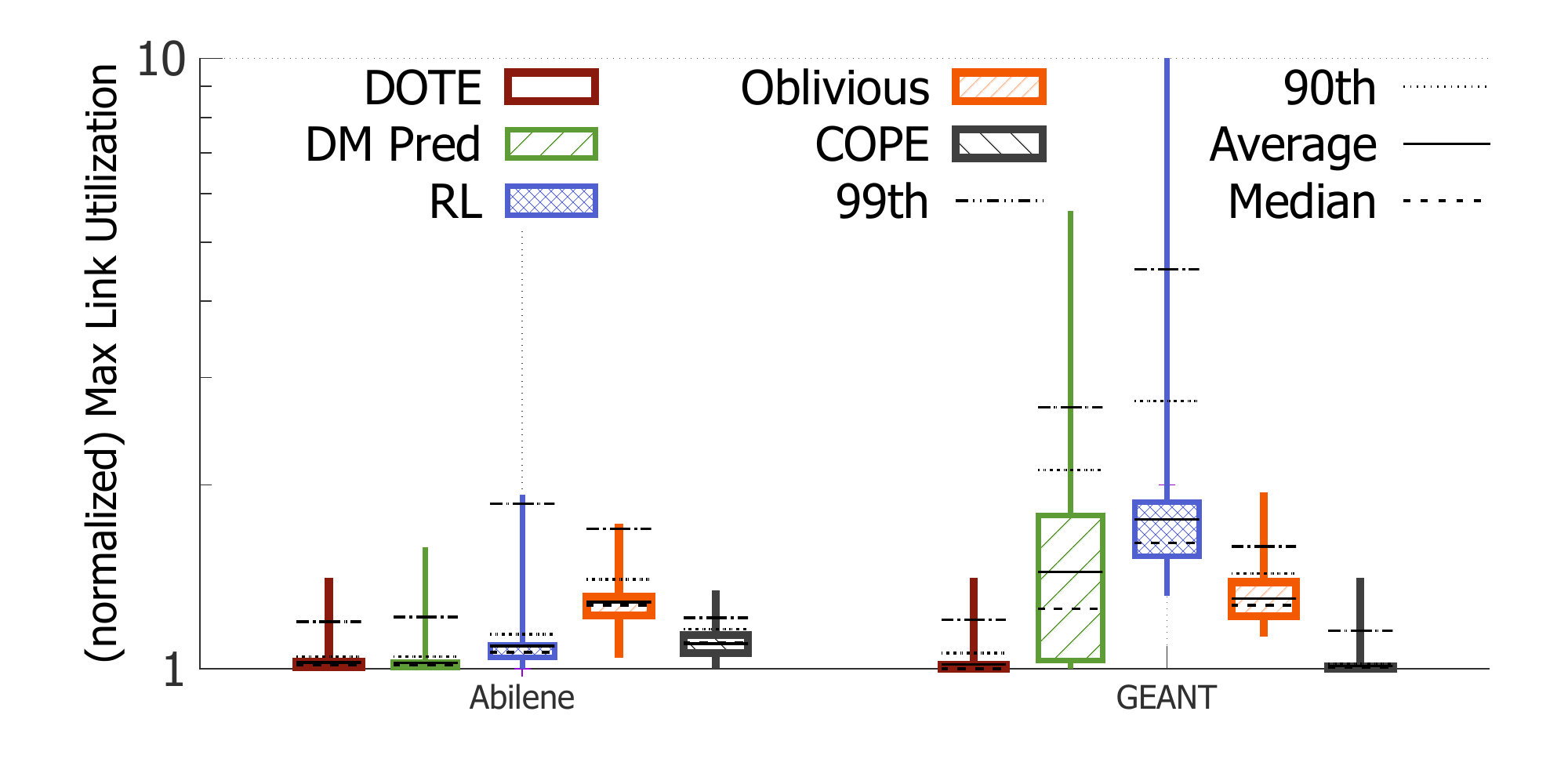}
\label{f:mlu_abilene_geant_disjoint}
  }
\vspace{-0.15in}
\caption{TE quality when aiming to minimize MLU with all possible edge-disjoint paths.\label{f:mlu_edge_disjoint}}
\end{figure*}
\begin{figure}[t!]
\centering
  \includegraphics[width=.99\linewidth]{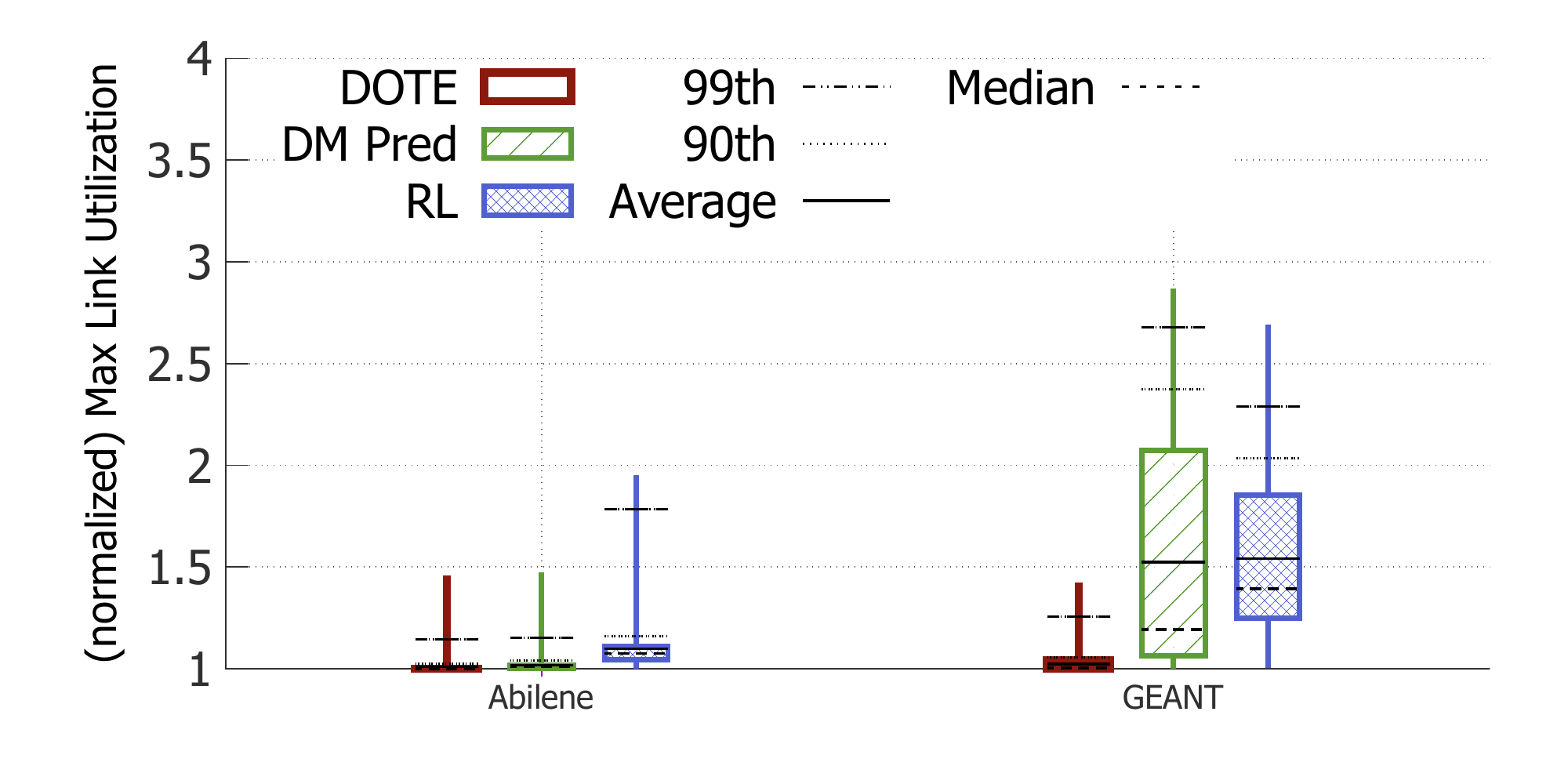}
\label{f:mlu_abilene_geant_smore}
\vspace{-0.2in}
\caption{TE quality when aiming to minimize MLU with routing trees chosen by SMORE.\label{f:mlu_smore_trees}}
\end{figure}

\vspace{0.05in}\noindent{\bf Runtimes.} \autoref{t:runtimes} presents a comparison of runtimes across TE schemes. The table presents the latency of applying each TE scheme to a new demand matrix and, wherever appropriate, the required precomputation time. Demand-oblivious schemes~\cite{Cohen2003} and COPE~\cite{cope} do not change the TE configuration online but involve very long precomputation latency and require very large memory. {\sysname} performs both precomputation on historical demands (training the DNN) and online computation (invoking the DNN). SMORE's online computation involves solving an LP to optimize over predicted demand matrices and so its latency is roughly as high as the LP's latency in the table. To compute Racke's routing trees, SMORE requires several hours on the larger topologies. 

The table shows that {\sysname}'s inference time is faster than the latency of using LPs to optimize over one (predicted) DM. The LP's latency is on par with results in recent studies~\cite{POP,NC}. {\sysname}'s online computation is short because it is effectively a few matrix multiplications.\footnote{Input is $12$ demand matrices and output is splitting ratios or one double per tunnel per demand. On the large {\pwan} network, both the input and output are a few tens of MBs.}  LP computation latency increases super-linearly with the network size and prior work notes that solver times can exceed several minutes on networks with thousands of nodes and edges~\cite{POP,NC}; {\sysname}'s inference latency on large WANs, such as KDL~(see~\autoref{tab:network_sizes}), is still within a few seconds. {\sysname}'s training time is less than $12$ hours for {\pwan} and can be accelerated using standard methods (e.g., by parallelization, SIMD and other model training enhancements).

COPE's precomputation latency is a few orders of magnitude higher than that of the demand-oblivious TE, which is, itself, a couple orders of magnitude higher than that of prediction-based TE. COPE also has much higher memory requirements~(over $256$GB on {\pwan}); in fact, on {\pwan}, COPE did not finish pre-computation even after four days on a $8$-core VM with $256$GB running Gurobi~\cite{gurobi} vers.~9.1, and hence \autoref{f:money} includes no results for COPE on {\pwan}. To understand COPE's runtime complexity better, we ran it on WAN topologies from Topology Zoo~\cite{topozoo} that are larger than GEANT and {\pwandc} but smaller than {\pwan}. On JanetBackbone which has $29$ nodes and $45$ edges, COPE ran for $1.5$ hours and on SurfNet ($50$ nodes, $68$ edges), COPE did not finish even in $10$ hours. These results suggest that COPE is inapplicable to large WANs.\footnote{Per Table 1 in~\cite{cope}, the previously published results on COPE are on much smaller topologies than considered here.}

\subsection{Generalizing to Other TE Objectives and Tunnel Choices}
\label{ss:e_generalization}
Here, we present results for two additional TE objectives -- maximizing multi-commodity-flow and maximizing concurrent flow-- as well as two other choices for tunnels. 

Note that some of the compared alternatives to \emph{DOTE}, namely, demand-oblivious TE~\cite{Cohen2003} and COPE~\cite{cope}, do not readily apply to these TE objectives (as both build on results from oblivious routing theory that provide provable guarantees for MLU minimization), and it is not clear how to extend them to other objectives. Our evaluation of \emph{DOTE} for these metrics is therefore restricted to benchmarking against the omniscient oracle and prediction-based TE.

\vspace{0.05in}\noindent{\bf Maximizing Total Flow: } \autoref{f:maxflow} compares {\sysname} with prediction-based TE on all four WANs for two different tunnel choices when the TE objective is to carry as much total flow as possible while respecting capacity constraints. Observe that {\sysname} carries substantially more flow and closely approximates the TE quality of the omniscient oracle.  As before, the gap between {\sysname} and prediction-based TE is larger on WANs where demands are less predictable (i.e., all WANs but Abilene) and at the higher percentiles. Generally, {\sysname} may be able to carry $10$\% to $20$\% more flow.

\vspace{0.05in}\noindent{\bf Maximizing Concurrent Flow: } \autoref{f:concurrentflow} compares {\sysname} with the omniscient oracle and prediction-based TE when the TE objective is to maximize the minimum fraction of demand satisfied across all demands. Observe that {\sysname} fully allocates almost all of the demands (the upper candlesticks are at $y=1$), whereas prediction-based TE allocates a smaller fraction of the demanded volume for many more demands.

\vspace{0.05in}\noindent{\bf Tunnel choice} does not qualitatively change our results for TE performance; contrast~\autoref{f:mlu_edge_disjoint} and \autoref{f:mlu_smore_trees} with~\autoref{f:money}. Note that when using Racke's routing trees (as in SMORE) prediction-based TE coincides with SMORE.

\begin{figure*}[t!]
\subfigure[8SP]{
  \includegraphics[width=.48\linewidth]{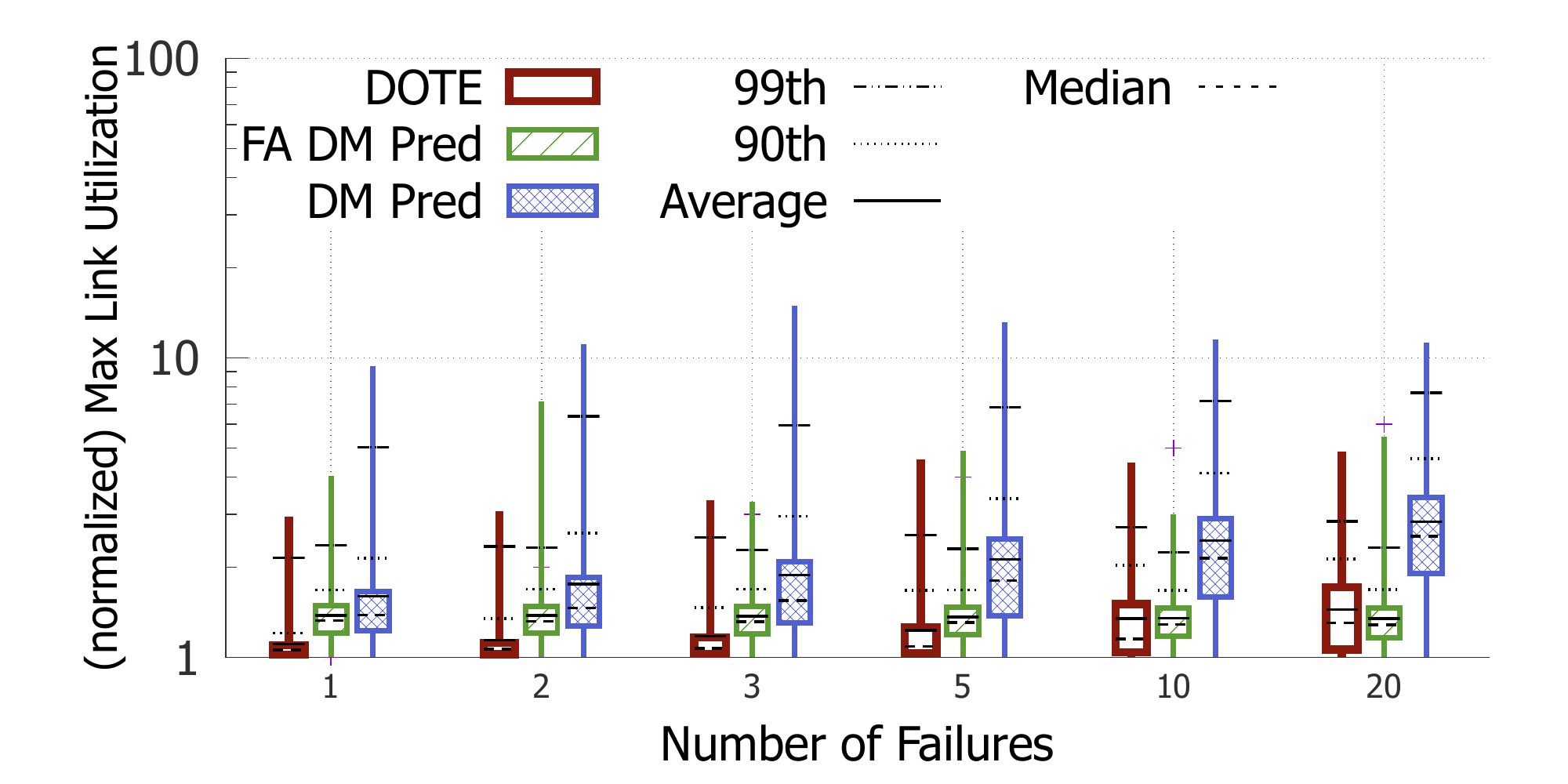}
  }
\subfigure[edge-disjoint]{
  \includegraphics[width=.48\linewidth]{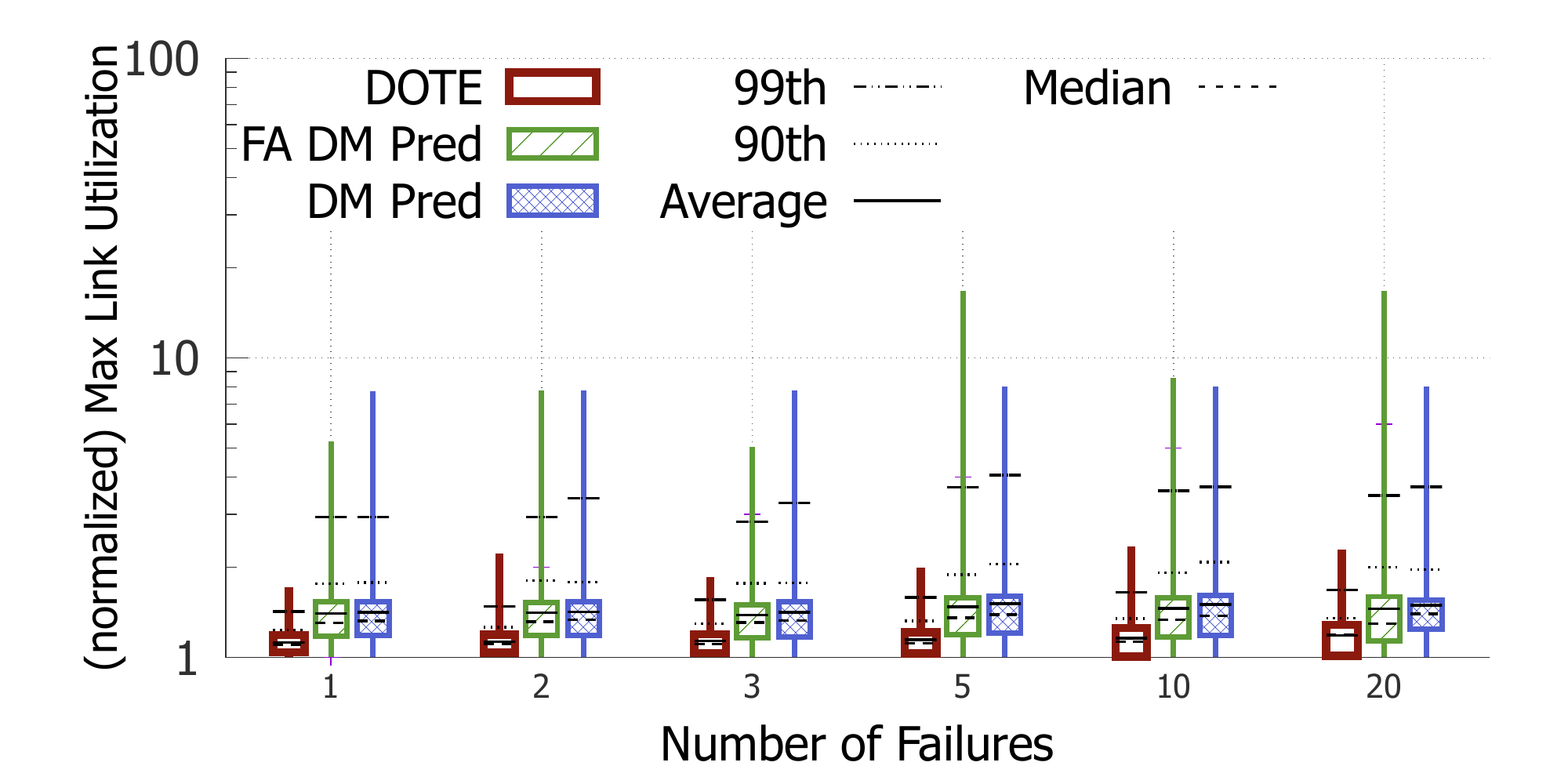}
  }
\vspace{-0.15in}
\caption{Coping with different numbers of random link failures on {\pwan}; the candlesticks show the distribution over $1700$ different randomly chosen failure cases.\label{f:failures_pwan}}
\end{figure*}

\subsection{Coping with Network Failures}
\label{ss:net_faults}

\autoref{f:failures_pwan} shows how {\sysname} performs, in terms of MLU, when different numbers of (randomly chosen) links fail in the {\pwan} topology. As noted in~\xref{ss:arch}, {\sysname} assumes that source nodes (or tunnel heads~\cite{MPLS-FR}) identify tunnels that fail and rebalance traffic proportionally among the surviving tunnels. The figure compares \emph{DOTE} with two variants of prediction-based TE: {\sf DM Pred.} which, similar to {\sysname}, has no a priori knowledge of future traffic demands or the link faults, and {\sf FA DM Pred.} which is identical to {\sf DM Pred.} except that it is fault-aware, i.e., knows the links that will fail. Our quality metric is still normalized MLU except that we now normalize based on an omniscient oracle that has perfect knowledge of \textit{both} future traffic demands \emph{and} the failures.

Our results show that \emph{DOTE} outperforms both demand-prediction-based TE~({\sf DM Pred.}) and demand-prediction-based TE with oracle access to future failures~({\sf FA DM Pred.}) for many concurrent link failures with different tunnel choices. We interpret this result as indicating that the error in demand predictions weights more heavily on attaining a good TE objective than the confusion induced by these link failures. Our results on other topologies (Abilene, GEANT, and {\pwandc}) and for the maximum-multicommodity-flow objective show a similar trend (\autoref{f:failures_other_topologies} and \autoref{f:failures_max_mcf}).

\cut{
The figure shows that the normalized MLU increases in spite of {\sysname}'s adaptation. The increase in MLU is larger when more links fail and when fewer tunnels are available (with edge-disjoint tunnel choice, demands generally have fewer than $8$ tunnels each). However, note also that the fault-aware oracle performs worse than {\sysname}. We believe that this is because the error in demand predictions weights more heavily on the TE objective than the confusion induced by link failures. Note that all production TE solvers~\cite{SWAN,B4} re-compute allocations periodically and so the effect of a fault lasts only until the next computation step -- typically five minutes -- at which time {\sysname} can optimize on the topology that remains after faults.}

\subsection{Robustness to Traffic Noise and Drift} 
\label{subsec:robust-traffic}
\label{ss:traffic_change}

\noindent{\bf Robustness to unexpected traffic changes.} To assess \emph{DOTE}'s robustness to noisy traffic, we evaluate \emph{DOTE} on the GEANT, Cogentco, and GtsCe WANs~\cite{topozoo}, where each demand in the realized DM is independently multiplied by a factor chosen uniformly at random from $[1-\alpha, 1+\alpha]$ for $\alpha \in \{0.1, 0.25, 0.35\}$. Our results (see~\xref{apx:robust}) show that under such traffic perturbations, the distance, in terms of MLU, from the omniscient oracle remains low across all evaluated WANs (e.g., $2$\%, $2.9$\%, and $3.8$\% for $\alpha=0.1,0.25,0.35$ for GEANT with edge-disjoint tunnels).

\vspace{0.05in}\noindent{\bf Robustness to natural traffic drift.} We investigate to what extent the quality of \emph{DOTE}'s TE configurations deteriorates when \emph{DOTE} is not frequently retrained. We quantify the distance from the omniscient oracle, in terms of both MLU and maximum-multicommodity-flow, of the average \textit{weekly} value achieved by {\sysname} on the Abilene and GEANT WANs over $4$ consecutive weeks (without retraining \emph{DOTE}). See~\autoref{t:weekly_dist_from_opt_mlu} and~\autoref{t:weekly_dist_from_opt_max_mcf} in the Appendix. Our results show that while the distance from the optimum increases over time, in general, {\sysname} remains close to the optimum~(within a few $\%$ on average) even weeks after the model is trained. This suggests that {\sysname} can provide high quality TE even if it was re-trained once every month. {\sysname}'s training time~(see~\autoref{t:runtimes}) allows for much more frequent retraining.

\section{Limitations and Future Research}\label{discussion}

We believe that our investigation of direct optimization for WAN TE has but scratched the surface and outline below current limitations of our approach, as well as intriguing directions for future research.

\vspace{0.05in}\noindent\textbf{Extending \emph{DOTE} to support latency-sensitive traffic.} To accommodate latency-sensitive traffic, the following strategy (similarly to~\cite{blastshield}) could be employed: reserve shortest paths (tunnels) for such traffic and always schedule short/latency-sensitive traffic flows to these paths.

\vspace{0.05in}\noindent\textbf{More expressive neural network architectures.} Our realization of $\emph{DOTE}$ uses a relatively simple neural network that does not leverage knowledge of the WAN topology. Consequently, the neural network has to (implicitly) learn the network topology during training. Directly incorporating the WAN structure into \emph{DOTE} using Graph Convolutional Networks~\cite{gcnsurvey} could potentially lead to faster training and/or better quality solutions.

\vspace{0.05in}\noindent\textbf{Extending \emph{DOTE} to incorporate data-driven tunnel selection.} Our discussion of \emph{DOTE} assumed an underlying tunnel-selection scheme. \emph{DOTE} can be extended to support \textit{data-driven} tunnel-selection by adding DNN output variables specifying a probability distribution over a finite set of candidate tunnels (e.g., shortest-path, edge disjoint, SMORE). At the beginning of each time epoch, the tunnels to be used in that time epoch would be chosen according to this probability distribution. \emph{DOTE}'s optimality results extend to this setting. We defer a more thorough study of data-driven tunnel selection (e.g., not limited to a finite set of predetermined candidate tunnels) to future research.

\vspace{0.05in}\noindent\textbf{Learning to contend with link failures.} We described (\S\ref{ss:arch}) an approach for dealing with link failures in the data plane. An alternative is incorporating fault tolerance into the DNN training process by introducing random link failures.

\section{Related Work} \label{sec:related}

\noindent{\bf (WAN) TE.} TE has been extensively studied~\cite{jiang2009cooperative,TExCP,mate, fortz2000internet, benson2011microte, conga, te_in_dc, SWAN, B4, calendaring, zhang2017guaranteeing, FFC, SMORE, TeaVaR, Arrow} in a broad variety of settings, including legacy networks~\cite{Fortz,coyote}, datacenter networks~\cite{hedera}, and backbone networks~\cite{TExCP}. SDN-controlled WAN TE has also received extensive attention~\cite{SWAN, B4, FFC, BWE, EBB, SMORE, TeaVaR, Arrow}.

\vspace{0.05in}\noindent{\bf TE via oblivious routing, COPE, and SMORE.} Oblivious routing optimizes \textit{worst-case MLU} across \emph{all} possible DMs~\cite{Cohen2003, Azar2003obliv, Racke2002obliv}. Since oblivious routing does not exploit \emph{any} information about past traffic demands, it naturally yields suboptimal solutions~\cite{Cohen2003,SMORE}. COPE~\cite{cope} optimizes \textit{MLU} across a set of DMs spanned by previously observed DMs, while retaining a worst-case performance guarantee. Since COPE both extends oblivious routing \textit{and} optimizes over \textit{ranges} of demand matrices, its optimization phase is extremely time-consuming (\S\ref{ss:e_mlu_one_tunnel_choice}). The key conceptual difference between \emph{DOTE} and such ``robust TE'' schemes is in the goal of the pre-computation. Instead of emitting a single TE configuration that minimizes some cost function (specifically, MLU) over some predetermined set of DMs, DOTE’s objective is to identify a mapping from a vector of DMs from the recent past to the next TE configuration. DOTE thus achieves higher flexibility by being able to emit different TE configurations on a case-by-case basis, and is also able to pick up on temporal patterns in traffic demands. SMORE~\cite{SMORE} employs Racke's oblivious routing trees~\cite{Racke2002obliv} to produce static tunnels that are robust to traffic uncertainty, with traffic splitting ratios still optimized with respect to the (inferred/predicted) future traffic demands. Thus, SMORE can be thought of as a instantiation of prediction-based TE.

\vspace{0.05in}\noindent{\bf Online TE}~\cite{TExCP,mate,REPLEX}, wherein traffic configurations (such as splitting ratios) adapt automatically and in short timescales to the observed demands is an enticing design point for TE, but is challenging to achieve. TexCP~\cite{TExCP} requires WAN routers to offer novel explicit feedback, while MATE~\cite{mate} relies on changes in end-to-end latency and hence takes much longer to react and converge and is also less stable~\cite{TExCP}. Recently deployed TE schemes~\cite{SWAN,B4} (see~\xref{s:prelim} and~\autoref{f:arch}) are simpler and easier to deploy because they replace such distributed, closed-loop, short-timescale control with centralized, open-loop and periodic adaptation. We view online TE as complementary to \emph{DOTE}; \emph{DOTE} could be used to \textit{periodically} compute a TE configuration while online TE could be \textit{continuously} used in between \emph{DOTE} updates to tweak this TE configuration in response to changes in network conditions.

\vspace{0.05in}\noindent{\bf Reinforcement-learning-based TE.} %
Demand-prediction-based and RL approaches to TE are contrasted in~\cite{l2r} in terms of MLU only on a small network ($12$ nodes and $32$ edges) for \textit{synthetic} traffic patterns and a model of hop-by-hop routing that does not capture routing along tunnels. Our theoretical and empirical results reveal that \emph{DOTE}'s stochastic optimization scheme outperforms both demand-prediction-based and RL-based TE.

\vspace{0.05in}\noindent{\bf Some recent work on TE~\cite{NC,POP}} speeds up the multicommodity flow computations that underpin TE optimization by effectively breaking the large LPs into smaller problems that can be solved in parallel. However, these approaches still rely on predicted demand matrices (unlike {\sysname}). {\sysname} offers an alternate way to speed up TE: replacing the LP solver with invocations of a fairly small DNN. This has the potential to be innately more efficient.

\section{Conclusion}\label{sec:conclusion}

We presented a new framework for WAN TE: data-driven end-to-end stochastic optimization using only historical information about traffic demands. Our theoretical and empirical results establish that this approach closely approximates the optimal TE configuration, significantly outperforming previously proposed TE schemes in terms of both solution quality and runtimes.

\newpage

\bibliographystyle{plain}
\bibliography{bib}

\begin{thebibliography}{10}

\bibitem{gc_ratelimits}
Google cloud armor: Rate limiting overview.
\newblock \url{https://bit.ly/3TnI1mO}.

\bibitem{repo_code}
{\em Github repo containing our code}.
\newblock 2022.
\newblock \url{https://github.com/PredWanTE/DOTE}.

\bibitem{abilene}
Abilene/Internet2.
\newblock \url{http://www.internet2.edu/}.

\bibitem{NC}
Firas Abuzaid, Srikanth Kandula, Behnaz Arzani, Ishai Menache, Matei Zaharia,
  and Peter Bailis.
\newblock Contracting wide-area network topologies to solve flow problems
  quickly.
\newblock In {\em 18th {USENIX} Symposium on Networked Systems Design and
  Implementation ({NSDI)}}, pages 175--200, 2021.

\bibitem{te_in_dc}
Ian~F. Akyildiz, Ahyoung Lee, Pu~Wang, Min Luo, and Wu~Chou.
\newblock A roadmap for traffic engineering in sdn-openflow networks.
\newblock {\em Comput. Netw.}, 71:1--30, October 2014.

\bibitem{hedera}
Mohammad Al-Fares, Sivasankar Radhakrishnan, Barath Raghavan, Nelson Huang, and
  Amin Vahdat.
\newblock Hedera: Dynamic flow scheduling for data center networks.
\newblock In {\em NSDI}, 2010.

\bibitem{conga}
Mohammad Alizadeh, Tom Edsall, Sarang Dharmapurikar, Ramanan Vaidyanathan,
  Kevin Chu, Andy Fingerhut, Vinh~The Lam, Francis Matus, Rong Pan, Navindra
  Yadav, and George Varghese.
\newblock Conga: Distributed congestion-aware load balancing for datacenters.
\newblock {\em SIGCOMM Comput. Commun. Rev.}, 44(4):503--514, August 2014.

\bibitem{Cohen2003}
David Applegate and Edith Cohen.
\newblock Making {I}ntra-{D}omain {R}outing {R}obust to {C}hanging and
  {U}ncertain {T}raffic {D}emands.
\newblock In {\em SIGCOMM}, 2003.

\bibitem{Azar2003obliv}
Yossi Azar, Edith Cohen, Amos Fiat, Haim Kaplan, and Harald Racke.
\newblock Optimal oblivious routing in polynomial time.
\newblock In {\em Proceedings of the Thirty-fifth Annual ACM Symposium on
  Theory of Computing}, STOC '03, pages 383--388, 2003.

\bibitem{benson2011microte}
Theophilus Benson, Ashok Anand, Aditya Akella, and Ming Zhang.
\newblock {MicroTE: Fine grained traffic engineering for data centers}.
\newblock In {\em Proceedings of the Seventh COnference on emerging Networking
  EXperiments and Technologies}, page~8. ACM, 2011.

\bibitem{TeaVaR}
Jeremy Bogle, Nikhil Bhatia, Manya Ghobadi, Ishai Menache, Nikolaj Bj{\o}rner,
  Asaf Valadarsky, and Michael Schapira.
\newblock {TEAVAR:} striking the right utilization-availability balance in
  {WAN} traffic engineering.
\newblock In {\em Proceedings of the {ACM} Special Interest Group on Data
  Communication, {SIGCOMM} 2019}, pages 29--43, 2019.

\bibitem{catapult1}
Adrian~M. Caulfield, Eric~S. Chung, Andrew Putnam, Hari Angepat, Jeremy Fowers,
  Michael Haselman, Stephen Heil, Matt Humphrey, Puneet Kaur, Joo-Young Kim,
  Daniel Lo, Todd Massengill, Kalin Ovtcharov, Michael Papamichael, Lisa Woods,
  Sitaram Lanka, Derek Chiou, and Doug Burger.
\newblock A cloud-scale acceleration architecture.
\newblock In {\em MICRO}, 2016.

\bibitem{coyote}
Marco Chiesa, G\'{a}bor R{\'e}tv\'{a}ri, and Michael Schapira.
\newblock Lying your way to better traffic engineering.
\newblock CoNEXT, 2016.

\bibitem{mate}
A.~Elwalid, C.~Jin, S.~Low, and I.~Widjaja.
\newblock Mate: Mpls adaptive traffic engineering.
\newblock In {\em Proceedings of IEEE {INFOCOM}}, volume~3, pages 1300--1309
  vol.3, 2001.

\bibitem{REPLEX}
Simon Fischer, Nils Kammenhuber, and Anja Feldmann.
\newblock Replex: Dynamic traffic engineering based on wardrop routing
  policies.
\newblock In {\em Proceedings of the 2006 ACM CoNEXT Conference}, 2006.

\bibitem{fortz2000internet}
Bernard Fortz and Mikkel Thorup.
\newblock Internet traffic engineering by optimizing ospf weights.
\newblock In {\em INFOCOM 2000. Nineteenth annual joint conference of the IEEE
  computer and communications societies. Proceedings. IEEE}, volume~2, pages
  519--528. IEEE, 2000.

\bibitem{Fortz}
Bernard Fortz and Mikkel Thorup.
\newblock Increasing internet capacity using local search.
\newblock {\em Computational Optimization and Applications}, 2004.

\bibitem{garg2007faster}
Naveen Garg and Jochen K{\"o}nemann.
\newblock Faster and simpler algorithms for multicommodity flow and other
  fractional packing problems.
\newblock {\em SIAM Journal on Computing}, 37(2):630--652, 2007.

\bibitem{gurobi}
Zonghao Gu, Edward Rothberg, and Robert Bixby.
\newblock {Gurobi Optimizer Reference Manual, Version 5.0}.
\newblock {\em Gurobi Optimization Inc., Houston, USA}, 2012.

\bibitem{hazan2015beyond}
Elad Hazan, Kfir Levy, and Shai Shalev-Shwartz.
\newblock Beyond convexity: Stochastic quasi-convex optimization.
\newblock {\em Advances in neural information processing systems}, 28, 2015.

\bibitem{henderson2018deep}
Peter Henderson, Riashat Islam, Philip Bachman, Joelle Pineau, Doina Precup,
  and David Meger.
\newblock Deep reinforcement learning that matters.
\newblock In {\em Proceedings of the AAAI conference on artificial
  intelligence}, volume~32, 2018.

\bibitem{SWAN}
Chi-Yao Hong, Srikanth Kandula, Ratul Mahajan, Ming Zhang, Vijay Gill, Mohan
  Nanduri, and Roger Wattenhofer.
\newblock Achieving high utilization with software-driven wan.
\newblock SIGCOMM, 2013.

\bibitem{b4_and_after}
Chi-Yao Hong, Subhasree Mandal, Mohammad Al-Fares, Min Zhu, Richard Alimi,
  Kondapa~Naidu B., Chandan Bhagat, Sourabh Jain, Jay Kaimal, Shiyu Liang,
  Kirill Mendelev, Steve Padgett, Faro Rabe, Saikat Ray, Malveeka Tewari, Matt
  Tierney, Monika Zahn, Jonathan Zolla, Joon Ong, and Amin Vahdat.
\newblock B4 and after: Managing hierarchy, partitioning, and asymmetry for
  availability and scale in google's software-defined wan.
\newblock {\em SIGCOMM '18}, pages 74--87, 2018.

\bibitem{B4}
Sushant Jain, Alok Kumar, Subhasree Mandal, Joon Ong, Leon Poutievski, Arjun
  Singh, Subbaiah Venkata, Jim Wanderer, Junlan Zhou, Min Zhu, Jon Zolla, Urs
  H\"{o}lzle, Stephen Stuart, and Amin Vahdat.
\newblock B4: Experience with a globally-deployed software defined wan.
\newblock SIGCOMM, 2013.

\bibitem{multicommodity}
William~S. Jewell.
\newblock {\em Multi-commodity Network Solutions}.
\newblock 1966.

\bibitem{jiang2009cooperative}
Wenjie Jiang, Rui Zhang-Shen, Jennifer Rexford, and Mung Chiang.
\newblock Cooperative content distribution and traffic engineering in an isp
  network.
\newblock In {\em ACM SIGMETRICS Performance Evaluation Review}, volume~37,
  pages 239--250. ACM, 2009.

\bibitem{TExCP}
Srikanth Kandula, Dina Katabi, Bruce Davie, and Anna Charny.
\newblock Walking the tightrope: Responsive yet stable traffic engineering.
\newblock In {\em SIGCOMM}. ACM, 2005.

\bibitem{calendaring}
Srikanth Kandula, Ishai Menache, Roy Schwartz, and Spandana~Raj Babbula.
\newblock Calendaring for wide area networks.
\newblock In {\em SIGCOMM}, 2014.

\bibitem{karakostas}
George Karakostas.
\newblock {Faster Approximation Schemes for Fractional Multicommodity Flow
  Problems}.
\newblock {\em ACM Trans. Algorithms}, 2008.

\bibitem{adam}
Diederik~P. Kingma and Jimmy Ba.
\newblock Adam: A method for stochastic optimization learning.
\newblock {\em arXiv preprint arXiv:1412.6980}, 2014.

\bibitem{topozoo}
S.~Knight, H.X. Nguyen, N.~Falkner, R.~Bowden, and M.~Roughan.
\newblock The internet topology zoo.
\newblock {\em IEEE Journal on Selected Areas in Communications}, 2011.

\bibitem{Kodialam2011}
M~Kodialam, T~V Lakshman, and S~Sengupta.
\newblock Traffic-oblivious routing in the hose model.
\newblock {\em IEEE/ACM Transactions on Networking}, 19(3):774 -- 787, 2011.

\bibitem{konnov2003convergence}
Igor~V Konnov.
\newblock On convergence properties of a subgradient method.
\newblock {\em Optimization Methods and Software}, 18(1):53--62, 2003.

\bibitem{blastshield}
Umesh Krishnaswamy, Rachee Singh, Nikolaj Bjørner, and Himanshu Raj.
\newblock Decentralized cloud wide-area network traffic engineering with
  {BlastShield}.
\newblock Technical Report MSR-TR-2021-31, Microsoft Research, 2021.

\bibitem{BWE}
Alok Kumar, Sushant Jain, Uday Naik, Nikhil Kasinadhuni, Enrique~Cauich
  Zermeno, C.~Stephen Gunn, Jing Ai, Björn Carlin, Mihai Amarandei-Stavila,
  Mathieu Robin, Aspi Siganporia, Stephen Stuart, and Amin Vahdat.
\newblock Bwe: Flexible, hierarchical bandwidth allocation for wan distributed
  computing.
\newblock In {\em Sigcomm '15}, 2015.

\bibitem{yates}
Praveen Kumar, Chris Yu, Yang Yuan, Nate Foster, Robert Kleinberg, and Robert
  Soul{\'e}.
\newblock Yates: Rapid prototyping for traffic engineering systems.
\newblock In {\em Proceedings of the Symposium on SDN Research}, SOSR '18,
  pages 11:1--11:7, New York, NY, USA, 2018. ACM.

\bibitem{SMORE}
Praveen Kumar, Yang Yuan, Chris Yu, Nate Foster, Robert Kleinberg, Petr
  Lapukhov, Chiun~Lin Lim, and Robert Soul{\'e}.
\newblock Semi-oblivious traffic engineering: The road not taken.
\newblock In {\em 15th {USENIX} Symposium on Networked Systems Design and
  Implementation ({NSDI} 18)}, pages 157--170, Renton, WA, 2018. {USENIX}
  Association.

\bibitem{EBB}
George Leopold.
\newblock {Building Express Backbone: Facebook’s new long-haul network}.
\newblock \url{http://code.facebook.com/posts/1782709872057497/}, 2017.

\bibitem{FFC}
Hongqiang~Harry Liu, Srikanth Kandula, Ratul Mahajan, Ming Zhang, and David
  Gelernter.
\newblock Traffic engineering with forward fault correction.
\newblock In {\em {ACM} {SIGCOMM} 2014 Conference, SIGCOMM'14, Chicago, IL,
  USA, August 17-22, 2014}, pages 527--538, 2014.

\bibitem{POP}
Deepak Narayanan, Fiodar Kazhamiaka, Firas Abuzaid, Peter Kraft, Akshay
  Agrawal, Srikanth Kandula, Stephen Boyd, and Matei Zaharia.
\newblock Solving large-scale granular resource allocation problems efficiently
  with {POP}.
\newblock In {\em Proceedings of the ACM SIGOPS 28th Symposium on Operating
  Systems Principles}, pages 521--537, 2021.

\bibitem{paszke2019pytorch}
Adam Paszke, Sam Gross, Francisco Massa, Adam Lerer, James Bradbury, Gregory
  Chanan, Trevor Killeen, Zeming Lin, Natalia Gimelshein, Luca Antiga, et~al.
\newblock Pytorch: An imperative style, high-performance deep learning library.
\newblock {\em Advances in neural information processing systems},
  32:8026--8037, 2019.

\bibitem{Racke2002obliv}
Harald R\"{a}cke.
\newblock Minimizing congestion in general networks.
\newblock In {\em Proceedings of the 43rd Symposium on Foundations of Computer
  Science}, FOCS '02, 2002.

\bibitem{senic}
Sivasankar Radhakrishnan, Yilong Geng, Vimalkumar Jeyakumar, Abdul Kabbani,
  George Porter, and Amin Vahdat.
\newblock Senic: Scalable nic for end-host rate limiting.
\newblock In {\em NSDI}, 2014.

\bibitem{MPLS-FR}
E.~Rosen, A.~Viswanathan, and R.~Callon.
\newblock {Multi-Protocol {L}abel {S}witching {A}rchitecture}.
\newblock RFC 3031.

\bibitem{gravity}
Matthew Roughan, Albert Greenberg, Charles Kalmanek, Michael Rumsewicz,
  Jennifer Yates, and Yin Zhang.
\newblock Experience in measuring backbone traffic variability: Models,
  metrics, measurements and meaning.
\newblock IMW, 2002.

\bibitem{Roughan-TE}
Matthew Roughan, Mikkel Thorup, and Yin Zhang.
\newblock Performance of estimated traffic matrices in traffic engineering.
\newblock In {\em SIGMETRICS}, 2003.

\bibitem{carousel}
Ahmed Saeed, Nandita Dukkipati, Vytautas Valancius, Vinh~The Lam, Carlo
  Contavalli, and Amin Vahdat.
\newblock Carousel: Scalable traffic shaping at end hosts.
\newblock In {\em SIGCOMM}, 2017.

\bibitem{Shahrokhi1990TheMC}
Farhad Shahrokhi and David~W. Matula.
\newblock The maximum concurrent flow problem.
\newblock {\em J. ACM}, 37:318--334, 1990.

\bibitem{shalev2014understanding}
Shai Shalev-Shwartz and Shai Ben-David.
\newblock {\em Understanding machine learning: From theory to algorithms}.
\newblock Cambridge university press, 2014.

\bibitem{shalev2009stochastic}
Shai Shalev-Shwartz, Ohad Shamir, Nathan Srebro, and Karthik Sridharan.
\newblock Stochastic convex optimization.
\newblock In {\em COLT}, volume~2, page~5, 2009.

\bibitem{shapiro2021lectures}
Alexander Shapiro, Darinka Dentcheva, and Andrzej Ruszczynski.
\newblock {\em Lectures on stochastic programming: modeling and theory}.
\newblock SIAM, 2021.

\bibitem{Joint-failure-TE}
Martin Suchara, Dahai Xu, Robert Doverspike, David Johnson, and Jennifer
  Rexford.
\newblock Network architecture for joint failure recovery and traffic
  engineering.
\newblock In {\em Proceedings of the 2011 ACM SIGMETRICS Conference}, 2011.

\bibitem{geant}
Steve Uhlig, Bruno Quoitin, Jean Lepropre, and Simon Balon.
\newblock Providing public intradomain traffic matrices to the research
  community.
\newblock {\em SIGCOMM Comput. Commun. Rev.}, 36(1):83–86, jan 2006.

\bibitem{l2r}
Asaf Valadarsky, Michael Schapira, Dafna Shahaf, and Aviv Tamar.
\newblock Learning to route.
\newblock In {\em Proceedings of the 16th ACM Workshop on Hot Topics in
  Networks}, HotNets-XVI, 2017.

\bibitem{cope}
Hao Wang, Haiyong Xie, Lili Qiu, Yang~Richard Yang, Yin Zhang, and Albert
  Greenberg.
\newblock Cope: Traffic engineering in dynamic networks.
\newblock In {\em SIGCOMM}, 2006.

\bibitem{gcnsurvey}
Zonghanu Wu, Shirui Pan, Fengwen Chen, Guodong Long, Chengqi Zhang, and
  Philip~S. Yu.
\newblock A comprehensive survey on graph neural networks.
\newblock arXiv preprint arXiv:1901.00596, 2019.

\bibitem{zhang2017guaranteeing}
Hong Zhang, Kai Chen, Wei Bai, Dongsu Han, Chen Tian, Hao Wang, Haibing Guan,
  and Ming Zhang.
\newblock Guaranteeing deadlines for inter-data center transfers.
\newblock {\em IEEE/ACM Transactions on Networking (TON)}, 25(1):579--595,
  2017.

\bibitem{estimating-DMs}
Yin Zhang, M.~Roughan, C.~Lund, and D.L. Donoho.
\newblock Estimating point-to-point and point-to-multipoint traffic matrices:
  an information-theoretic approach.
\newblock {\em IEEE/ACM Transactions on Networking}, 13(5):947--960, 2005.

\bibitem{Arrow}
Zhizhen Zhong, Manya Ghobadi, Alaa Khaddaj, Jonathan Leach, Yiting Xia, and
  Ying Zhang.
\newblock {ARROW:} restoration-aware traffic engineering.
\newblock In {\em {ACM} {SIGCOMM} 2021 Conference, Virtual Event, USA, August
  23-27, 2021}, pages 560--579, 2021.

\end{thebibliography}

\newpage
\section*{Appendix}
\appendix

\section{Predictability of WAN TE Traffic}

\begin{figure*}[t]
\subfigure[Inter-data-center traffic]{ 
\label{fig:bb1_predict}\includegraphics[width=0.42\linewidth]{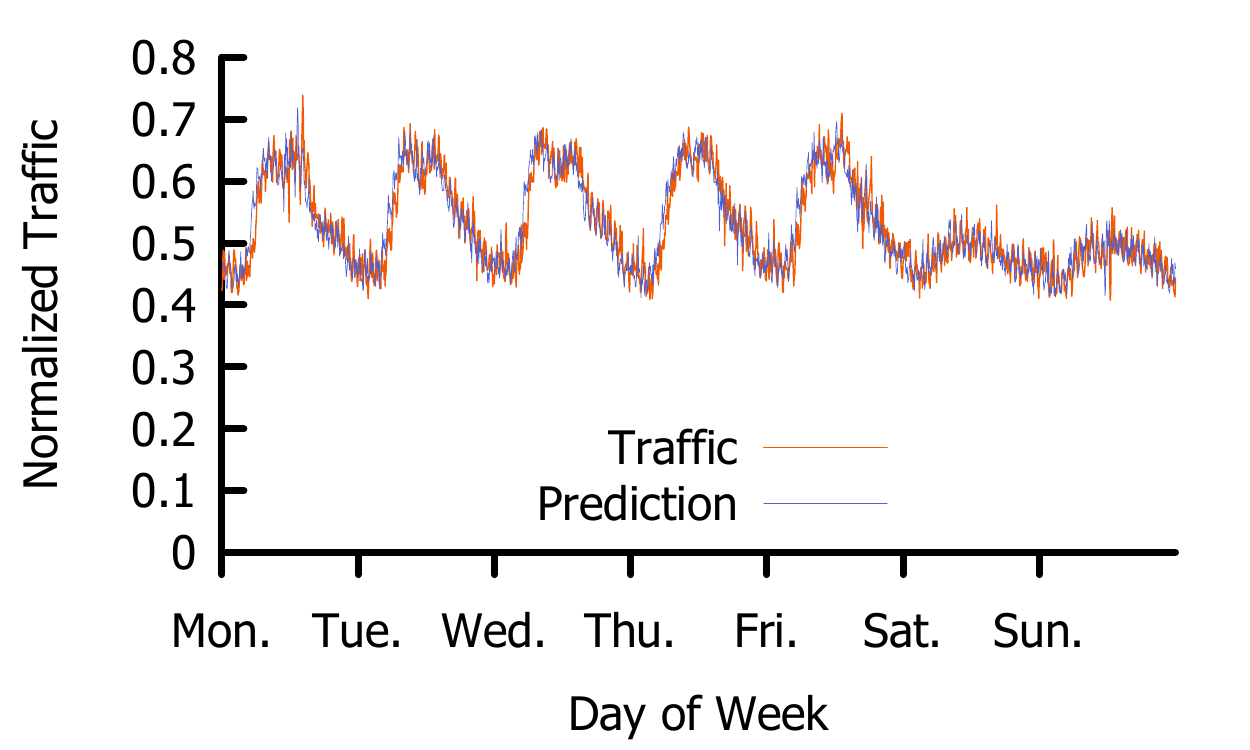}}
\hfill
\subfigure[Customer-facing traffic]{ 
\label{fig:bb2_predict}\includegraphics[width=0.42\linewidth]{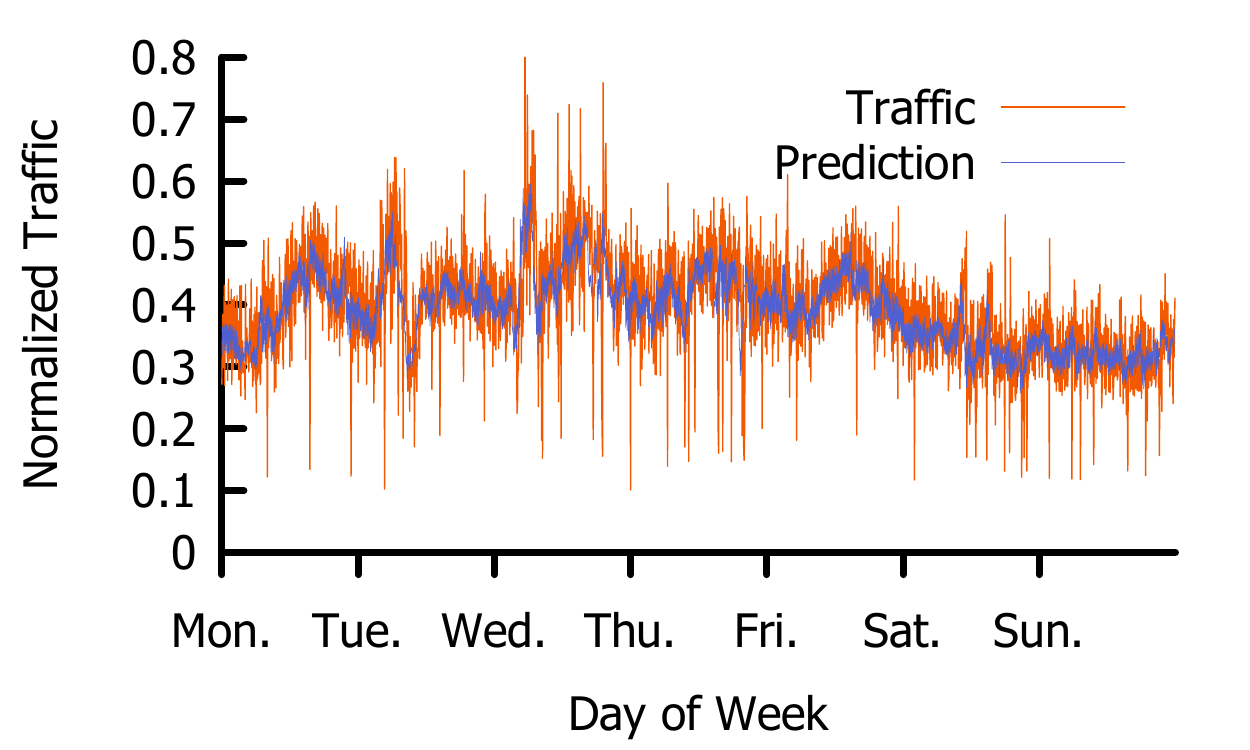}}
\caption{Inter-data-center traffic and customer-facing traffic over the course of a week, along with the predictions of a linear regression model for the time-series.}
\label{fig:bb_predict}
\end{figure*}

\autoref{fig:bb1_predict} plots the inter-data-center traffic demand between the pair of data centers with the highest average demand over the course of a week. Similarly, \autoref{fig:bb2_predict} plots the normalized volume of custo\-mer-facing traffic for the pair of nodes with the largest highest demand over the course of a week. Demands are shown at $5$-minute granularity and are normalized by the peak demand. As shown in \autoref{fig:bb1_predict}, inter-data-center traffic demands exhibit very distinct diurnal and hourly patterns. Indeed, the figure also presents the predictions of a linear regression model trained on data from the $3$ preceding weeks, which takes as input the traffic demands observed in the previous hour (at $5$-minute granularity), and outputs the predicted traffic demand for the upcoming $5$ minutes. In contrast, the predictions of a linear regressor for customer-facing traffic, as shown in \autoref{fig:bb2_predict}, are quite often far from the actual traffic demands.

\section{Analytical Results}\label{apx:DOTE-theory}

\subsection{Minimizing Max-Link Utilization}

We next prove that, for an infinitely expressive TE function, i.e., when each history of DMs can be \textit{independently} mapped to a TE configuration, and in the limit of infinite empirical data sampled from the underlying Markov process' stationary distribution, \emph{DOTE} attains optimal performance. This establishes that our approach is \textit{fundamentally sound}, and so high performance in practice can be achieved by acquiring sufficient empirical data and employing a sufficiently expressive decision model (e.g., a deep enough neural network).

For the sake of analysis, we make the following simplifying assumptions. We first assume that the set of possible history realizations, which we denote by $\histset$, is finite. Let $\dmax$ denote an upper bound on the maximum traffic demand between a source-destination pair, $\cmin$ denote the minimum link capacity, and $\pmax$ denote the maximum number of tunnels interconnecting a source-destination pair. Note that any valid TE configuration specifies, for each source-destination pair, a point in the $\pmax$-dimensional simplex (specifying its splitting ratios across at most $\pmax$ tunnels); let $\teset$ denote the space of valid TE configurations.
Let $\policy:\histset \to \teset$ denote a mapping from history to TE configuration. $\policy$ can be represented as a vector with $|\histset|\times n^2 \times (\pmax-1)$ components.\footnote{Note that we dropped the subscript $\theta$ in $\pi$, as in our analysis we consider the space of all possible TE configurations, and not a specific parametrization.} Since each element in this vector is itself a vector in the $\pmax$-dimensional simplex, we have that $\|\policy\|\leq \sqrt{|\histset|  n^2 (\pmax-1)} \doteq \pibound$, where $\|\cdot\|$ is the Euclidean norm.
We make the following observation.

\begin{proposition}
The loss function $\obj(\pi(D_{t-1},\dots,D_{t-\hor}),D_t)$ is convex in $\policy$ and $\rho$-Lipschitz, with $\rho=\dmax/\cmin$.
\end{proposition}
\begin{proof}
$f_e$ is, by definition, linear in the traffic splitting ratios and so in $\policy$. Since the $\max$ is a convex function, we have that $\obj$ is convex in $\policy$. 
Similarly, since each component in 
$\frac{f_e}{c(e)}$ is 
$\dmax / \cmin$-Lipschitz, 
the maximum is also $\dmax/\cmin$-Lipschitz.
\end{proof}

\sloppy We now consider an idealized stochastic gradient descent (SGD) algorithm where at each iteration $k$ we sample $D_t, D_{t-1},\dots,D_{t-\hor}$ from the probability distributions $P(D_{t-1},\dots,D_{t-\hor})$ and $P(D_t|D_{t-1},\dots,D_{t-\hor})$, 
and update $\pi_{k+1} = \project \left\{ \pi_{k} - \eta \grad_k \right\}$, 
where $\grad_k \in \partial \obj(\pi_{k}(D_{t-1},\dots,D_{t-\hor}),D_t)$ denotes a subgradient of the objective function\footnote{The objective is not necessarily differentiable everywhere because of the $\max$, but the subgradient exists for every $\pi$.}, and $\project$ denotes a projection onto the simplex for each $(s,d)$ pair. The final output after $K$ iterations is $\bar{\pi} = \frac{1}{K}\sum_{k=1}^K \pi_k$. 

The next theorem, based on Theorem 14.12 in \cite{shalev2014understanding}, bounds the loss of this algorithm. Let $\bar{\obj}(\pi) =  \mathbb{E}\left[{\obj(\pi(D_{t-1},\dots,D_{t-\hor}),D_t)}\right]$ denote the expected loss of a TE function, and let
$\pi^* \in \argmin_{\pi} \bar{\obj}(\pi)$ denote the optimal TE function. 
\begin{theorem}\label{thm:sgd-apx}
For every $\epsilon > 0$, if SGD is run for $K \geq \frac{B^2 \rho^2}{\epsilon^2}$ iterations with $\eta = \sqrt{\frac{B^2}{\rho^2 K}}$, then the output of SGD satisfies
$$
\mathbb{E}\left[\bar{\obj}(\bar{\pi})\right] \leq \bar{\obj}(\pi^*) + \epsilon,
$$
where the expectation is w.r.t. the sampling by the algorithm.
\end{theorem}

Theorem \ref{thm:sgd-apx} shows that without function approximation (the TE function space spans \textit{all} possible mappings from history to TE configuration), and with infinite data (the algorithm continuously samples from the true demand distribution), SGD converges to the optimal TE function with arbitrary precision. In practice, we relax both assumptions. In \emph{DOTE} we sample from a large, but \textit{finite}, dataset of historical demands, and use a \textit{parametric} model (specifically, a neural network) to map from an \textit{infinite} set of possible histories to valid TE configurations. Our empirical results show that, with enough data and a deep enough neural network, the approximate TE function \emph{DOTE} learns is still very close to optimal.

\subsection{Maximum-Multicommodity-Flow and Maximum-
Concurrent-Flow}

We begin by stating a general convergence result for quasi-convex functions that satisfy certain assumptions. We then proceed to show that both maximum-multicommodity-flow and maximum-concurrent-flow indeed satisfy these assumptions, implying their convergence.

\subsubsection{General results}

We begin by providing an analysis of stochastic quasi-convex optimization, under general assumptions. In the next section, we will show that maximum-multicommodity-flow and maximum-
concurrent-flow are special cases of this setting.\footnote{While we present results for quasi-convexity, the extension of these results to quasi-concave problems is immediate.} Our analysis builds on two studies -- the analysis of stochastic normalized subgradient of~\cite{hazan2015beyond}, which is for smooth and unconstrained problems, and the study of \cite{konnov2003convergence}, which considered non-smooth quasi-convex optimization. 

A quasi-convex function $f(x)$ satisfies that its level sets, $L(f;\alpha) = \left\{ x | f(x) \leq \alpha \right\}$, are convex sets for all $\alpha$.

We first define a normalized subgradient in the context of quasi-convex functions, following \cite{konnov2003convergence}. The normal cone to a convex set $X$ at point $x$ is defined by $N(X,x) = \left\{q\in \mathbb{R}^n | \langle q,y-x \rangle \leq 0 \quad \forall y \in X \right\}. $ The set of subgradients at a point $x$ are given by $N(L(f;f(x)),x)$. The set of normalized subgradients, $Q(f;x)$, at a point $x$, are given by $Q(f;x) = \mathrm{S}(0,1) \cap N(L(f;f(x)),x)$, where $\mathrm{S}(0,1)$ is the $n$-dimensional sphere of radius $1$. These are directions of descent -- normalized vectors such that taking an infinitely small step in their direction is guaranteed to not increase the function.

In the following, we consider a general stochastic optimization problem:
\begin{equation}\label{eq:so_quasi}
    \min_{x \in X} \mathbb{E}_{D \sim P(D)} \left[ f(x,D)\right],
\end{equation}
where $f$ is quasi-convex in $x$ for every $D$.

We will further assume the following. Let $\mathbb{B}(z, r)$ denote the $n$-dimensional ball centered on $z$ with radius $r$. 
\begin{assumption}\label{ass:qc_sum}
Set $X$ is convex and bounded by $\mathbb{B}(0, \bar{B})$. The function $f$ is bounded by $B$. It is also $G$-Lipschitz and quasi-convex in $x$ for every $D$. Furthermore, for every $D_1,\dots,D_M$, we have that $\frac{1}{M}\sum_{i=1}^M f(x,D_i)$ is quasi-convex in $x$.
\end{assumption}
Note that the last requirement in Assumption \ref{ass:qc_sum} is not immediate, as the sum of quasi-convex functions is not necessarily quasi-convex.

The stochastic normalized subgradient method we consider works as follows~\cite{hazan2015beyond}. At each iteration $k$ we sample a minibatch $\left\{D_i\right\}_{i=1}^b \sim P(D)$ and define $f_k = \frac{1}{b}\sum_{i=1}^b f(x,D_i)$. We then
update $x_{k+1} = \project \left\{ x_{k} - \eta \grad_k \right\}$, 
where $\grad_k \in Q(f_k; x_k)$ denotes the subgradient of the minibatch, and $\project$ denotes a projection onto the set $X$. The final output after $K$ iterations is $\bar{x}_K = \argmin_{x_1,\dots,x_K} f_k(x_k)$. 

The analysis in \cite{hazan2015beyond} bounds the error of the normalized subgradient method, for smooth and unconstrained functions. We next adapt it to our setting.

The next definition adapts a central definition from \cite{hazan2015beyond} to our non-smooth setting.

\begin{definition}\label{def:slqc} (SLQC) Let $x,x^* \in \mathbb{R}^n$, $\kappa,\epsilon >0$. We say that $f$ is $(\epsilon,\kappa,x^*)$-strictly-locally-quasi-convex (SLQC) in $x$ if at least one of the following applies. (1) $f(x) - f(x^*) \leq \epsilon$. (2) For any $\Delta \in Q(f; x)$, and every $y\in \mathbb{B}(x^*, \frac{\epsilon}{\kappa})$, it holds that $\langle \Delta, y-x\rangle \leq 0$.
\end{definition}

We next show that the Lipschitz and quasi-convex properties in Assumption \ref{ass:qc_sum} suffice to establish SLQC.
\begin{lemma}\label{lem:SLQC}
Let $f$ satisfy Assumption \ref{ass:qc_sum}. Fix $D$, and let $x^* \in \argmin_{x\in X}f(x;D)$. Then $f$ is $(\epsilon,G,x^*)$-SLQC for all $x \in X$.
\end{lemma}
\begin{proof}
Assume $f(x; D) - f(x^*; D) > \epsilon$. Let $Z$ denote the $f(x; D)$-level set of $f(x; D)$.
Let $\partial Z$ be the boundary of $Z$. By definition of the level set, for every $z \in \partial Z$,  $f(z) - f(x^*) > \epsilon$. From the Lipschitz property then, for every $z \in \partial Z$ we must have $\|z - x^*\| \geq \frac{\epsilon}{G}$. Since $Z$ is convex, we therefore have that $\mathbb{B}(x^*, \frac{\epsilon}{G}) \subset Z$. From the definition of $Q(f; x)$, we have that for every $y\in \mathbb{B}(x^*, \frac{\epsilon}{G})$, if $\Delta \in Q(f; x)$ then $\langle \Delta, y-x\rangle \leq 0$.
\end{proof}

We next show that with high probability, the subgradient of each minibatch is a descent direction for the expected objective in \eqref{eq:so_quasi}. 

\begin{lemma}\label{lem:s_SLQC}
Let Assumption \ref{ass:qc_sum} hold, and let $x^* \in \argmin_{x \in X} \mathbb{E}_{D \sim P(D)} \left[ f(x,D)\right]$. Assume that the minibatch size satisfies $b = \mathcal{O}\left( \frac{n B^2 \log (G \bar{B}/\delta)}{2 \epsilon ^2}\right)$. Then, with probability at least $1-\delta$, we have that the minibatch average $f_k = \frac{1}{b}\sum_{i=1}^b f(x,D_i)$ is $(\epsilon,2G,x^*)$-SLQC in $x_k$.
\end{lemma}
\begin{proof}
Let 
$$
\xi = \frac{1}{b}\sum_{i=1}^b f(x^*,D_i) - \mathbb{E}_{D \sim P(D)} \left[ f(x^*,D)\right].
$$
From Hoeffding's inequality, we have that
$$
P(|\xi| > t) \leq 2 \exp \left( - \frac{2 b t^2}{B^2}\right).
$$
Thus, if $b \geq \frac{B^2 \log (2/\delta)}{2 t^2}$ we have that with probability $1-\delta$, $|\xi| < t$.

Let $x^*_k \in \argmin_{x \in X} \frac{1}{b}\sum_{i=1}^b f(x,D_i)$. Let 
$$
\xi' = \frac{1}{b}\sum_{i=1}^b f(x^*_k,D_i) - \mathbb{E}_{D \sim P(D)} \left[ f(x^*_k,D)\right].
$$
Then, using a covering number argument\cite{shalev2009stochastic}, we have that for $b \geq \frac{n B^2 \log (G \bar{B}/\delta)}{2 t^2}$, with probability $1-\delta$, $|\xi'| < t$.
We have that
$$
    \frac{1}{b}\sum_{i=1}^b f(x^*_k,D_i) \leq \frac{1}{b}\sum_{i=1}^b f(x^*,D_i) \leq \mathbb{E}_{D \sim P(D)} \left[ f(x^*,D)\right] + \xi,
$$
and
$$
\mathbb{E}_{D \sim P(D)} \left[ f(x^*,D)\right] - \xi' \leq \mathbb{E}_{D \sim P(D)} \left[ f(x^*_k,D)\right] - \xi' \leq \frac{1}{b}\sum_{i=1}^b f(x^*_k,D_i).
$$
Therefore, 
$$
\frac{1}{b}\sum_{i=1}^b f(x^*,D_i) - \frac{1}{b}\sum_{i=1}^b f(x^*_k,D_i) \leq \xi + \xi'.
$$
Now, similarly to the proof of Lemma \ref{lem:SLQC}, assume that $\frac{1}{b}\sum_{i=1}^b f(x_k,D_i) - \frac{1}{b}\sum_{i=1}^b f(x^*_k,D_i) > \epsilon$. We choose $b = \mathcal{O}\left( \frac{n B^2 \log (G \bar{B}/\delta)}{2 \epsilon ^2}\right)$ such that with probability $1-\delta$, $\xi + \xi' \leq \epsilon/2$.

We therefore have:
$$
\frac{1}{b}\sum_{i=1}^b f(x_k,D_i) - \frac{1}{b}\sum_{i=1}^b f(x^*,D_i) > \epsilon - (\xi + \xi') \geq \frac{\epsilon}{2}.
$$

For simplicity, we denote $\bar{f}(x) = \frac{1}{b}\sum_{i=1}^b f(x,D_i)$. Note that $\bar{f}$ is quasi-convex, by Assumption \ref{ass:qc_sum}. Let $Z$ denote the $\bar{f}(x_k)$-level set of $\bar{f}(x)$.
Let $\partial Z$ be the boundary of $Z$. By definition of the level set, for every $z \in \partial Z$,  $f(z) - f(x^*) > \epsilon / 2$. From the Lipschitz property then, for every $z \in \partial Z$ we must have $\|z - x^*\| \geq \frac{\epsilon}{2G}$. Since $Z$ is convex, we therefore have that $\mathbb{B}(x^*, \frac{\epsilon}{2G}) \subset Z$. From the definition of $Q(f; x)$, we have that for every $y\in \mathbb{B}(x^*, \frac{\epsilon}{2G})$, if $\Delta \in Q(f; x)$ then $\langle \Delta, y-x\rangle \leq 0$.
\end{proof}

We are finally ready to present the converge result. 
\begin{theorem}\label{thm:norm_sgd}
    Let Assumption \ref{ass:qc_sum} hold. Suppose we run the stochastic normalized subgradient method for $K \geq \frac{4G^2 \|x_1 - x^*\|^2}{\epsilon^2}$ iterations, $\eta = \epsilon/2G$, and the minibatch size satisfies $b = \mathcal{O}\left( \frac{n B^2 \log (K G \bar{B}/\delta)}{2 \epsilon ^2}\right)$. Then with probability $1-2\delta$, we have that $f(\bar{x}_K) - f(x^*) \leq 3 \epsilon$.
\end{theorem}
\begin{proof}
This is a direct application of Theorem 5.1 of \cite{hazan2015beyond}, where we used Lemma \ref{lem:s_SLQC} to guarantee that at each iteration the minibatch is SLQC, as required in \cite{hazan2015beyond}. We note that by our Definition \ref{def:slqc}, the proof in \cite{hazan2015beyond} holds without change to the non-smooth setting. The projection onto the set $X$ requires a straightforward modification to the proof of \cite{hazan2015beyond}, where the first equality in their proof of Theorem 4.1 should be a $\leq$. The rest of the proofs remain unchanged. 
\end{proof}

\subsubsection{Results for Maximum-Multicommodity-Flow}
We formally define the problem as follows.

for each tunnel $T$, let $x_T$ denote the flow on that tunnel, and let $x_e = \sum\limits_{T:e \in T}{x_T}$, for each edge $e$, denote the total flow on edge $e$. We define
$$\gamma = \max\left(\max\limits_{e}{\frac{x_e}{C_e}},1\right),$$
and normalize the flows by $\gamma$, yielding normalized flows on a tunnel,
$$y_T = \frac{x_T}{\gamma},$$
and correspondingly, total normalized flows from source $s$ to target $t$, $y_{s,t} = \sum\limits_{T \in P_{st}}{y_T}$. Let $x = \{ x_T \}$ denote our decision variables. Given a demand matrix $D$, the Max-MCF objective is
$$
f_{MMCF}(x,D)=\sum\limits_{s,t}{\min{(D_{s,t},y_{s,t})}}.
$$

We next show that $f_{MMCF}$ is Lipschitz.

\begin{lemma}\label{lem:cmax_bound}
For any tunnel $T$ and $x \ge 0$, $\frac{x_T}{\gamma(x)} \le C_{max}$.
\end{lemma}
\begin{proof}
Let $e \in T$, then by the definitions of $\gamma(x)$ and $x_e$, $\gamma(x) \ge \frac{x_e}{c_e} \ge \frac{x_T}{C_{max}}$.
\end{proof}

\begin{lemma}\label{lem:xt_gamma_lipschitz}
$f_T(x)=\frac{x_T}{\gamma(x)}$ is Lipschitz on $\mathbb{R}_{+}^n$, and its Lipschitz constant is at most $K=2 \cdot \frac{C_{max}}{C_{min}}$.
\end{lemma}

\begin{proof}
Assume, without loss of generality, that $f(x) \ge f(y)$.\\
Case 1: $\gamma(y) = 1$\\
\begin{equation*}
\begin{split}
&|f(x)-f(y)|\\
&=f(x)-f(y)\\
&=\frac{x_T}{\gamma(x)}-\frac{y_T}{\gamma(y)}\\
&=\frac{x_T}{\gamma(x)}-y_T\\
&\le x_T-y_T\\
&\le |x_T-y_T|\\
&\le \| x - y \|_{1}\\
&\le 2 \cdot \frac{C_{max}}{C_{min}} \cdot \| x - y \|_{1},\\
\end{split}
\end{equation*}
where the first inequality is since $\gamma(x) \ge 1$, and the third inequality is by the definition of $\lVert x \rVert_{1}$.\\
Case 2: $\gamma(y) = \frac{y_{e_0}}{C_{e_0}} > 1$, for some edge $e_0$.\\
\begin{equation*}
\begin{split}
&|f(x)-f(y)|\\
&=f(x)-f(y)\\
&=\frac{x_T}{\gamma(x)}-\frac{y_T}{\gamma(y)}\\
&=\frac{x_T}{\gamma(x)}-\frac{y_T}{\gamma(x)}+\frac{y_T}{\gamma(x)}-\frac{y_T}{\gamma(y)}\\
&=\frac{1}{\gamma(x)}\cdot(x_T-y_T)+\frac{y_T}{\gamma(y)} \cdot \frac{1}{\gamma(x)}(\gamma(y)-\gamma(x))\\
&\le \frac{1}{\gamma(x)}\cdot\left(x_T-y_T\right)+\frac{y_T}{\gamma(y)} \cdot \frac{1}{\gamma(x)}\left(\frac{y_{e_0}}{C_{e_0}}-\frac{x_{e_0}}{C_{e_0}}\right)\\
&\le \left|\frac{1}{\gamma(x)}\cdot(x_T-y_T)+\frac{y_T}{\gamma(y)} \cdot \frac{1}{\gamma(x)}\left(\frac{y_{e_0}}{C_{e_0}}-\frac{x_{e_0}}{C_{e_0}}\right)\right|\\
&\le \frac{1}{\gamma(x)}\cdot |x_T-y_T|+\frac{y_T}{\gamma(y)} \cdot \frac{1}{\gamma(x)}\left|\frac{y_{e_0}}{C_{e_0}}-\frac{x_{e_0}}{C_{e_0}}\right|\\
&\le |x_T-y_T| + \frac{C_{max}}{C_{min}} \cdot |y_{e_0}-x_{e_0}|\\
&\le 2 \cdot \frac{C_{max}}{C_{min}} \cdot \| x - y \|_{1},
\end{split}
\end{equation*}
where the first inequality is since $\gamma(y) = \frac{y_{e_0}}{C_{e_0}}$, $\gamma(x) \ge \frac{x_{e_0}}{C_{e_0}}$, $y_T \ge 0$, and $\gamma > 0$, the third inequality is since $|a+b| \le |a| + |b|$, the fourth inequality is by Lemma \ref{lem:cmax_bound} and since $\gamma(x) \ge 1$, and the last inequality is by the definitions of $\| x \|_{1}$, $x_e$ and since $|a+b| \le |a| + |b|$.\\
\end{proof}

\begin{proposition}\label{prop:fmmc_lipschitz}
The function $f_{MMCF}$ is Lipschitz, and its Lipschitz constant is at most
$\sum_{s,t}\sum_{p \in P_{st}}2 \cdot \frac{C_{max}}{C_{min}}$.
\end{proposition}
\begin{proof}
By Lemma \ref{lem:xt_gamma_lipschitz} and as a sum and minimum of Lipschitz functions.
\end{proof}

We next state two lemmas that we will use in our analysis.

\begin{lemma}\label{lem:claim1}
For any $a,b \ge 0$, $c,d > 0$ and $\lambda \in [0,1]$, we have that  $\min{\left(\frac{a}{c}, \frac{b}{d}\right)} \le \frac{\lambda a + (1-\lambda) b}{\lambda c + (1-\lambda) d}$.
\end{lemma}
\begin{proof}
Let $f(\lambda)=\frac{\lambda a + (1-\lambda) b}{\lambda c + (1-\lambda) d}$. Then,
\begin{equation*}
  \begin{split}
      f'(\lambda) &= \frac{(a-b)(\lambda c + (1-\lambda) d) - (c-d)(\lambda a + (1-\lambda) b)}{(\lambda c + (1-\lambda) d)^2} \\
      &= \frac{ad-bc}{(\lambda c + (1-\lambda) d)^2}.
  \end{split}  
\end{equation*}
Also, $f(0)=\frac{b}{d}, f(1)=\frac{a}{c}$, and $f'(\lambda)$ has a fixed sign for any $\lambda \in [0,1]$. Therefore, $f(\lambda) \ge \min{(\frac{a}{c}, \frac{b}{d})}.$
\end{proof}

\begin{lemma}\label{lem:claim2}
Let $x=\{x_T\}$, $x'=\{x'_T\}$, and $\lambda \in [0,1]$. Let $x''=\{\lambda x_T + (1-\lambda)x'_T\}$, and let $\gamma$, $\gamma'$ and $\gamma''$ be the respective normalization constants. Then $\gamma'' \le \lambda\gamma + (1-\lambda)\gamma'$.
\begin{proof}
We have that

\begin{equation*}
    \begin{split}
        \gamma''&=\max\left(\max\limits_{e}{\frac{x''_e}{C_e}},1\right)\\
        &=\max\left(\max\limits_{e}{\frac{\lambda x_e + (1 - \lambda)x'_e}{C_e}},1\right)\\
        &\le \max\left(\max\limits_{e}{\frac{\lambda x_e}{C_e}},\lambda\right) + \max\left(\max\limits_{e}{\frac{(1-\lambda) x'_e}{C_e}},1-\lambda\right)\\
        &= \lambda\max\left(\max\limits_{e}{\frac{x_e}{C_e}},1\right) + (1-\lambda)\max\left(\max\limits_{e}{\frac{x'_e}{C_e}},1\right)\\
        &= \lambda\gamma + (1-\lambda)\gamma'.
    \end{split}
\end{equation*}

\end{proof}
\end{lemma}

We next show that Max-MCF satisfies Assumption \ref{ass:qc_sum}.

\begin{proposition}\label{prop:mmcf_qc}
The function $f_{MMCF}$ is Lipschitz and bounded. Its maximum is obtained inside a convex set $X$. Furthermore, for every $D^1,\dots,D^M$, we have that $\frac{1}{M}\sum_{i=1}^M f(x,D^i)$ is quasi-concave in $x$
\end{proposition}
\begin{proof}
By definition, $x_T \geq 0$ for all $T$.
Let $C_{max} = \max_{e} C_e$, and consider $T$-dimensional hypercube $X = [0,C_{max}]^T$. By definition, for every $x\geq 0$ that is outside $X$, there is an $x' \in X$ with an equivalent objective value. To see this, let $\gamma$ the normalizing constant for $x$, and set $x' = x / \gamma$. Then,
$$
x'_T = \frac{x_T}{\max\left(\max_{e}\frac{x_e}{C_e},1\right)} \leq \frac{x_T}{\max\left(\max_{e}\frac{x_e}{C_{max}},1\right)} \leq \frac{x_T}{\frac{x_T}{C_{max}}} = C_{max}.
$$
But the normalizing factor for $x'$ is $1$, so $x$ and $x'$ have the same objective value.

Clearly, $f_{MMCF}$ is bounded by $\sum_{s,t}\sum_{T:e \in T}C_{max}$.

The function is Lipschitz by proposition \ref{prop:fmmc_lipschitz}.

Let $\bar{f}_{MMCF}(x) = \frac{1}{M}\sum_{i=1}^M f_{MMCF}(x,D^i)$. We shall now show that for any $x, x'\in X$, and $\lambda \in [0,1]$, $\bar{f}_{MMCF}(\lambda x + (1-\lambda)x') \geq \min\{\bar{f}_{MMCF}(x), \bar{f}_{MMCF}(x') \}$, proving that $\bar{f}_{MMCF}$ is quasi-concave. We denote by $\gamma'$ and $y'$ the respective normalization constant and normalized flows corresponding to $x'$. We also denote $x'' = \lambda x + (1-\lambda)x'$, and let $\gamma''$ and $y''$ denote its corresponding normalization constant and normalized flows, respectively.
\begin{small}
\begin{equation*}
\begin{split}
&\min{(\sum\limits_{i=1}^M\sum\limits_{s,t}\min{(D^i_{s,t},y_{s,t})}, \sum\limits_{i=1}^M\sum\limits_{s,t}\min{(D^i_{s,t},y'_{s,t})})} \\
=& \min{(\sum\limits_{i=1}^M\sum\limits_{s,t}\min{(D^i_{s,t},\sum\limits_{T \in P_{st}}\frac{x_T}{\gamma})}, \sum\limits_{i=1}^M\sum\limits_{s,t}\min{(D^i_{s,t},\sum\limits_{T \in P_{st}}\frac{x'_T}{\gamma'})})}\\
=& \min{(\frac{1}{\gamma}\sum\limits_{i=1}^M\sum\limits_{s,t}\min{(\gamma D^i_{s,t},\sum\limits_{T \in P_{st}}x_T)}, \frac{1}{\gamma'}\sum\limits_{i=1}^M\sum\limits_{s,t}\min{(\gamma' D^i_{s,t},\sum\limits_{T \in P_{st}}x'_T)})}\\
\le& \frac{\lambda \sum\limits_{i=1}^M\sum\limits_{s,t}\min{(\gamma D^i_{s,t},\sum\limits_{T \in P_{st}}x_T)} + (1 - \lambda) \sum\limits_{i=1}^M\sum\limits_{s,t}\min{(\gamma' D^i_{s,t},\sum\limits_{T \in P_{st}}x'_T)}}{\lambda \gamma + (1 - \lambda) \gamma'}
\\
=& \frac{\sum\limits_{i=1}^M\sum\limits_{s,t}\min{(\lambda\gamma D^i_{s,t} , \lambda\sum\limits_{T \in P_{st}}X_T)} + \min{((1 - \lambda)\gamma' D^i_{s,t} , (1 - \lambda)\sum\limits_{T \in P_{st}}X'_T)}}{\lambda \gamma + (1 - \lambda) \gamma'}\\
\le& \frac{1}{\lambda \gamma + (1 - \lambda) \gamma'}\sum\limits_{i=1}^M\sum\limits_{s,t}\min \bigg(\lambda\gamma D^i_{s,t} + (1 - \lambda)\gamma' D^i_{s,t}, \\
& \lambda\sum\limits_{T \in P_{st}}x_T + (1 - \lambda)\sum\limits_{T \in P_{st}}x'_T\bigg)\\
=&\sum\limits_{i=1}^M\sum\limits_{s,t}\min{\bigg(D^i_{s,t}, \frac{1}{\lambda \gamma + (1 - \lambda) \gamma'}\sum\limits_{T \in P_{st}}(\lambda x_T + (1 - \lambda)x'_T)\bigg)}\\
\le& \sum\limits_{i=1}^M\sum\limits_{s,t}\min{(D^i_{s,t}, \sum\limits_{T \in P_{st}}\frac{x''_T}{\gamma''})}\\
=& \sum\limits_{i=1}^M\sum\limits_{s,t}\min{(D^i_{s,t}, {y''_{s,t}})},
\end{split}
\end{equation*}
\end{small}
where the first inequality is by Lemma \ref{lem:claim1}, the second inequality is since $\min(a,b)+\min(c,d)\leq \min(a+c,b+d)$, and the third inequality is by Lemma \ref{lem:claim2}.
\end{proof}

Since Assumption \ref{ass:qc_sum} holds, Theorem \ref{thm:norm_sgd} guarantees that the stochastic normalized subgradient method will converge to an optimal solution of the Max-MCF objective.

\subsubsection{Results for Maximum-Concurrent-Flow}
Given a demand matrix $D$, the Max-Concurrent-Flow objective is
$$
f_{MCONC}(x,D)=\min{(\{\frac{y_{s,t}}{D_{s,t}}\}_{s,t \in V, D_{s,t} > 0} \cup \{1\})}.
$$
We assume that when $D_{s,t}\neq 0$, there is a minimal value $\varepsilon$ for $D_{s,t}$, corresponding, e.g., to a single packet. 
We next show that Max-Concurrent-Flow satisfies Assumption \ref{ass:qc_sum}.

\begin{proposition}\label{prop:fmconc_lipschitz}
The function $f_{MCONC}$ is Lipschitz, and its Lipschitz constant is at most
$\max\limits_{s,t}\left(\sum_{p \in P_{st}}\frac{2 \cdot C_{max}}{\epsilon \cdot C_{min}}\right)$.
\end{proposition}
\begin{proof}
By Lemma \ref{lem:xt_gamma_lipschitz} and as a sum, minimum and multiplication by a constant of Lipschitz functions.
\end{proof}

\begin{proposition}
The function $f_{MCONC}$ is Lipschitz and bounded. Its maximum is obtained inside a convex set $X$. Furthermore, for every $D^1,\dots,D^M$, we have that $\frac{1}{M}\sum_{i=1}^M f(x,D^i)$ is quasi-concave in $x$
\end{proposition}

\begin{proof}
The claims in the beginning of proposition \ref{prop:mmcf_qc} hold for $f_{MCONC}$, and therefore its maximum is obtained inside a convex set.

Clearly, $f_{MCONC}$ is bounded by $1$.

The function is Lipschitz by proposition \ref{prop:fmconc_lipschitz}.

Let $\bar{f}_{MCONC}(x) = \frac{1}{M}\sum_{i=1}^M f_{MCONC}(x,D^i)$. We shall now show that for any $x, x'\in X$, and $\lambda \in [0,1]$, $\bar{f}_{MCONC}(\lambda x + (1-\lambda)x') \geq \min\{\bar{f}_{MCONC}(x), \bar{f}_{MCONC}(x') \}$, proving that $\bar{f}_{MCONC}$ is quasi-concave. We denote by $\gamma'$ and $y'$ the respective normalization constant and normalized flows corresponding to $x'$. We also denote $x'' = \lambda x + (1-\lambda)x'$, and let $\gamma''$ and $y''$ denote its corresponding normalization constant and normalized flows, respectively.

\begin{small}
\begin{equation*}
\begin{split}
&\min{(\sum\limits_{i=1}^M\min{(\left\{\frac{y_{s,t}}{D^i_{s,t}}\right\} \cup \left\{1\right\})}, \sum\limits_{i=1}^M\min{(\left\{\frac{y'_{s,t}}{D^i_{s,t}}\right\} \cup \left\{1\right\})})}\\
=& \min{(\sum\limits_{i=1}^M\min{(\left\{\frac{\sum\limits_{T \in P_{st}}\frac{x_T}{\gamma}}{D^i_{s,t}}\right\} \cup \left\{1\right\})}, \sum\limits_{i=1}^M\min{(\left\{\frac{\sum\limits_{T \in P_{st}}\frac{x'_T}{\gamma'}}{D^i_{s,t}}\right\} \cup \left\{1\right\})})}\\
=& \min{(\frac{1}{\gamma}\sum\limits_{i=1}^M\min{(\left\{\frac{\sum\limits_{T \in P_{st}}x_T}{D^i_{s,t}}\right\} \cup \left\{\gamma\right\})}, \frac{1}{\gamma'}\sum\limits_{i=1}^M\min{(\left\{\frac{\sum\limits_{T \in P_{st}}x'_T}{D^i_{s,t}}\right\} \cup \left\{\gamma'\right\})})}\\
\le& \frac{\lambda\sum\limits_{i=1}^M\min{(\left\{\frac{\sum\limits_{T \in P_{st}}x_T}{D^i_{s,t}}\right\} \cup \left\{\gamma\right\})} + (1-\lambda)\sum\limits_{i=1}^M\min{(\left\{\frac{\sum\limits_{T \in P_{st}}x'_T}{D^i_{s,t}}\right\} \cup \left\{\gamma'\right\})}}{\lambda\gamma + (1-\lambda)\gamma'}\\
=& \frac{\sum\limits_{i=1}^M(\min{(\left\{\frac{\sum\limits_{T \in P_{st}} \lambda x_T}{D^i_{s,t}}\right\} \cup \left\{\lambda\gamma\right\})} + \min{(\left\{\frac{\sum\limits_{T \in P_{st}}(1-\lambda)x'_T}{D^i_{s,t}}\right\} \cup \left\{(1-\lambda)\gamma'\right\})})}{\lambda\gamma + (1-\lambda)\gamma'}\\
\le& \frac{\sum\limits_{i=1}^M\min{(\left\{\frac{\sum\limits_{T \in P_{st}} \lambda x_T + \sum\limits_{T \in P_{st}}(1-\lambda)x'_T}{D^i_{s,t}}\right\} \cup \left\{\lambda\gamma + (1-\lambda)\gamma'\right\})}}{\lambda\gamma + (1-\lambda)\gamma'}\\
=& \sum\limits_{i=1}^M\min{(\left\{\frac{\sum\limits_{T \in P_{st}} \frac{\lambda x_T + (1-\lambda)x'_T}{\lambda\gamma + (1-\lambda)\gamma'}}{D^i_{s,t}}\right\} \cup \left\{1\right\})}\\
\le& \sum\limits_{i=1}^M\min{(\left\{\frac{\sum\limits_{T \in P_{st}} \frac{x''}{\gamma''}}{D^i_{s,t}}\right\} \cup \left\{1\right\})}\\
=& \sum\limits_{i=1}^M\min{(\left\{\frac{y''_{s,t}}{D^i_{s,t}}\right\} \cup \left\{1\right\})},
\end{split}
\end{equation*}
\end{small}
where the first inequality is by Lemma \ref{lem:claim1}, the second inequality is since $\min(a,b)+\min(c,d)\leq \min(a+c,b+d)$, and the third inequality is by Lemma \ref{lem:claim2}.
\end{proof}

Since Assumption \ref{ass:qc_sum} holds, Theorem \ref{thm:norm_sgd} guarantees that the stochastic normalized subgradient method will converge to an optimal solution of the Max-Concurrent-Flow objective.

\section{A Closer Look at Demand Prediction}\label{sec:demand-prediction}

\begin{figure}[t!]
\centering
\includegraphics[width=.99\linewidth]{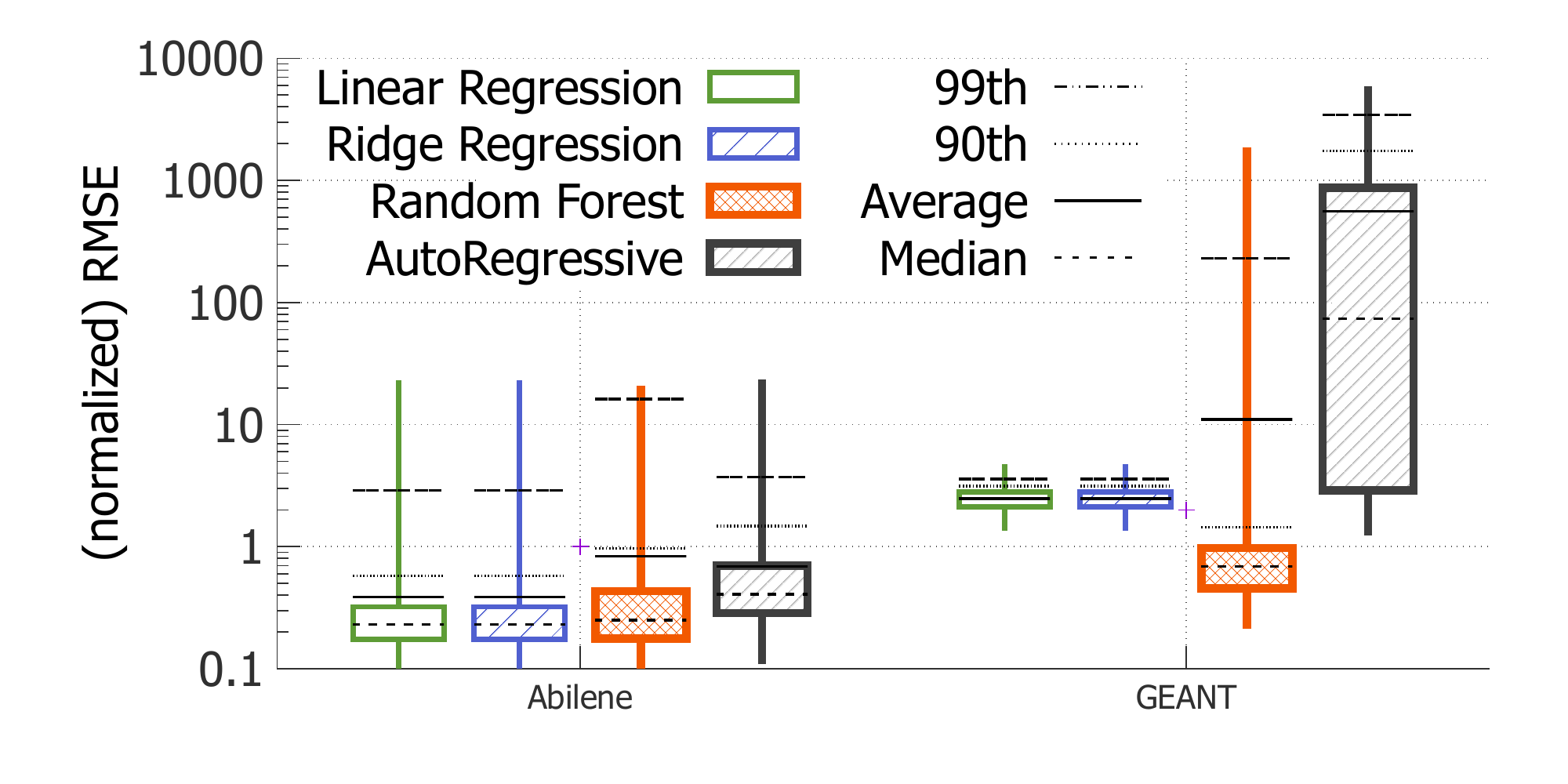}
\vspace{-0.15in}
\caption{Accuracy of predicting demands; results from different prediction methods.}
\label{fig:sl_accuracy}
\end{figure}

\begin{figure}[t!]
\centering
\includegraphics[width=.99\linewidth]{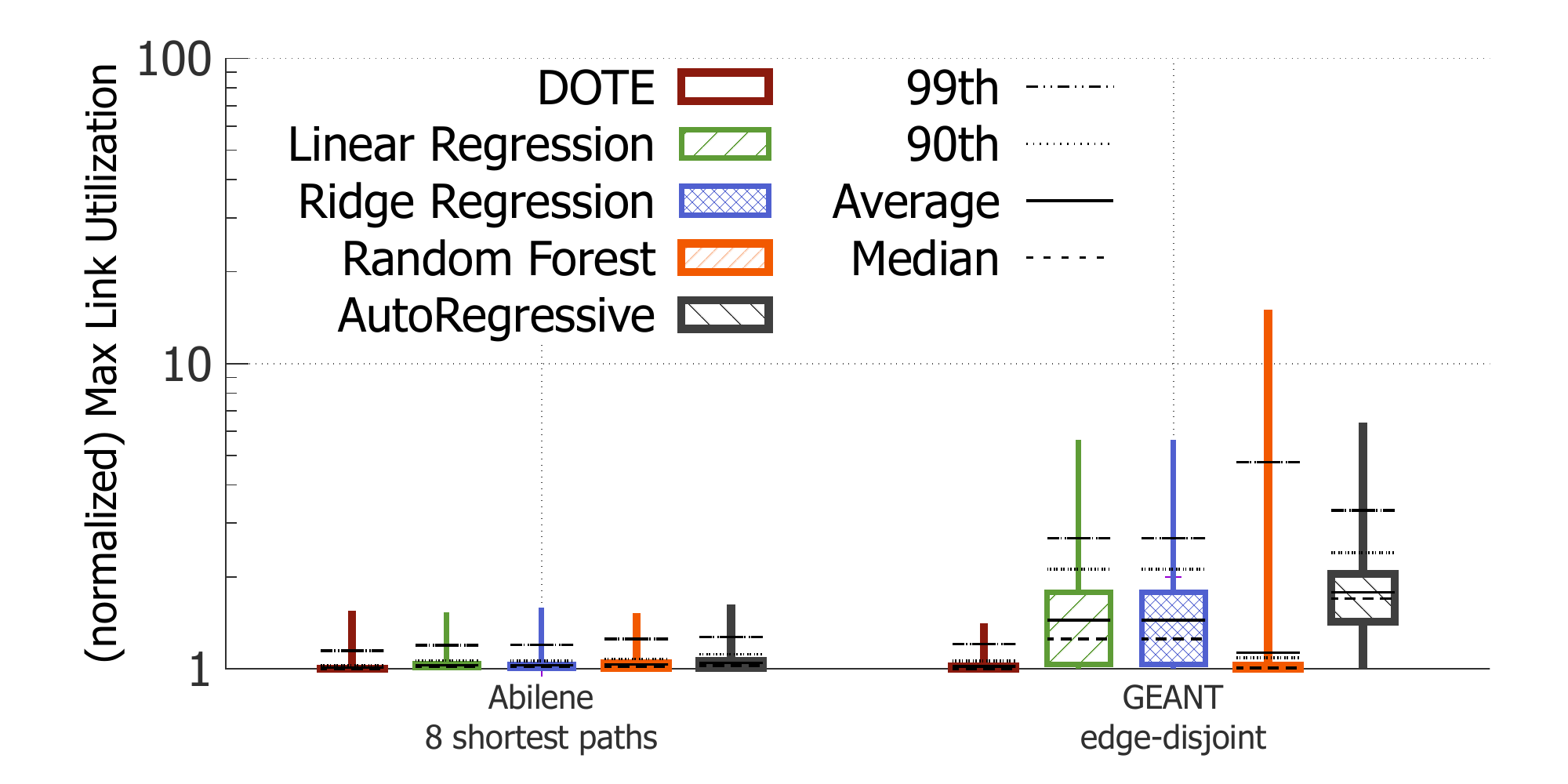}
\vspace{-0.15in}
\caption{Impact of demand prediction accuracy on max-link-utilization.}
\label{fig:sl_performance}
\end{figure}

Our results in~\xref{sec:eval} considered a demand-prediction-based scheme that utilizes linear regression. We next contrast linear regression with other prediction methods on our datasets. Specifically, we consider the following prediction methods: linear regression, ridge regressing, random forrest, and autoregressive model. With the exception of the autoregressive model, each of these schemes predicts the next traffic demand for each source-destination pair using only that specific pair's recently observed $12$ most traffic demands, \textit{i.e.}, the prediction for each pair is independent from the prediction for other pairs (as in SWAN~\cite{SWAN}). The autoregressive model, in contrast, predicts the entire next DM from the $12$ most recently observed DMs, to allow for detecting correlations between different pairs that might be conducive for prediction.

\autoref{fig:sl_accuracy} plots the accuracy of the different predictors, as quantified by the root-mean-squared-error, for the two publicly available WAN datasets. The accuracy is normalized by the average traffic demand for the dataset and presented in log-scale. Our results for \emph{PWAN} and \emph{PWAN$_{DC}$} exhibit similar trends. As shown in the figure, linear regression and ridge regression achieve the best results on average on both WANs. We also considered a DNN-based predictor with a single hidden layer with $128$ neurons and ReLU activation functions, but its performance was strictly dominated by linear regression on the test data (results omitted). Moreover, treating source-destination pairs individually attains better accuracy than that provided by the autoregressive model. We believe that this is because, on the one hand, the previous traffic demands for a single pair already contain a lot of valuable information and, on the other hand, the much larger input and output of the autoregressive model (entire DMs \textit{vs.} single demands) makes effective learning more difficult.

\autoref{fig:sl_performance} plots the implications of choosing different predictors for TE performance, as quantified by the max-link-utilization, benchmarked against \emph{DOTE}. Observe that \emph{DOTE} outperforms all considered flavors of demand-based-prediction TE, and also that accuracy in demand prediction does not always translate to better TE performance, exemplifying the potential objective mismatch between the two, discussed in the Introduction.

\section{Robustness to Unexpected Traffic Changes} \label{apx:robust}

We consider the GEANT, Cogentco, and GtsCe network topologies with edge-disjoint tunnels. For Cogentco, and GtsCe we use the gravity model to generate demands for both train and test. To evaluate the implications of unexpected traffic changes, we add noise to the test set by multiplying each demand independently by a factor sampled uniformly at random from the range $[1-\alpha, 1+\alpha]$ for $\alpha \in \{0.1, 0.25, 0.35\}$. 

Recall that for GEANT, \emph{DOTE} generates TE configurations that are extremely close to the optimum (less than $2\%)$. Our results show that even under random traffic perturbations, the distance from the omniscient oracle remains low; $2$\%, $2.9$\%, and $3.8$\% for $\alpha=0.1,0.25,0.35$, respectively. For $\alpha=0.35$, the distance from the omniscient oracle was $0.01$\% in the median, $13$\% in the 90th percentile, and no higher than $28$\% even in the 99th percentile.

For both Cogentco and GtsCe, \emph{DOTE}'s trained model is roughly $0.5\%$ from the omniscient oracle on the test demands are perturbed. This is because traffic is generated using the gravity model naturally does not reflect the intricate \textit{temporal} patterns and complexity of real-world traffic. Even after perturbing the traffic in our experiments \emph{DOTE} achieved near-optimal performance. Specifically, on Cogentco, the average distance from the omniscient oracle was $0.54$\%, $0.57$\%, and $0.6$\% for $\alpha=0.1,0.25,0.35$, respectively. For $\alpha=0.35$, the distance from the omniscient oracle was $0.56$\% in the median, $1$\% in the 90th percentile, and $1.4$\% in the 99th percentile. On GtsCe, the average distance from the omniscient oracle was $0.51$\%, $0.56$\%, and $0.61$\% for $\alpha=0.1,0.25,0.35$ respectively. For $\alpha=0.35$, the distance from the omniscient oracle was $0.57$\% in the median, $1$\% in the $90$th percentile, and $1.4$\% in the $99$th percentile.

\begin{table}[h]
    \centering
{\scriptsize
    \begin{tabular}{c|c||c|c|c|c}
          & {\bf Tunnels} & {\bf Week 1} & {\bf Week 2} & {\bf Week 3} & {\bf Week 4}\\
    \hline
         {\bf Abilene} & $8$ SP & $0.7$ & $0.3$ & $1.0$ & $1.5$ \\
    \hline
         {\bf Abilene} & edge-disjoint & $2.1$ & $2.4$ & $2.4$ & $2.0$ \\
    \hline
         {\bf GEANT} & $8$ SP & $1.4$ & $2.7$ & $2.9$ & $3.1$ \\
    \hline
         {\bf GEANT} & edge-disjoint & $0.7$ & $1.6$ & $2.0$ & $2.5$ \\
    \end{tabular}
}
    \caption{\label{t:weekly_dist_from_opt_mlu}Average weekly distance from the omniscient oracle achieved by {\sysname} for MLU across $4$ consecutive weeks}
    \label{tab:network_sizes}
\end{table}

\begin{table}[h]
\centering
{\scriptsize
    \begin{tabular}{c|c||c|c|c|c}
          & {\bf Tunnels} & {\bf Week 1} & {\bf Week 2} & {\bf Week 3} & {\bf Week 4}\\
    \hline
         {\bf Abilene} & $8$ SP & $1.6$ & $2.1$ & $3.9$ & $6.2$ \\
    \hline
         {\bf Abilene} & edge-disjoint & $1.1$ & $1.4$ & $3.1$ & $5.5$ \\
    \hline
         {\bf GEANT} & $8$ SP & $4.9$ & $4.7$ & $5.0$ & $4.8$ \\
    \hline
         {\bf GEANT} & edge-disjoint & $6.3$ & $6.8$ & $6.9$ & $6.4$ \\
    \end{tabular}
}
    \caption{\label{t:weekly_dist_from_opt_max_mcf}Average weekly distance from the omniscient oracle achieved by {\sysname} for maximum-multicommodity-flow across $4$ consecutive weeks}
    \label{tab:network_sizes}
\end{table}

\section{Stochastic Optimization Loss Function Pseudocode}\label{sec:description_loss}

\begin{algorithm}\label{description_loss}
\floatname{algorithm}{Function}
\caption{Stochastic Optimization Loss Function Pseudocode}
\begin{algorithmic}

  \State $G = (V, E, c)$ \ // \textit{capacitated directed graph that models the WAN topology}
  \State $U = \{(i,j) | i \in V, j \in V, i \neq j \}$\ // \textit{all pairs of nodes}
  \State $T = \cup_{(s,t) \in U}P_{s,t}$ \ // \textit{the set of all tunnels}
  \State $A^{|U| \times |T|}$ \ // \textit{specifies, for each pair of nodes $i\in U$ and tunnel $j\in T$ whether tunnel $j$ interconnects the nodes in $i$}

  \State $A_{i,j} =$
    $\begin{cases}
      1 & j\in P_i\\
      0 & \text{otherwise}
    \end{cases}$

    \State $B^{|T|\times|E|}$ \ // \textit{specifies, for each tunnel $i$ and edge $j$, whether tunnel $i$ contains edge $e$}
    
  \State $B_{i,j} =$
    $\begin{cases}
      1 & j\in i\\
      0 & \text{otherwise}
    \end{cases}$
  
  \State $C^{|E|\times1}$ // \textit{vector representing WAN link capacities} 
  \State $C_{i,1} = c(i)$
\\
\Function{Loss}{$DNN_{output}, DM_{next}$}
  \State $DNN_{output}^{|T| \times 1}$ // \textit{the output of the DNN}
  \State $DM_{next}^{|U| \times 1}$ // \textit{the (actual) next demand matrix}
  \State
  \State // $\times$ and $/$ are element-wise operations
  \State // 1. Compute the splitting ratios
  \State $PathsSplit^{|T| \times 1} = DNN_{output} \times (A^T(1.0 / A\times DNN_{output}))$
  \State // 2. Calculate the flow on each edge
  \State $FlowOnEdges^{|E| \times 1} = B^T((A^T\times DM_{next}) \times PathsSplit)$
  \State // 3. Compute the maximum-link-utilization
  \State $MaxLoad = max(FlowOnEdges / C)$
  \State \Return $MaxLoad$

\EndFunction
\end{algorithmic}
\end{algorithm}

\begin{figure*}[t!]
\subfigure[Abilene]{
  \includegraphics[width=.32\linewidth]{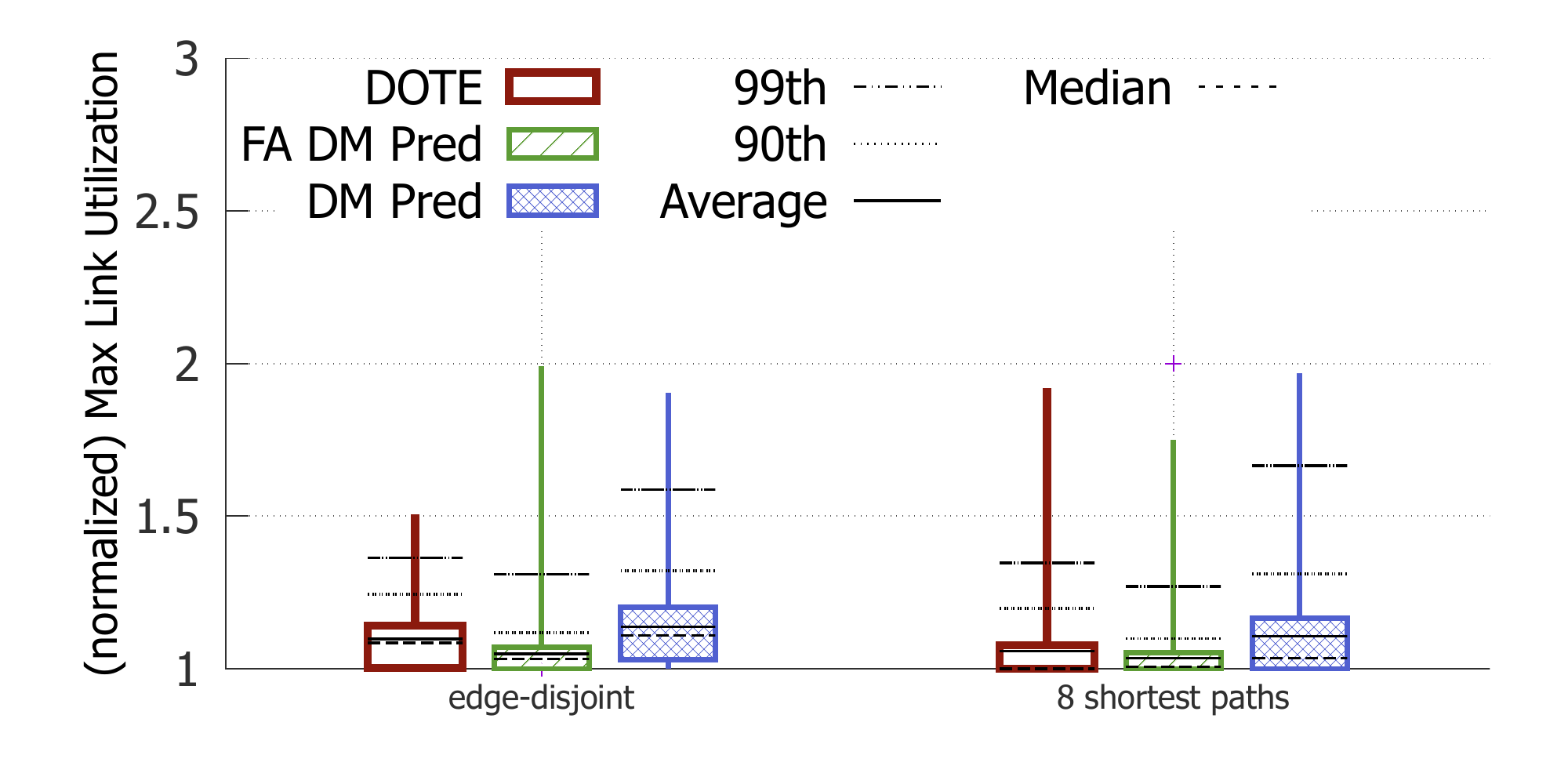}
  }
\subfigure[GEANT]{
  \includegraphics[width=.32\linewidth]{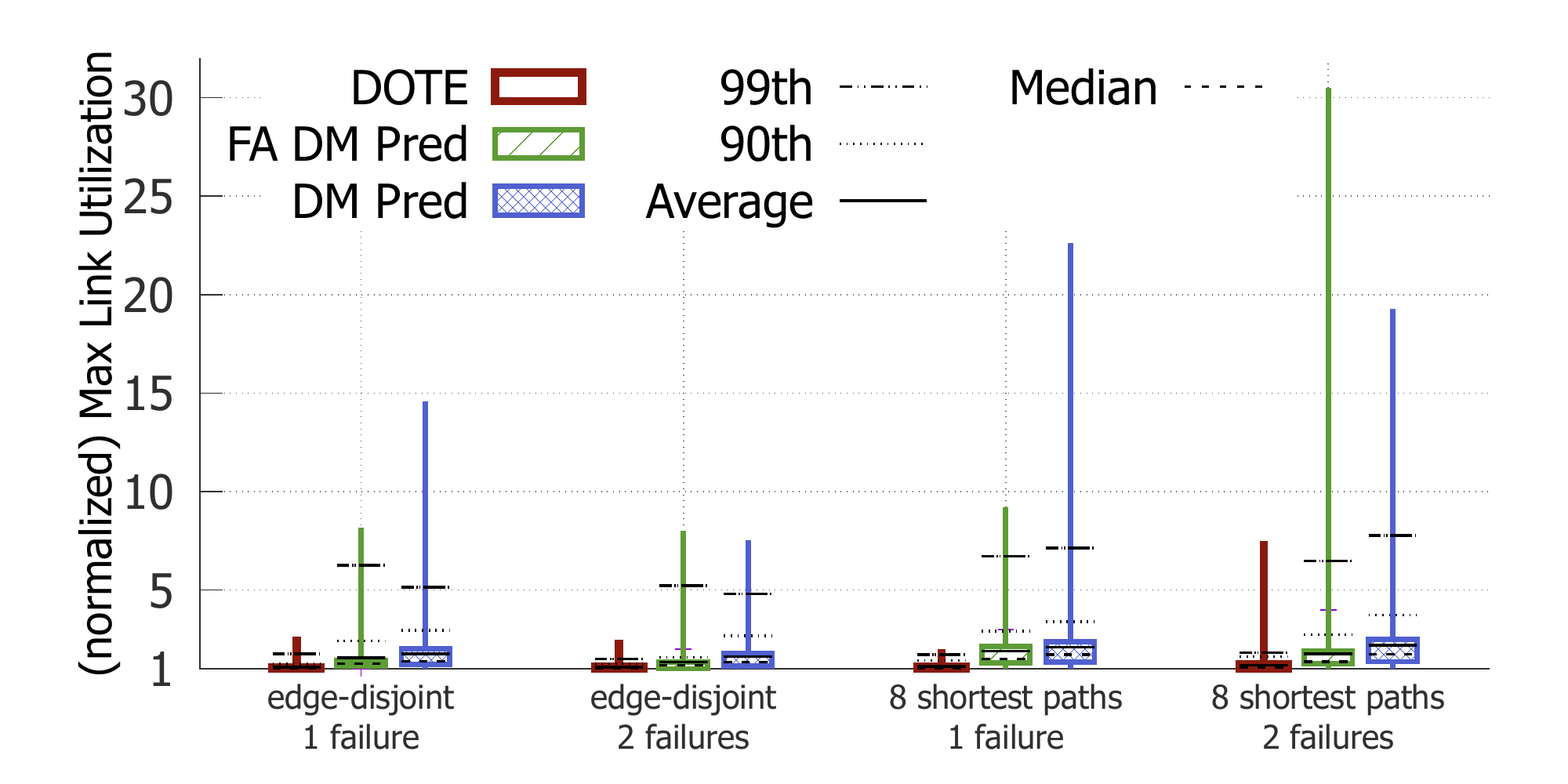}
  }
\subfigure[{\pwandc}]{
  \includegraphics[width=.32\linewidth]{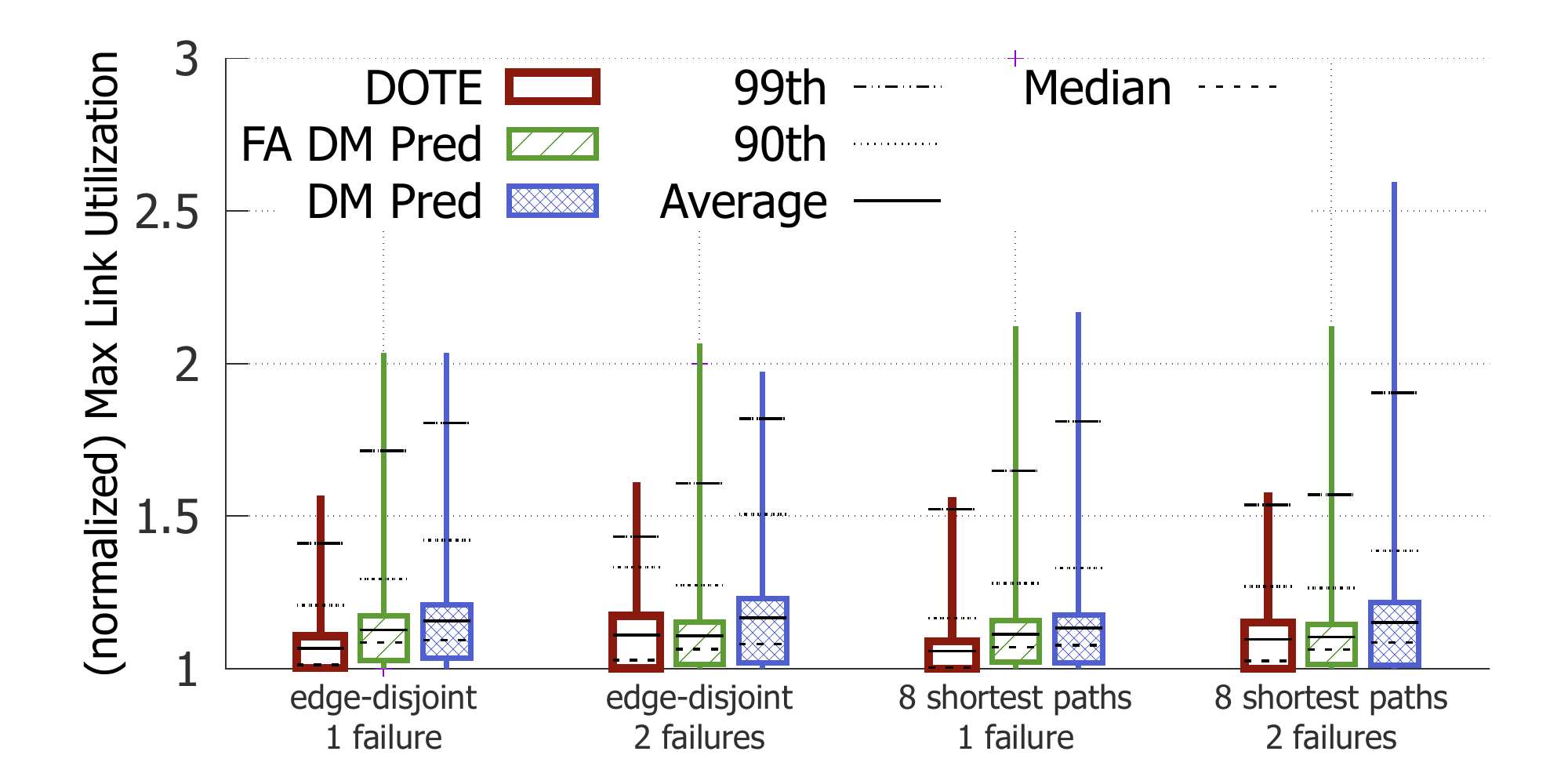}
  }
\vspace{-0.1in}
\caption{Understanding the behavior of {\sysname} under failures on different WAN datasets. The results are qualitatively similar to~\autoref{f:failures_pwan}.\label{f:failures_other_topologies}}
\end{figure*}

\begin{figure*}[t!]
\subfigure[8SP]{
  \includegraphics[width=.47\linewidth]{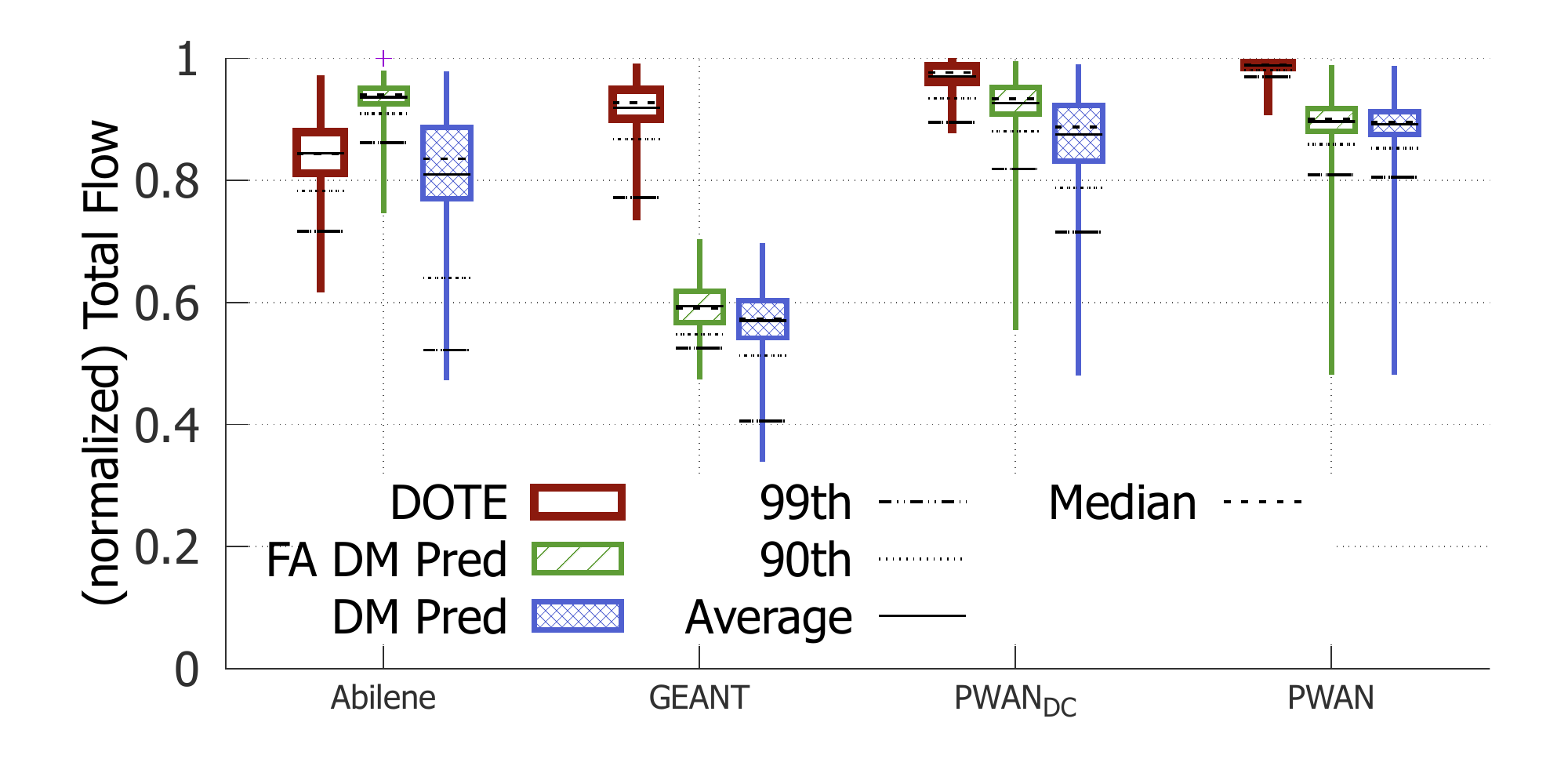}
  }
\subfigure[edge-disjoint]{
  \includegraphics[width=.47\linewidth]{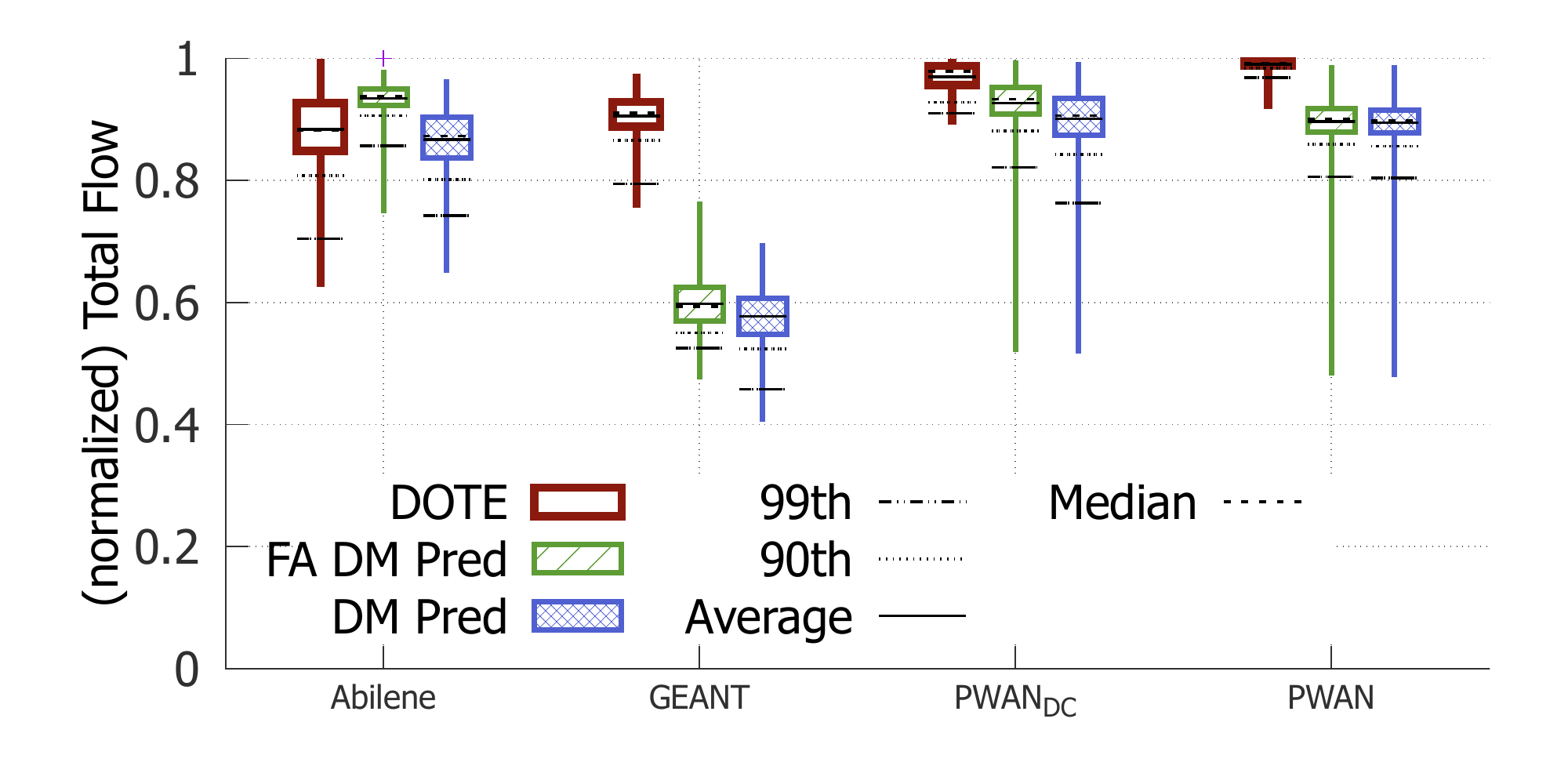}
  }
\vspace{-0.1in}
\caption{Coping with a random link failure when aiming to maximize the total flow for two different tunnel choices.}
\label{f:failures_max_mcf}
\end{figure*}

\section{Additional Failure Results}
\label{ss:additional_failure}
Analogous to~\autoref{f:failures_pwan}, \autoref{f:failures_other_topologies} shows the behavior under faults for the Abilene, GEANT and {\pwandc} topologies respectively.
\autoref{f:failures_max_mcf} shows the results for maximum-multicommodity-flow.

\end{document}